\documentclass{article}

\usepackage[text={173 mm,25cm},centering,letterpaper]{geometry}
\usepackage{graphicx}
\usepackage[margin = 10pt, font=small, labelfont=bf, labelsep = endash]{caption}
\usepackage[pdfpagemode=UseOutlines, hidelinks]{hyperref}
\usepackage{apacite}

\begin{document}

\title{Multivariate Estimations of Equilibrium Climate Sensitivity from Short Transient Warming Simulations}

\author{Robbin Bastiaansen\thanks{Institute for Marine and Atmospheric research Utrecht, Department of Physics, Utrecht University, Utrecht, The Netherlands (r.bastiaansen@uu.nl, h.a.dijkstra@uu.nl, a.s.vonderheydt@uu.nl)} , Henk A. Dijkstra\footnotemark[1] \thanks{Centre for Complex System Studies, Department of Physics, Utrecht University, Utrecht, The Netherlands} , Anna S. von der Heydt\footnotemark[1] \footnotemark[2] }

\maketitle

\begin{abstract}
	One of the most used metrics to gauge the effects of climate change is the equilibrium climate sensitivity, defined as the long-term (equilibrium) temperature increase resulting from instantaneous doubling of atmospheric CO$_2$. Since global climate models cannot be fully equilibrated in practice, extrapolation techniques are used to estimate the equilibrium state from transient warming simulations. Because of the abundance of climate feedbacks -- spanning a wide range of temporal scales -- it is hard to extract long-term behaviour from short-time series; predominantly used techniques are only capable of detecting the single most dominant eigenmode, thus hampering their ability to give accurate long-term estimates. Here, we present an extension to those methods by incorporating data from multiple observables in a multi-component linear regression model. This way, not only the dominant but also the next-dominant eigenmodes of the climate system are captured, leading to better long-term estimates from short, non-equilibrated time series.
\end{abstract}

\section{Introduction}
The use of (equilibrium) climate sensitivity to assess the impact of changes in atmospheric CO$_2$ dates back at least a century \cite{Arrhenius1896, aboutArrhenius}. First, estimations of its value were made with rudimentary computations \cite{Arrhenius1896,Charney1979}; nowadays, improved knowledge of the climate system is used to infer climate sensitivity from observational data, proxy data, and global climate models \cite{knutti2008equilibrium, anna2016lessons, rohling2018comparing, lunt2010earth, knutti2017beyond}. However, the reported values (still) vary much between studies~\cite{IPCCreport} and the current consensus is that climate sensitivity is between $2.3 K$ and $4.7 K$ \cite<5\%-95\% ranges,>{Sherwood2020}. On top of that, recent results of the new generation of global climate models show even higher sensitivities, possibly due to better representation of cloud formation when using finer spatial grids~\cite{Bacmeister2020, Zelinka2020, Andrews2019, bony2015clouds, duffy2003high, govindasamy2003high, haarsma2016high}. Still, even these state-of-the-art climate models report significantly different climate sensitivities~\cite{flynn2020climate, Zelinka2020, Forster2020}; moreover, estimates for a single model tend to have large uncertainties further hampering accurate pinpointing of the climate sensitivity~\cite{Rugenstein2020, Dai2020}.

For conceptual models and earth system models of intermediate complexity, it is possible to let a simulation run until the system is fully equilibrated~\cite{holden2014plasim}. However, for more refined models, including contemporary and future state-of-the-art global climate models, this is not viable~\cite{Rugenstein2019}; equilibrating those models simply takes too much computing power. With the current trend -- and need -- to build models with higher temporal and spatial resolutions~\cite{eyring2016overview, duffy2003high, govindasamy2003high, haarsma2016high}, this is not expected to resolve itself in the near future. Hence, the equilibrium climate sensitivity of these models is instead estimated by extrapolating transient warming simulations -- way before these models have reached equilibrium~\cite{knutti2008equilibrium, Rugenstein2020, Dai2020, knutti2017beyond}. There are several techniques to perform such extrapolation that use different physical and mathematical properties of the system to give sensible estimates for the true equilibrium climate sensitivity of a model~\cite{knutti2008equilibrium, Dai2020, gregory2004new, geoffroy2013transient, proistosescu2017slow}.

The main problem with equilibrium estimations lies with the abundance of feedbacks present in the climate system~\cite{anna2016lessons, annaquantification}. These feedbacks are quite diverse and span a wide range of spatial and temporal scales; these include, for example, the very fast Planck feedback, the slower ice-albedo feedback and the even slower ocean circulation feedbacks. Estimation techniques deal differently with this problem, for instance by incorporating multiple time scales directly in the estimation method~\cite{proistosescu2017slow}, by explicit modelling of long-term (ocean) heat uptake~\cite{geoffroy2013transient} or more indirectly by ignoring initial fast warming behaviour~\cite{Rugenstein2020,Dai2020}.

The most predominantly used estimation technique is the one developed by \citeA{gregory2004new}. In this technique, the top-of-atmosphere radiative imbalance ($\Delta R$) is fitted using a linear regression against the temperature increase ($\Delta T$). However, recently, it has become clear that $\Delta R$ and $\Delta T$ do not always adhere to such linear relationship~\cite{Andrews2012, armour2017energy, knutti2015feedbacks}. Typically, there is an initial fast warming which is followed by one or several slower additional (less substantial) warming processes. Hence, estimates made by this method depend heavily on the time period used in the regression and typically underestimate the equilibrium warming. Most of the times, this problem is largely circumvented by ignoring the first part of a simulation that contains the initial fast processes; the regression is then applied only on the last part of the simulation.

The thus ignored data does however still contain information about the dynamics of the system -- even beyond the initial fast warming. The issue here is that this information cannot be extracted using a one-dimensional linear regression; that kind of fit will only ever recover the one process (i.e. one eigenmode of decay to equilibrium) that is most dominantly present on the time scale of the regression data. In this paper, we present an extension to this technique that is capable of capturing multiple eigenmodes by incorporating additional observables into a multi-component linear regression (abbreviated as MC-LR) model. Subsequently, we show the potential efficiency of this technique using both low dimensional conceptual models and modern global climate models.

\section{Method: a Multi-Component Linear Regression Model}

In the linear regime of the decay to equilibrium, the evolution of any observable $O$ (e.g. global mean temperature increase or top-of-atmosphere radiative imbalance) is given by the sum of exponentials (i.e. the eigenvalue decomposition), capturing the behaviour on different time scales based on the different eigenmodes of the system. Specifically, denoting the equilibrium value of an observable by $O_*$, this evolution follows

\begin{equation}
	O(t) -  O_* = \sum_j \beta^{[O]}_j e^{\lambda_j t}, \hspace{2cm} \left( \sum_j \beta^{[O]}_j = O(0) - O_* \right)
\end{equation}

where $\lambda_j$ denote the eigenvalues and $\beta^{[O]}_j$ the contributions of each eigenmode to the evolution of the observable $O$.

If only one eigenmode would be present (or relevant, as other eigenmodes are exponentially small on the time scale of the data), the evolution of the global mean surface temperature increase $\Delta T$ and the top-of-atmosphere radiative imbalance $\Delta R$ can be combined into the linear relation

\begin{equation}
	\Delta R - \Delta R_* = \frac{ \beta^{[\Delta R]}_1 }{ \beta^{[\Delta T]}_1 } \left( \Delta T - \Delta T_* \right).
\end{equation}

Since $\Delta R_* = 0$, this readily gives rise to the commonly used regression model by \citeA{gregory2004new},

\begin{equation}
	\Delta R = a \Delta T + f
\end{equation}

where $a := \frac{\beta_1^{[\Delta R]}}{\beta_1^{[\Delta T]}}$ and $f:= - a \Delta T_*$ are to be determined from the used regression data. In this case, the equilibrium warming is estimated by $\Delta T_*^\mathrm{est} := - \frac{1}{a} f$.

If multiple eigenmodes are relevant, there no longer is such linear relationship between $\Delta R$ and $\Delta T$ (as time $t$ cannot be eliminated from the equations anymore) and this technique breaks down. It is, however, possible to extend the technique by taking additional observables into account: if $N$ eigenmodes are relevant, one must use two sets of $N$ observables, denoted here by $\vec{X}$ and $\vec{Y}$; using a similar procedure, the equations for their evolutions can be combined together (e.g. using basic matrix computations) to obtain the linear relation

\begin{equation}
	 \vec{Y} - \vec{Y}_* = A \left( \vec{X} - \vec{X}_* \right),
\end{equation}

where $A$ is a $N \times N$ matrix. If the set of observables in $\vec{Y}$ only contains observables that tend to zero in equilibrium (i.e. $Y_* = 0$), this gives rise to a new multi-component linear regression model

\begin{equation}
	\vec{Y} = A \vec{X} + \vec{F},
\end{equation}

with $A$ and $\vec{F} := - A \vec{X}_*$ to be determined by the regression data. Here, equilibrium estimates are given by the vector $\vec{X}_*^\mathrm{est} := - A^{-1} \vec{F}$ and contain equilibrium estimates for all observables in $\vec{X}$. 

The method by \citeA{gregory2004new} is a special example of this regression model, where $N = 1$, $\vec{X} = \Delta T$ and $\vec{Y} = \Delta R$. Here, this model is extended by adding one or two observables to the data vectors $\vec{X}$ and $\vec{Y}$ (i.e. $N = 2$ or $N=3$, lining up with previous studies by~\citeA{caldeira2013projections, tsutsui2017quantification, proistosescu2017slow}). Specifically, the mean global effective top-of-atmosphere short-wave albedo $\alpha$ and long-wave emissivity $\varepsilon$ are considered as additional observables. Their values are added to the data vector $\vec{X}$ and the values of their (numerical) time-derivatives -- that tend to zero in equilibrium -- to $\vec{Y}$. The fits in this study are all made using standard least squares regression.

A different and much more extensive take on the rationale behind the technique can be found in Supporting Information Text S1.

\section{Results: Conceptual Models}

First, we present the results on a variant of the conceptual Budyko-Sellers energy balance model for global mean surface temperature~\cite{budyko1969effect,sellers1969global}. This model has been extended such that albedo and emissivity are no longer instantaneous processes, but will settle slowly over time. Moreover, white noise has been added to simulate climate variability. Thus, a three-component stochastic ordinary differential equation is created, which has been simulated in MATLAB with an Euler-Maruyama scheme. A more extensive description of the model can be found in Supporting Information Text S2.

\begin{figure}

\centering
	\includegraphics[width=0.475 \textwidth]{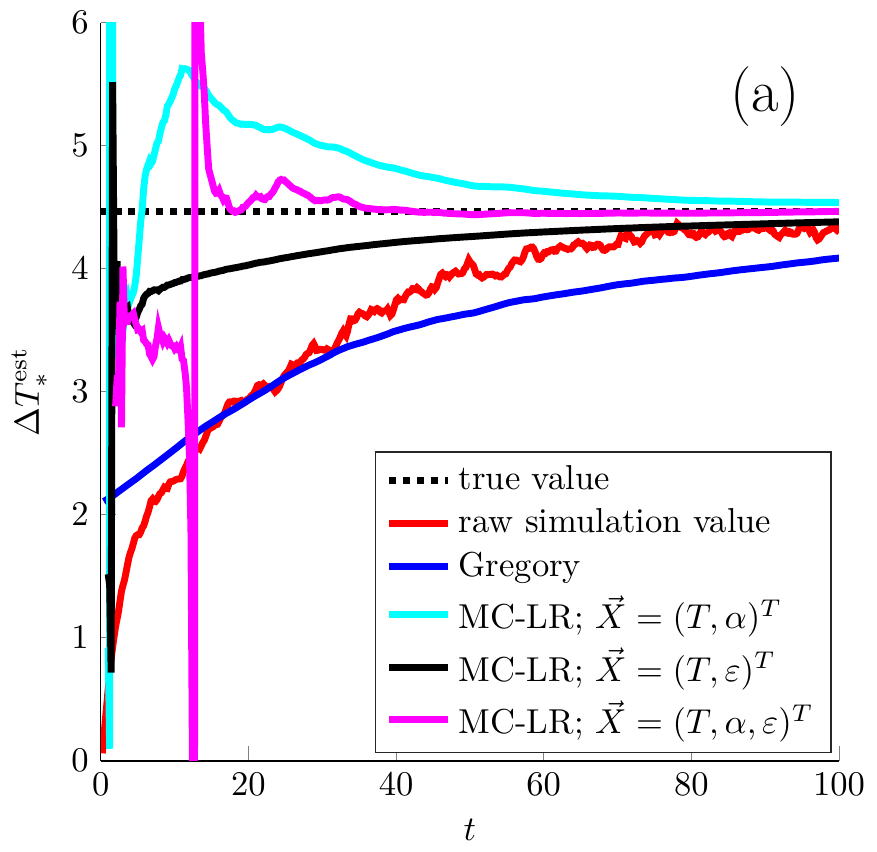}
~
	\includegraphics[width=0.475 \textwidth]{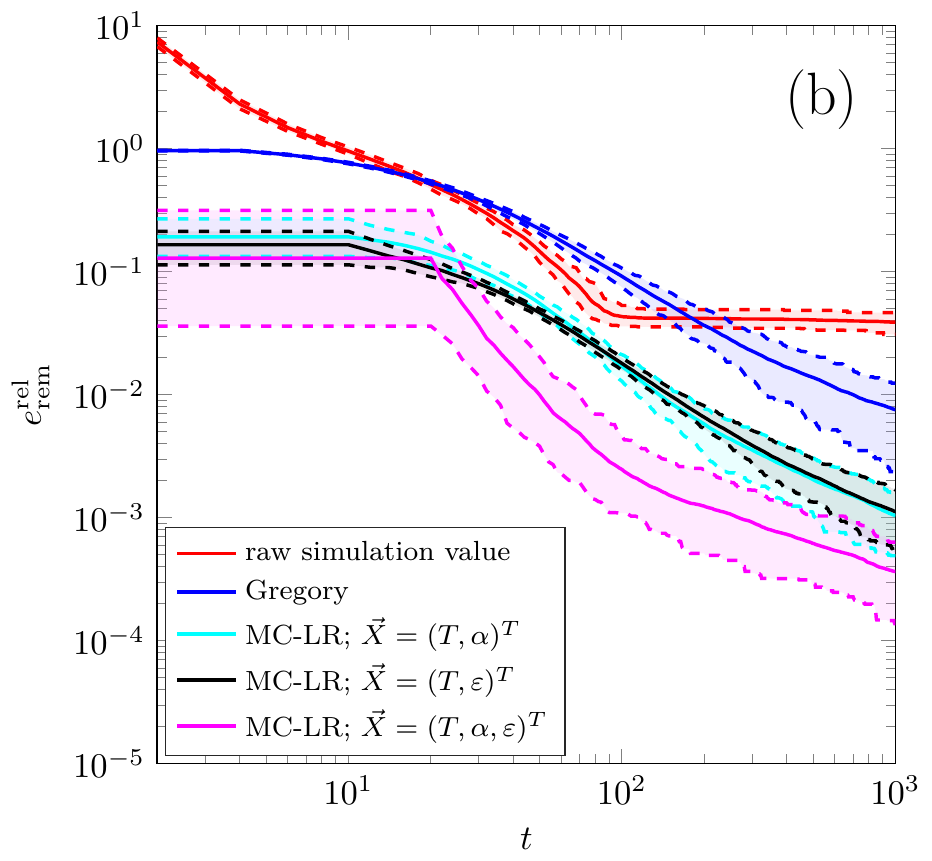}
\caption{Results from various estimation techniques on a conceptual ($3$-component) global energy budget model with moderate noise (noise strength $\nu = 0.5$ -- see Supporting Information Text S2 for details). (a) Estimates $\Delta T_*^\mathrm{est}(t)$ for a single model realisation; the value at time $t$ gives the estimation when only data up to time $t$ is used. (b) Evolution of expected remaining error $e_\mathrm{rem}^\mathrm{rel}(t)$ over time based on an ensemble of one hundred runs; solid lines indicate mean values and dashed lines the $5$ and $95$ percentile values. Results on models with different noise levels can be found in Supporting Information Figures S3 and S4.}
\label{fig:conceptualModelResults}
\end{figure}

Output of this model has been analyzed using the previously described MC-LR technique with the use of some or all of the observables. The resulting estimates for the equilibrium climate increase $\Delta T^\mathrm{est}_*(t)$ are given in Figure~\ref{fig:conceptualModelResults} for simulation runs with moderate noise (figures for other noise levels can be found in Supporting Information Figures S3 and S4). These estimates are given as functions of model time: the value for time $t$ indicates the estimate is made with model output up to time $t$ only. To evaluate the various estimation techniques and track their accuracy depending on the amount of data used, the remaining relative error is computed: the maximum in relative error of the estimates occurring after the current time (i.e. when more data points are used). This gives a better impression of the kind of error to expect when using data up to time $t$. Mathematically, the remaining error is defined as

\begin{equation}
	e_\mathrm{rem}^\mathrm{rel}(t) := \max_{s \geq t} \left| \frac{ \Delta T_*^\mathrm{est}(s) - \Delta T_*}{ \Delta T_*^\mathrm{est}(s) } \right|,
\end{equation}
where $\Delta T_*$ is the true equilibrium warming (determined numerically via Newton's method).

For this kind of low-dimensional models, it is clear that the multi-component linear regression leads to better estimations of the real equilibrium warming than conventional techniques (Figure~\ref{fig:conceptualModelResults}). Although the estimations for very short time series are not very accurate, estimations for slightly longer time series quickly pick up and are much better compared to the linear `Gregory' fit  (Figure~\ref{fig:conceptualModelResults}a), because also the longer time dynamics are taken into account (and are accurately fitted; see Supporting Information Text S2 and Figure S5). It takes some tens of (arbitrary) time units for the new estimates to get within $0.1 K$ of the actual equilibrium value, whereas hundreds of time units are needed for the conventional technique (Figure~\ref{fig:conceptualModelResults}b). Moreover, it also seems that the MC-LR technique still works reliable in case of noise.

\section{Results: LongRunMIP Models}

The MC-LR technique has also been tested on more detailed global climate models. Specifically, data is taken from abrupt $4\times\mathrm{CO}_2$ forcing experiments of models participating in LongRunMIP, a model intercomparison project that focuses on millennia-long simulation runs~\cite{Rugenstein2019}. Because of these long time series, a relative accurate value for the true equilibrium temperature can be determined, which is needed to adequately assess the performance of the estimation techniques.

For these climate models, global data on near-surface atmospheric temperature ($T$ = `tas') and top-of-atmosphere radiative fluxes (incoming short-wave, `rsdt', outgoing short-wave, `rsut' and outgoing long-wave, `rlut') has been downloaded from the LongRunMIP data server~\cite{Rugenstein2019}. These datasets have been used to compute top-of-atmosphere radiative imbalance ($R$ = `rsdt' - `rsut' - `rlut'), effective short-wave albedo ($\alpha$ = `rsut' / `rsdt') and effective long-wave emissivity ($\varepsilon \sigma$ = `rlut' / (`tas')$^4$; where $\sigma$ is the Stefan–Boltzmann constant). Initial, non-forced values were defined as means of piControl runs and changes $\Delta T$, $\Delta R$, $\Delta \alpha$ and $\Delta \varepsilon \sigma$ were computed from the abrupt $4\times$CO$_2$ forcing runs. The real equilibrium warming $\Delta T_*$ for these models was estimated from the last warming of the forcing experiments, following the approach taken in~\citeA{Rugenstein2020}. A more detailed description of these procedures, including minor practical variants, can be found in Supporting Information Text S3.

\begin{figure}[t!]
\centering
\includegraphics[width = 0.85\textwidth]{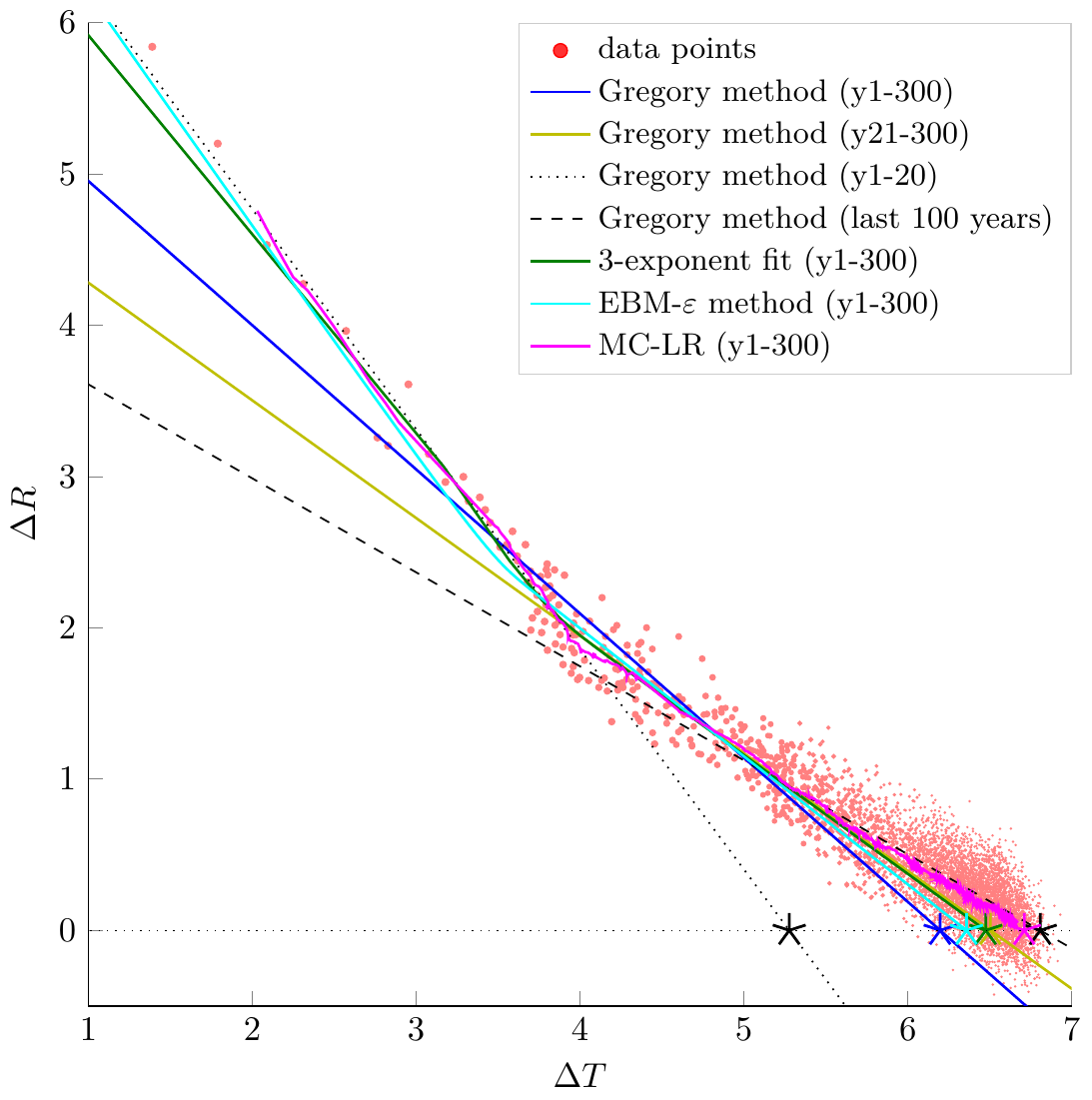}
\caption{Gregory plot of $\Delta R$ as function of $\Delta T$ for a $5,900$ year abrupt CO$_2$ quadrupling experiment in the CESM 1.0.4 model along with results of various common equilibrium estimation methods and the here introduced multi-component linear regression (MC-LR) method when used on data up to model year $300$. In the plot, red dots denote all data points (the later in the run, the smaller in size). The blue line shows the linear `Gregory' fit when all data from years $1$ to $300$ is used and the yellow line the Gregory fit when the first $20$ years are ignored. The green line shows the $3$ exponent fit \protect\cite{proistosescu2017slow}. The cyan line indicates a fit to the EBM-$\varepsilon$ model that includes ocean heat uptake \protect\cite{geoffroy2013transient}. The magenta line visualises the newly introduced multi-component linear regression that, in this case, utilises both albedo and emissivity (for this visualisation only -- and not for any of the fits in this paper -- averaged data from the experiment are used). The stars ($\star$) are the estimated equilibrium warming values from the different methods. Finally, dotted and dashed black lines indicate linear Gregory fits for the first and last part of the simulation that can be used for comparison -- and that show how the various estimation methods capture dynamics on multiple time scales.}
\label{fig:GregoryPlotNonLinear}
\end{figure}

With the use of the model output, various techniques have been used to estimate equilibrium warming for all models. In Figure~\ref{fig:GregoryPlotNonLinear}, a Gregory $(\Delta T, \Delta R)$-plot is given along with results of commonly used estimation techniques for one of the models (CESM 1.0.4) when applied on data up to model year $300$. This illustrates the capabilities of the various techniques in capturing the behaviour of the model system over different time scales. Clearly, the classical Gregory method mainly captures initial fast warming from the data. Hence, it is common practice to ignore an arbitrary number of years from the start of the simulation run -- that show the initial fast warming -- in a Gregory fit~\cite{Rugenstein2020,Dai2020}. That technique has also been tested here, where the initial $20$ years have been excluded. In contrast, the multi-component linear regression technique does not rely on such arbitrary choices for data selection and outperforms both of these classical methods. Certainly, there also exist other alternative estimation techniques that aim to extract long-term behaviour from short simulation runs (of which two have been added to Figure~\ref{fig:GregoryPlotNonLinear}). However, these often amount to fitting an explicit low-dimensional model to transient simulations \cite<e.g.>{geoffroy2013transient} and/or a non-linear regression \cite<e.g.>{proistosescu2017slow}. The proposed MC-LR method does neither -- and furthermore seems to perform similar or better than the mentioned other methods.

\begin{figure}
\centering
	\includegraphics[width=0.475\textwidth]{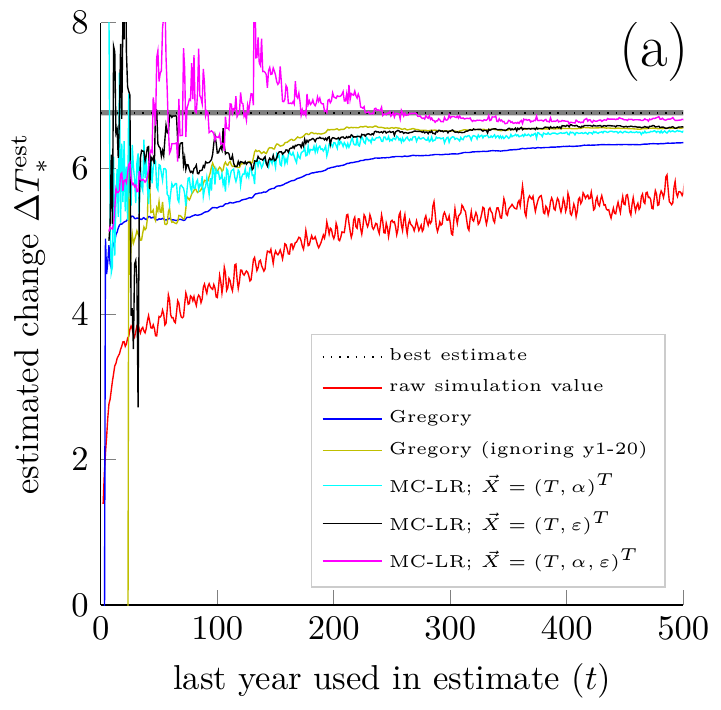}
~
	\includegraphics[width=0.475\textwidth]{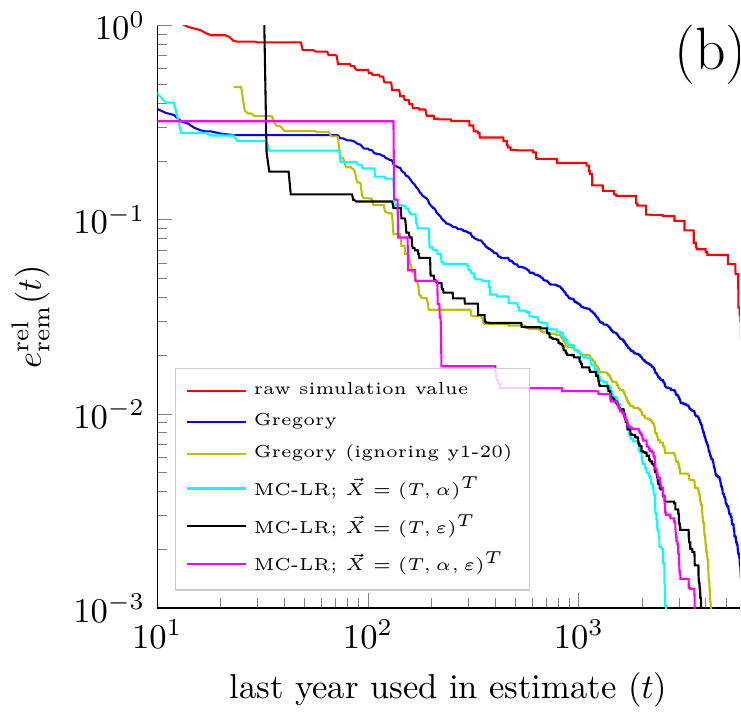}
\caption{Performance of equilibrium warming estimation techniques on the model CESM 1.0.4. (a) Estimated equilibrium temperature increase $\Delta T_*^\mathrm{est}(t)$ for various estimators as function of time $t$, where only model output data up to time $t$ has been used. The shaded region indicates the range of likely values for the system's true equilibrium warming $\Delta T_*$, along with the best estimate for this (dashed line), based on end-of-simulation data (see the Supporting Information Text S3.4). Only the first 500 years of the 5900 year simulation are shown. (b) Plot of the remaining relative error for estimation methods based on the whole simulation run. This shows the kind of error to expect when using a certain estimation technique on a given time scale. Full set of results for this and the other models can be found in Supporting Information Figures S8-S18.}
\label{fig:longrunmipResults1}
\end{figure}

The results for other time frames are shown in Figure~\ref{fig:longrunmipResults1}. Here, as before, estimates $\Delta T_*^\mathrm{est}(t)$ are functions of time, which only use data up to a given time $t$ for the estimation, and remaining relative errors have been computed as well. These results show that the MC-LR method also performs better on other time frames; in particular, when data for more than 150 years is being used, a multi-component linear regression that utilises both albedo and emissivity leads to better estimates compared to the classical Gregory methods. Especially on a century time scale this leads to significant improvements. Detailed results for all models can be found in Supporting Informaiton Figures S8-S18.

\begin{figure}
\centering

	\includegraphics[width=0.8\textwidth]{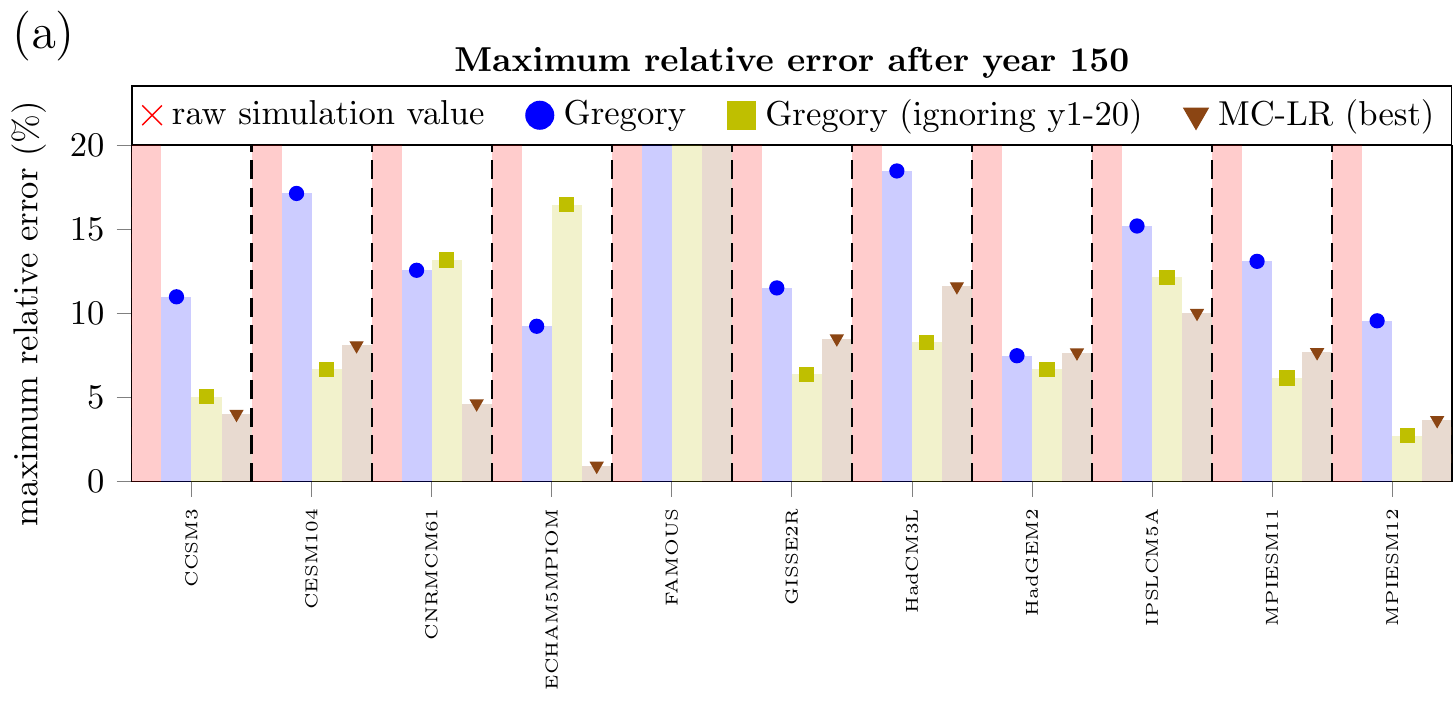}

	\includegraphics[width=0.8\textwidth]{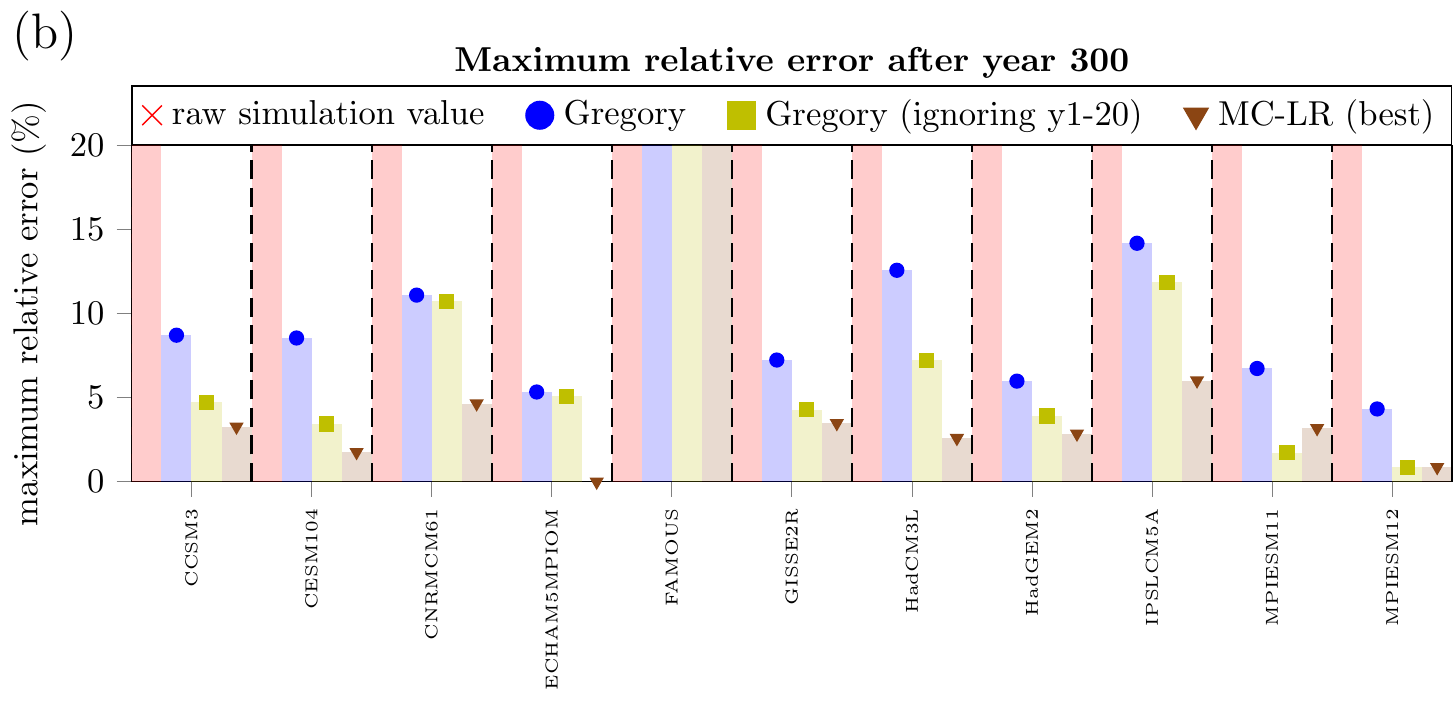}
	
	\includegraphics[width=0.8\textwidth]{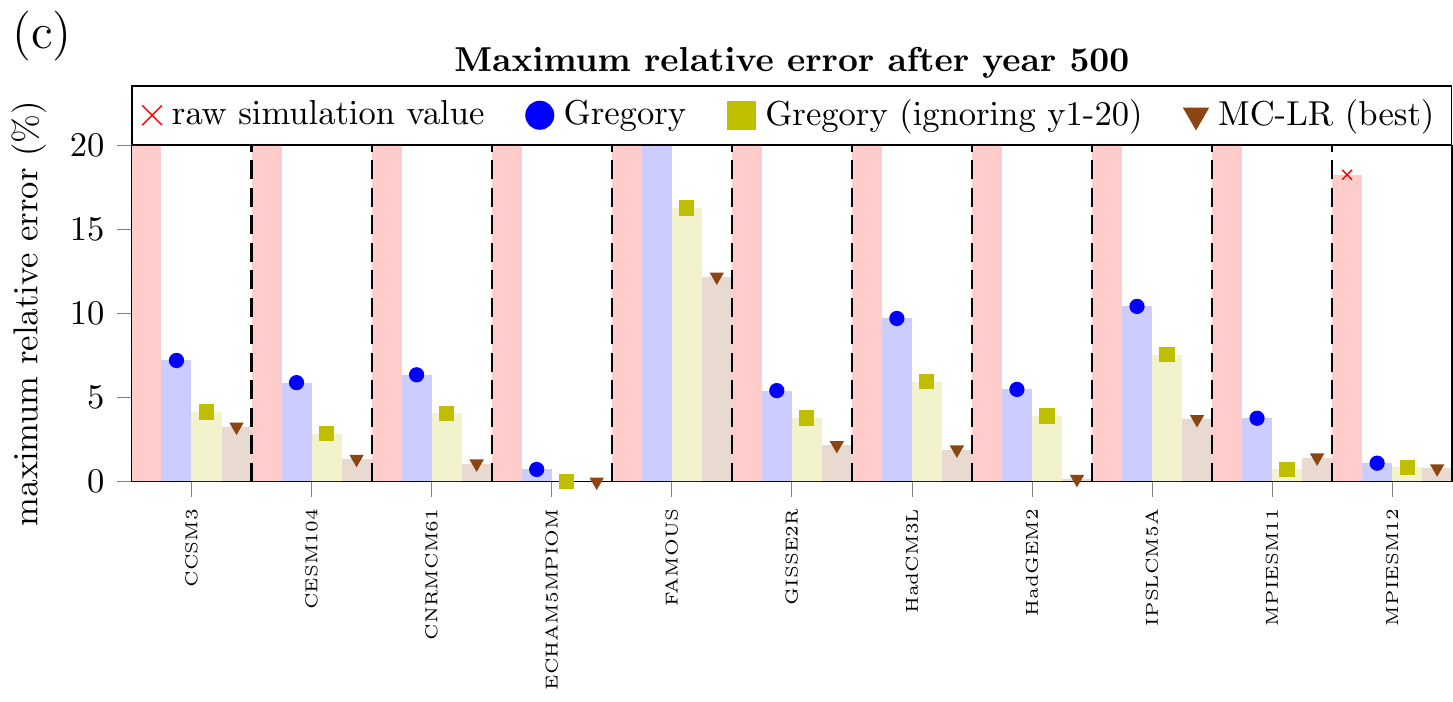}

\caption{Remaining relative error for various estimation techniques when used on model output up to (a) $t = 150 y$, (b) $t = 300 y$ and (c) $t = 500 y$. Here, only the best MC-LR method is depicted for each model (because of differences in model dynamics, which observables yield the best estimates differs per model). A complete list for all variants of the estimation techniques can be found in Supporting Information Figure S6, and a scatter plot of the results in this figure can be found in Supporting Information Figure S5.}
\label{fig:longrunmipResults2}
\end{figure}

To further disseminate the results and to assess the effectiveness over the range of models, in Figure~\ref{fig:longrunmipResults2} the remaining errors are given for all considered models at given times $t = 150$ years (CMIP protocol,~\citeA{eyring2016overview}), $t = 300$ years and $t = 500$ years. These results indicate that the MC-LR method can lead to more accurate equilibrium warming estimates. This new approach also better captures the long-term dynamics than the classical Gregory method when used on all data (with the HadGEM2 model for $t = 150$ years being the exception, where performance is similar). Moreover, the MC-LR method also tends to outperform the Gregory method that ignores the first $20$ years of data when $t > 150$ years. For $t = 150$ years, results vary much per model. This is closely related to the difference in model behaviour: if dynamics happen on two dominant time scales, and the Gregory plot has an inflection point around (the arbitrarily chosen) year $20$, this Gregory method works well (see for example the model MPI-ESM 1.1); otherwise, the MC-LR method will (eventually) outperform it. A more in-depth discussion per model is included in Supporting Information Text S3.5.

\section{Discussion}

In this paper, we have introduced a new equilibrium climate sensitivity estimation technique -- the multi-component linear regression (MC-LR) -- that better captures the long-term behaviour compared to conventional techniques. This MC-LR method has one prime rationale: a perturbed climate system evolves according to a linear system (given that the radiative perturbation is small). This linear evolution is recovered through the multi-component linear regression (i.e. regression to $\vec{Y} = A \vec{X} + \vec{F}$). Although, here, only data from one transient simulation is used in the fits, data from multiple runs (with the same radiative forcing) can also be put together -- possibly leading to even better estimates. As the goal of the method is to recover the eigenmodes in the linear regime of the system, such combination of runs seems extremely beneficial if runs follow the evolution of different eigenmodes. Indeed, it seems plausible -- and an interesting direction for further research -- that a small ensemble of short runs, each with a different perturbation of the initial system state, will better estimate the coefficients of the fitted linear system (i.e. $A$ and $\vec{F}$) without compromising in terms of total computing power.

The most difficult -- and the most important -- aspect of the MC-LR method is the choice of the observables used in the regression data. It is key that this data well-represents the different eigenmodes of the system. If too few are used, not all eigenmodes are found; if too many (or redundant ones) are used, estimates become unusable (as data becomes linearly dependent, which causes the fitted matrix $A$ to become near singular). In this study, we have focused on the use of (effective top-of-atmosphere) albedo and/or emissivity -- observables that can be computed from datasets that are already normally used for climate sensitivity~\cite{gregory2004new, geoffroy2013transient, proistosescu2017slow}. However, the use of other, more curated observables might -- and probably will -- work better. For instance, the very long-term ocean dynamics might be better represented in data on ocean heat uptake~\cite{geoffroy2013transient, geoffroy2013transientP1, raper2002role, li2013deep}. It also seems natural to capture the known climate feedbacks, e.g. surface albedo, water vapour and lapse rate~\cite{annaquantification}. One should beware though that all these (feedback) processes together combine to the system's eigenmodes in non-straightforward ways. For example, summing feedbacks -- like is commonly done in climate literature -- only makes sense in systems that only have one component; in systems with multiple components, processes and eigenmodes are not linked directly like this. Nevertheless, a careful inclusion of these feedbacks might lead to even better estimates and may further shorten the needed length of simulation runs.

The method described in this study does not only lead to better estimates for the equilibrium climate sensitivity, but can also be used to develop extensions of climate sensitivity, by incorporating other observables. Regression of the multi-component model $\vec{Y} = A \vec{X} + \vec{F}$ leads to equilibrium estimates for all the observables in $\vec{X}$ as $\vec{X}_*^\mathrm{est} = - A^{-1} \vec{F}$. This estimate can be seen as a multi-variate metric for climate sensitivity, in contrast to classical uni-variate metrics that focus only on changes in global temperature. Such multivariate metrics can better describe and quantify the changes that occur to the climate system due to changes in radiative forcing. In fact, many -- if not all -- climate subsystems and ecosystems do not depend critically on the global mean surface temperature, but on other observables such as the amount of precipitation or ocean heat transport~\cite{lenton2008tipping, scheffer2009early, rockstrom2009safe}. Estimating those directly -- rather than considering them enslaved to the global mean surface temperature -- will possibly lead to better projections for those (sub)systems.

Accurate estimations of equilibrium climate sensitivity are hard to come by, mostly due to the lengthy computation times needed to fully equilibrate modern global climate models. Going forward, it seems the more and more realistic state-of-the-art models will only take longer and longer to equilibrate (even considering developments in computer hardware). In particular, for high-resolution simulations with ultra fine numerical grids such equilibration runs are just not a practical option. For these kind of simulations it is vital to have extrapolation techniques that only need relatively short transient simulations to estimate the system's long-term behaviour. Once fully developed, such methods -- the one introduced in this study being a first step towards them -- can help to design the kind, amount and length of the experiments performed with these high-resolution models, indicating an optimum between accurate (multi-variate) climate sensitivity estimation and computing time.

\section*{Acknowledgments}
All MATLAB and Python codes are made available on \url{github.com/Bastiaansen/MCLR-ECS-estimation}. Data is available through LongRunMIP~\cite{Rugenstein2019}.

This project is TiPES contribution \#43: this project has received funding from the European Union’s Horizon 2020 research and innovation programme under grant agreement No 820970.

\bibliographystyle{apacite}
\bibliography{sources}
\end{document}


\maketitle

\tableofcontents

\listoffigures

\listoftables

\newpage
\doublespacing
\linenumbers

\section{Rationale}

\subsection{Mathematical Rationale Behind the Gregory Method}\label{sec:GregoryMethod}

Ultimately, warming of Earth's climate can only happen if more energy enters than leaves the Earth system. The amount of energy that enters or leaves is given by the (top-of-atmosphere) radiative imbalance $\Delta R$, which is the incoming short-wave solar radiation minus outgoing short- and long-wave radiation ($\Delta R = R_{SW, \downarrow} - R_{SW, \uparrow} - R_{LW,\uparrow}$). As the amount of outgoing radiation is determined by many (feedback) processes and the Earth's current temperature, the radiative imbalance can be seen as function of these. That is,
\begin{equation}
	\Delta R = \Delta R(T,\vec{\beta}),
	\label{eq:imbalance1}
\end{equation}
where $\vec{\beta} := (\beta_1,\ldots,\beta_N)$ indicate other state variables that influence the radiative fluxes (e.g., sea ice, surface albedo, ocean heat content, cloud formation). Note that, for the generality of this reasoning, neither the exact processes nor the precise relevant state variables, need to be known.

When the system is close to equilibrium -- a typical assumption when studying equilibrium climate sensitivity -- expression~\eqref{eq:imbalance1} can be linearised via a Taylor expansion to obtain
\begin{equation}
	\Delta R = \Delta R(T_*, \vec{\beta}_*) + \frac{\partial \Delta R}{\partial T}(T_*,\vec{\beta}_*) \left[T - T_*\right] + \sum_{j=1}^N \frac{\partial \Delta R}{\partial \beta_j}(T_*, \vec{\beta}_*) \left[\beta_j - \beta_j^*\right] + h.o.t.
	\label{eq:imbalance2}
\end{equation}
Here, $T_*$ is the equilibrium temperature and $\vec{\beta}_* := \left( \beta_1^*, \ldots, \beta_N^*\right)$ denotes the equilibrium values of the other state variables. Since there is, by definition, no radiative imbalance in equilibrium, $\Delta R_* := \Delta R(T_*, \vec{\beta}_*) = 0$. Hence this expression reduces to
\begin{equation}
	\Delta R = \frac{\partial \Delta R}{\partial T}(T_*,\vec{\beta}_*) \left[T - T_*\right] + \sum_{j=1}^N \frac{\partial \Delta R}{\partial \beta_j}(T_*, \vec{\beta}_*) \left[\beta_j - \beta_j^*\right] + h.o.t.
	\label{eq:imbalance3}
\end{equation}

\cite{gregory2004new} argued that, even closer to equilibrium, the radiative imbalance can be seen as function of temperature alone. Put differently, close to equilibrium the other state variables are a function of temperature, i.e. $\vec{\beta} = \vec{\beta}(T)$. Mathematically, this is true only if feedbacks are virtually instantaneous -- which is generically the case only very, very close to the equilibrium state. Under this assumption another Taylor expansion reveals
\begin{equation}
	\vec{\beta} = \vec{\beta}(T_*) + \frac{\partial \vec{\beta}}{\partial T}(T_*) \left[ T - T_* \right] + h.o.t.
	\label{eq:feedbackLinearisation}
\end{equation}
Substitution in \eqref{eq:imbalance3} yields
\begin{equation}
	\Delta R = \left\{ \frac{\partial \Delta R}{\partial T} (T_*, \vec{\beta}(T_*)) + \sum_{j=1}^N \frac{\partial \Delta R}{\partial \beta_j}(T_*, \vec{\beta}(T_*)) \frac{\partial \beta_j}{\partial T}(T_*) \right\} \left[ T - T_* \right] + h.o.t.
\end{equation}
Since, $T = T_0 + \Delta T$ ($T_0$ being the initial, non-forced temperature), a slight rewriting of this expression gives
\begin{equation}
	\Delta R = \left\{ \frac{\partial \Delta R}{\partial T} (T_*, \vec{\beta}(T_*)) + \sum_{j=1}^N \frac{\partial \Delta R}{\partial \beta_j}(T_*, \vec{\beta}(T_*)) \frac{\partial \beta_j}{\partial T}(T_*) \right\} \left[ \Delta T - \Delta T_* \right] + h.o.t.
\end{equation}
Thus, there is a linear relationship between $\Delta R$ and $\Delta $T of the form
\begin{equation}
	\Delta R = a \Delta T + f,
	\label{eq:GregoryLinear}
\end{equation}
with
\begin{align}
	a &:= \frac{\partial \Delta R}{\partial T} (T_*, \vec{\beta}(T_*)) + \sum_{j=1}^N \frac{\partial \Delta R}{\partial \beta_j}(T_*, \vec{\beta}(T_*)) \frac{\partial \beta_j}{\partial T}(T_*) \\
	f &:= - a \Delta T_*.
\end{align}
Even without explicit knowledge of the various functions in this section, this gives a recipe to determine equilibrium warming: close to equilibrium, there is a linear relationship between the time series for $\Delta R$ and $\Delta T$. Linear regression of those gives approximate values for constants $a$ and $f$ in \eqref{eq:GregoryLinear} and an estimation for the equilibrium warming is given by $\Delta T_*^\mathrm{est} := - \frac{1}{a}f$.

One of the reasons this method is used so often is its relative simplicity: almost no system knowledge is needed to perform this procedure and results are good provided the system is close enough to equilibrium. However, it is often not clear if the system is close enough to equilibrium: in fact, the relation between $\Delta R$ and $\Delta T$ often is not linear but nonlinear -- and this is not always obvious from short time series~\citep{Andrews2012, armour2017energy, knutti2015feedbacks}. A practical work-around is to only use the last part of the time series; this way, the initial non-linear parts are not taken into account and only the more close to equilibrium data is used, which better satisfies the assumptions of this technique and gives better long-term predictions. Yet, in practice, it is not always easy to determine how many years of data should be ignored. Moreover, the signal-to-noise ratio tends to be smaller in the later parts of the data, leading to imprecise estimations if too much data is ignored. 

It is also possible to automatise the detection of the inflection point in the Gregory plot; for instance, a double linear function can be fitted:
\begin{equation}
	\Delta R = 
		\begin{cases}
			a_1 \Delta T + f_1, & \mbox{if $\Delta T < - \frac{f_2-f_1}{a_2-a_1}$;}\\
			a_2 \Delta T + f_2, & \mbox{if $\Delta T > - \frac{f_2-f_1}{a_2-a_1}$.}
		\end{cases}
\end{equation}
This technique, dubbed the `double Gregory method' here, is able to find the inflection point of a Gregory plot if the data clearly shows behaviour on two time scales (and estimates the equilibrium warming as $\Delta T_*^\mathrm{est} = -\frac{1}{a_2} f_2$). However, if there are multiple time scales, these time scales overlap, there is a lot of noise or anything else that obscures the two-part splitting of the Gregory plot, this automatic detection tends to fail at detecting the long-term trend as well. Also note that this `double Gregory' method is similar to the mode decomposition in~\cite{proistosescu2017slow}; in fact, in case of (only) two very distinct modes, both techniques are essentially equivalent.

\subsection{General Dynamical Systems Properties}

The heart of the previously described method by \cite{gregory2004new} constitutes of two approximations: (i) linearisation of the radiative imbalance (i.e.~\eqref{eq:imbalance2}), and (ii) linearisation of the feedback processes (i.e.~\eqref{eq:feedbackLinearisation}). Both of these hold true `close to equilibrium'. However, the requirement for the second approximation is much stricter: it only holds when the system is much closer to equilibrium and the decay to equilibrium happens approximately on one eigenmode only (i.e. feedback processes are virtually instantaneous). In this section, this will be explained in more detail by considering general properties of dynamical systems.

Consider the dynamical system
\begin{equation}
	\vec{y}' = \vec{f}(\vec{y}),\quad \left( \vec{y}(t) \in \mathbb{R}^n \right)
	\label{eq:dynamical_system_general}
\end{equation}
(the prime denoting the derivative with respect to time) and let $\vec{y}_*$ be a fixed point of the system (i.e. $\vec{f}(\vec{y}_*) = 0$). For $\vec{y}$ `close to' the fixed point $\vec{y}_*$, \eqref{eq:dynamical_system_general} can be linearized to capture the leading order dynamics of the difference $\vec{z} := \vec{y} - \vec{y}_*$:
\begin{equation}
	\vec{z}' = A \vec{z},
	\label{eq:dynamical_system_linearized}
\end{equation}
where $A = Df(y_*) \in \mathbb{R}^{n \times n}$. From~\eqref{eq:dynamical_system_linearized} it follows that
\begin{equation}
	\vec{z}(t) = e^{A t} \vec{z}(0).
\end{equation}
The dynamics close to equilibrium are thus determined by the eigenvalues and eigenvectors of matrix $A$. For illustrative purposes, assume that $A$ has $n$ distinct real eigenvalues that are ordered as $\lambda_1 < \lambda_2 < \ldots < \lambda_n$ (other possibilities do not alter the reasoning, but give rise to distracting bookkeeping that are non-essential here). In this case, the evolution of $\vec{z}$ is equivalently given by
\begin{equation}
	\vec{z}(t) = \sum_{j=1}^n c_j \vec{v}_j e^{\lambda_j t},
\end{equation}
with $\{ \lambda_j \}$ being the eigenvalues, $\{ \vec{v}_j \}$ the corresponding eigenvectors and $\{ c_j\}$ constants such that $\sum_{j=1}^n c_j \vec{v}_j = \vec{z}(0)$.

\begin{figure}[t]
	\centering
	\includegraphics[width=0.45 \textwidth]{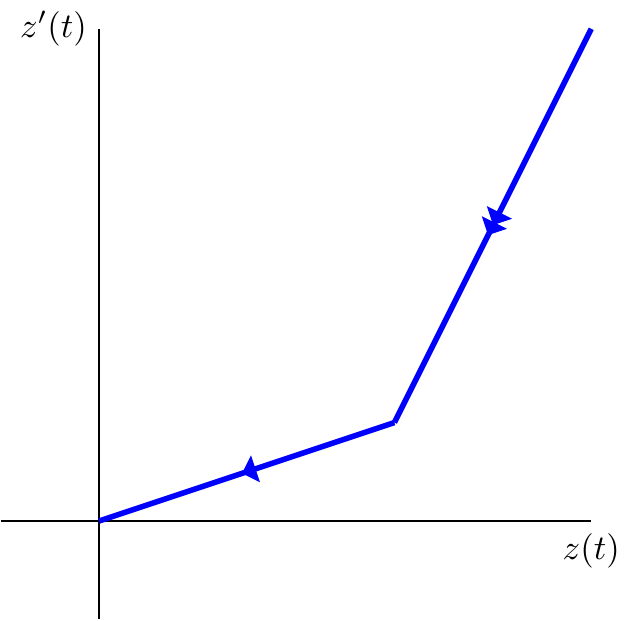}
	\caption[Sketch of typical evolution of one component of a linear system]{Exaggerated sketch of a possible evolution of $z_1$ as in \eqref{eq:z1_evolution}, for $\lambda_1 < \lambda_2 < 0$.}
	\label{fig:z1_nonlinearity}
\end{figure}

If only part of the state $\vec{z}$ is observed (or considered), only a projection of the dynamics on that part can be found. For example, if $n = 2$ and only the first component of $\vec{z} = (z_1,z_2)^T$ can be observed, then $z_1$ evolves as
\begin{equation}
	z_1(t) = c_1 v_{1,1} e^{\lambda_1 t} + c_2 v_{2,1} e^{\lambda_2 t}.
	\label{eq:z1_evolution}
\end{equation}
Here, $v_{j,1}$ denote the first component of the eigenvector $\vec{v}_j$. This expression indicates a non-linearity in the evolution of $z_1(t)$ -- see Figure~\ref{fig:z1_nonlinearity}. In general, linear multi-component systems will result in non-linear behaviour of the system's individual components.

The nonlinearity is the result of the presence of multiple eigenmodes that act on different time scales (i.e. are associated with different eigenvalues). When the system has evolved for long -- and thus is very close to equilibrium -- the exponent $e^{\lambda_1 t}$ is much smaller than $e^{\lambda_2 t}$ (i.e. for $t \gg 1$, $e^{\lambda_1 t} \ll e^{\lambda_2 t}$). Thus, $z_1(t) \approx c_2 v_{2,1} e^{\lambda_2 t}$, which is equivalent to $z_1' = \lambda_2 z_1$. So, only in this situation, there is a linear relationship between $z_1'$ and $z_1$ -- and does the linearisation of the feedback processes in~\eqref{eq:feedbackLinearisation} hold true. That is, for $t \gg 1$, the system dynamics can generically be captured by a $1$-component linear system; otherwise, multiple components are necessary to accurately track the system dynamics.

Shifting attention back to equilibrium estimations, these general considerations also give insight. Similar to before, since $\vec{y} = \vec{y}_0 + \Delta \vec{y}$ (with $\vec{y}_0$ the initial, unforced state), it follows that $\vec{z} = \Delta\vec{y} - \Delta \vec{y}_*$. Hence, \eqref{eq:dynamical_system_linearized} becomes
\begin{equation}
	\Delta \vec{y}\, ' = A \Delta \vec{y}+ A \Delta \vec{y}_*.
	\label{eq:systemRegression}
\end{equation}
If the whole system state $\vec{y}$ can be observed, a linear fit of $\Delta \vec{y}\, '$ and $\Delta \vec{y}$ to \eqref{eq:systemRegression} gives approximate values of $A$ (and thus to all of its eigenvalues and eigenvectors) and $F := A \Delta \vec{y}_*$, such that the equilibrium system state is estimated by $\Delta y_*^\mathrm{est} := A^{-1} F$. 

\begin{figure}[t]
	\centering
	\includegraphics[width=0.45 \textwidth]{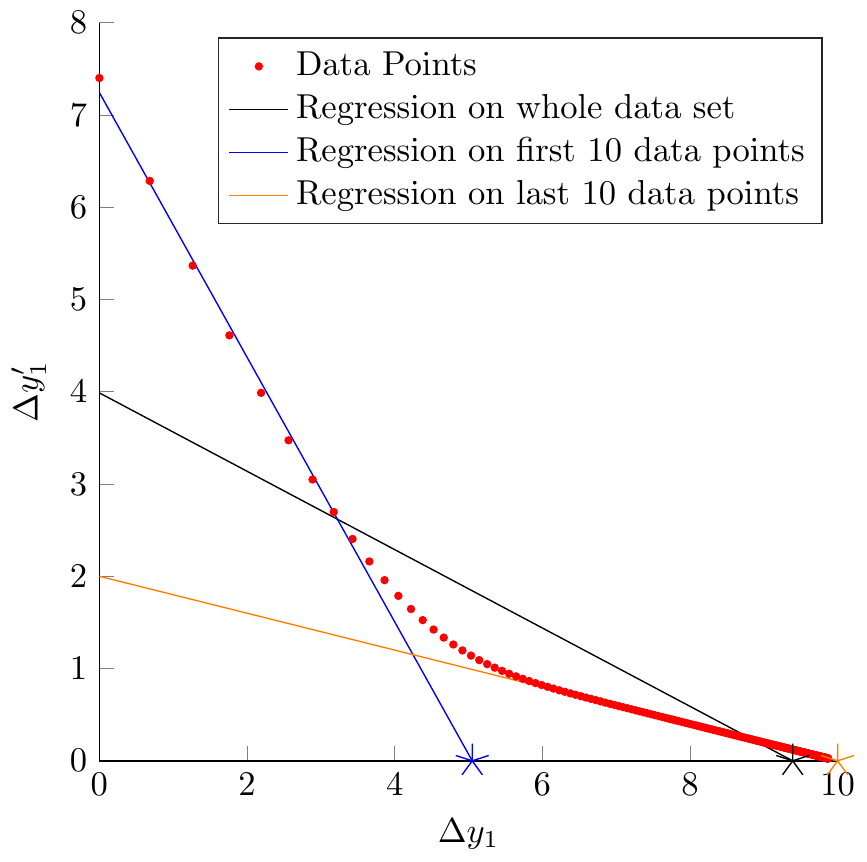}
	\caption[Dynamics of single component of linear system with linear regression examples]{Example of dynamics of a single component of a linear system, along with linear regressions on parts of the full time series. Clearly, quite different lines can be fitted through the nonlinear graph -- with corresponding very different estimates for the equilibrium value (stars in the figure). Plot here is the graph $\left\{ \left(\Delta y_1, \Delta y_1' \right) \right\}$ for $\Delta y_1(t) = 10 - 3 e^{-2t} - 7 e^{-0.2 t}$.}
	\label{fig:dy1_example}
\end{figure}

If only one component of $\Delta \vec{y}$ can be observed, a linear regression only finds the projection of the full dynamics; that is, only the most dominant dynamics present in the data is captured. Going back to the example system, a linear regression of $\Delta y_1'$ and $\Delta y_1$ to $\Delta y_1' = a \Delta y_1 + f$ gives different results over time:
\begin{itemize}
	\item When only considering data from times $t \ll -1$, $\Delta y_1 \approx c_1 v_{1,1} e^{\lambda_1 t}$, and hence $a = \lambda_1$ and $f = 0$ are found.
	\item When only considering data from times $t \gg 1$, $\Delta y_1 \approx \Delta y_1^* + c_2 v_{2,1} e^{\lambda_2 t}$, and hence $a = \lambda_2$ and $f = -\lambda_2 \Delta y_2^*$.
	\item In between these asymptotic cases, the found slope will change from $\lambda_1$ to $\lambda_2$ (the intercept will change from $0$ to $\lambda_1 c_1 v_{1,1} + \lambda_2 c_2 v_{1,2}$ for negative times $t$, and then, for positive $t$ change to $-\lambda_2 \Delta y_1^*$).
\end{itemize}
To illustrate this, in Figure~\ref{fig:dy1_example}, an example is given of a trajectory in $(\Delta y_1, \Delta y_1')$-space, along with linear regression results when applied on various subsets of the data.

So, in essence, by only considering one component of the state space $\Delta \vec{y}$, one thus only finds the \emph{one} eigenmode that is dominant on the specific time scale one considers. This can be extended to observations of multiple components: if one considers $m$ components of the state space $z$, one expects to find the first $m$ eigenmodes that are dominant on that time scale -- provided the $m$ components sufficiently represent the associated eigenmodes!

Therefore, in Gregory plots one throws away the first part of the data -- because the dominant eigenmode on those time scales only captures the short-time behaviour; to obtain the information about the evolution on longer time scales, the next eigenmode is needed. By ignoring the first part of the data, this feat is achieved. However, as argued above, it is also possible to capture these other eigenmodes by employing a larger set of observables, from which the relevant eigenmodes -- and thus the long-time dynamics -- can be derived from this early data (that is typically ignored in Gregory plots).

Also note that it is not necessary to observe the state space directly, nor to use the same observables on the left and right hand sides (as is also not done in the Gregory plots either). That is, it is possible to use any $m$ observables $\vec{Y} = ( Y_1, \ldots, Y_m)^T$ that tend to $0$ in equilibrium, and any $m$ observables $\vec{X} = (X_1, \ldots, X_m)^T$, some of which need to be estimated in equilibrium. Provided these form a good representation of the system's dynamics, these can be regressed to
\begin{equation}
	\vec{Y} = A  \vec{X} + \vec{F}
	\label{eq:SystemRegression2}
\end{equation}
to obtain the equilibrium estimate $\vec{X}_*^\mathrm{est} := - A^{-1} \vec{F}$. In principle, this should work for any observable if the system is close enough to the equilibrium (as transforming~\eqref{eq:systemRegression} to~\eqref{eq:SystemRegression2} requires another Taylor expansion for general observables). However, if observables are linear combinations of the components of the state space -- such as globally averaged surface temperature and radiative imbalance -- there is no such additional caveat.

Finally, we note that this method, and specifically~\eqref{eq:SystemRegression2}, can also be derived directly from the equation $\vec{Y} = \vec{Y}(\vec{X}, \vec{\beta})$ using a generalisation of the reasoning in section~\ref{sec:GregoryMethod}. The only relevant change is that now one assumes the (other) feedbacks $\vec{\beta}$ can be captured adequately by the $m$ observables in $\vec{X}$ -- which is a weaker assumption, that should hold further from the equilibrium state, if the chosen observables represent the relevant dynamics of the system.

\newpage

\section{Additional Details on Conceptual Model}

To test the effectiviness of the new estimation technique, a low-order conceptual global energy budget model has been considered. Since behaviour of this kind of model can be understood almost completely from mathematical analysis alone, tests on these models readily give insight in its use and allow for straightforward quantification of the errors made by estimation methods.

Specifically, we consider the following global energy budget model, that models the dynamics of global mean surface temperature $T$, effective top-of-atmosphere global mean short wave albedo $\alpha$, and effective top-of-atmosphere global mean long wave emissivity $\varepsilon$ as
\begin{equation}
	\begin{cases}
		C_T \frac{dT}{dt} & = Q_0 \left[1 - \alpha\right] - \varepsilon \sigma T^4 + \mu + \nu \xi(t); \\
		\frac{d\alpha}{dt} & = - \varepsilon_\alpha \left[\alpha - \alpha_0(T)\right]; \\
		\frac{d\varepsilon}{dt} & = - \varepsilon_\varepsilon \left[ \varepsilon - \varepsilon_0(T)\right].
	\end{cases}
	\label{eq:simple_model_3}
\end{equation}
Here, $\alpha_0(T) = \alpha_1 + \frac{\alpha_2 - \alpha_1}{2} \left(1 + \tanh\left(K_\alpha \left[T - \frac{T_1+T_2}{2}\right]\right)\right)$ and $\varepsilon_0(T) = \varepsilon_1 + \frac{\varepsilon_2 - \varepsilon_1}{2} \left(1 + \tanh\left(K_\varepsilon \left[T - T_\varepsilon\right]\right)\right)$ indicate the (ultimate) albedo and emissivity for a given temperature $T$, but these values are approached slowly -- not instantaneously. Furthermore, $Q_0$ is incoming solar radiation, $-Q_0 \alpha$ the reflected solar radiation, $-\varepsilon \sigma T^4$ the outgoing long wave blackbody `Planck' radiation and $\mu$ models the effect of CO$_2$ in the atmosphere, where we set $\mu = \mu_0 + A_0 \log\left(C/C_0\right)$ (with $C/C_0$ the factor of the considered CO$_2$ increase). Finally $\nu \xi(t)$ is Gaussian white noise, representing natural climate variability.

\subsection{Details of Numerical Simulations}

In the simulations we have used the following parameter values: $\alpha_1 = 0.7$, $\alpha_2 = 0.289$, $K_\alpha = 0.1$, $T_1 = 260$, $T_2 = 289$, $\sigma = 5.67 \cdot 10^{-8}$, $\varepsilon_1 = 0.7$, $\varepsilon_2 = 0.6$, $K_\varepsilon = 0.05$, $T_\varepsilon = 288$, $Q_0 = 341.3$, $C_T = 10$, $\varepsilon_\alpha = 0.05$, $\varepsilon_\varepsilon = 0.1$ and $A_0 = 5.35$. As initial conditions we have taken $T_0 = 289$, $\alpha_0 = \alpha_0(T_0)$, $\varepsilon_0 = \varepsilon_0(T_0)$, $\mu_0 = \varepsilon_0 \sigma T_0^4 - Q_0 \left[1 - \alpha_0\right]$ and we have taken $C/C_0 = 4$ (i.e. a quadruppling of the CO$_2$ concentration). Noise strength $\nu$ has been varied. 

Numerical simulations were performed via discretisation of~\eqref{eq:simple_model_3} with a standard Euler-Maruyama scheme and numerical time step $\Delta t = 0.001$ was used. Output of the model includes time series $T(t)$, $\alpha(t)$, and $\varepsilon(t)$. From these, time series for the increment of the observables have been computes: $\Delta T(t) = T(t) - T_0$, $\Delta \alpha(t) = \alpha(t) - \alpha_0$, $\Delta \varepsilon(t) = \varepsilon(t) - \varepsilon_0$ and $\Delta R(t) = Q_0 \left[1 - \alpha(t)\right] - \varepsilon(t) \sigma T(t)^4 + \mu$. Derivatives needed for some of the estimation techniques have been computed numerically as forward differences.

\subsection{Estimation Techniques}
For each individual realisation of the model, estimations for the real equilibrium temperature increase, $\Delta T_*^\mathrm{est}$, have been made using the following different techniques:
\begin{itemize}
	\item[(1)] Use of the raw time series, i.e. $\Delta T_*^\mathrm{est}(t) := \Delta T(t)$;
	\item[(2)] Fit $\Delta R = a \Delta T+ f$, and take $\Delta T_*^\mathrm{est}(t) = - a^{-1} f$ (`Gregory method');
	\item[(3)] Fit a double-linear function to $\Delta R$ and $\Delta $T, and take $\Delta T_*^\mathrm{est}(t) = - a^{-1} f$ for the slope and intercept of the latter linear part (a `double Gregory method').
	\item[(4)] Fit the linear system $\left[ \begin{array}{c} \Delta R \\ \frac{d \Delta \alpha}{dt} \end{array}\right] = A \left[ \begin{array}{c} \Delta T \\ \Delta \alpha \end{array}\right] + \vec{F}$. $\Delta T_*^\mathrm{est}(t)$ is the first component of $-A^{-1} \vec{F}$ (a MC-LR method).
	\item[(5)] Fit the linear system $\left[ \begin{array}{c} \Delta R \\ \frac{d \Delta \varepsilon}{dt} \end{array}\right] = A \left[ \begin{array}{c} \Delta T \\ \Delta \varepsilon \end{array}\right] + \vec{F}$. $\Delta T_*^\mathrm{est}(t)$ is the first component of $-A^{-1} \vec{F}$ (a MC-LR method).
	\item[(6)] Fit the linear system $\left[ \begin{array}{c} \Delta R \\ \frac{d \Delta \alpha}{dt} \\ \frac{d \Delta \varepsilon}{dt} \end{array}\right] = A \left[ \begin{array}{c} \Delta T \\ \Delta \alpha \\ \Delta \varepsilon \end{array}\right] + \vec{F}$.  $\Delta T_*^\mathrm{est}(t)$ is the first component of $-A^{-1} \vec{F}$ (a MC-LR method).
\end{itemize}

\begin{figure}[h]
	\centering
		\begin{subfigure}[t]{0.48\textwidth}
			\centering
				\includegraphics[width = \textwidth]{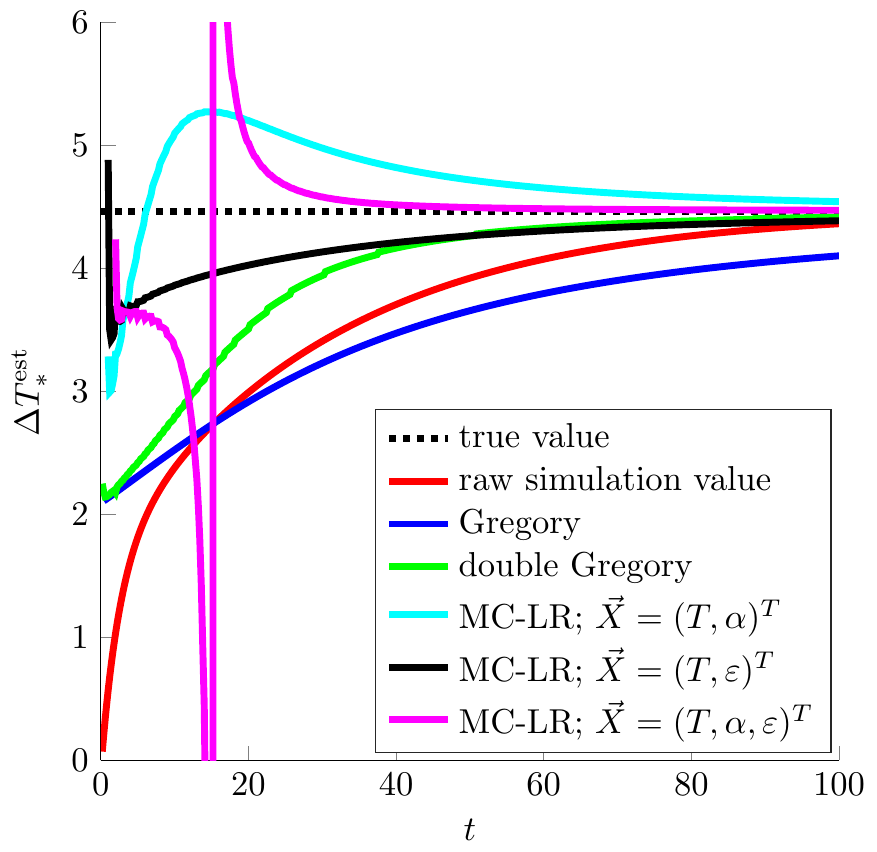}
			\caption{$\nu = 0$}
		\end{subfigure}
		~
		\begin{subfigure}[t]{0.48\textwidth}
			\centering
				\includegraphics[width = \textwidth]{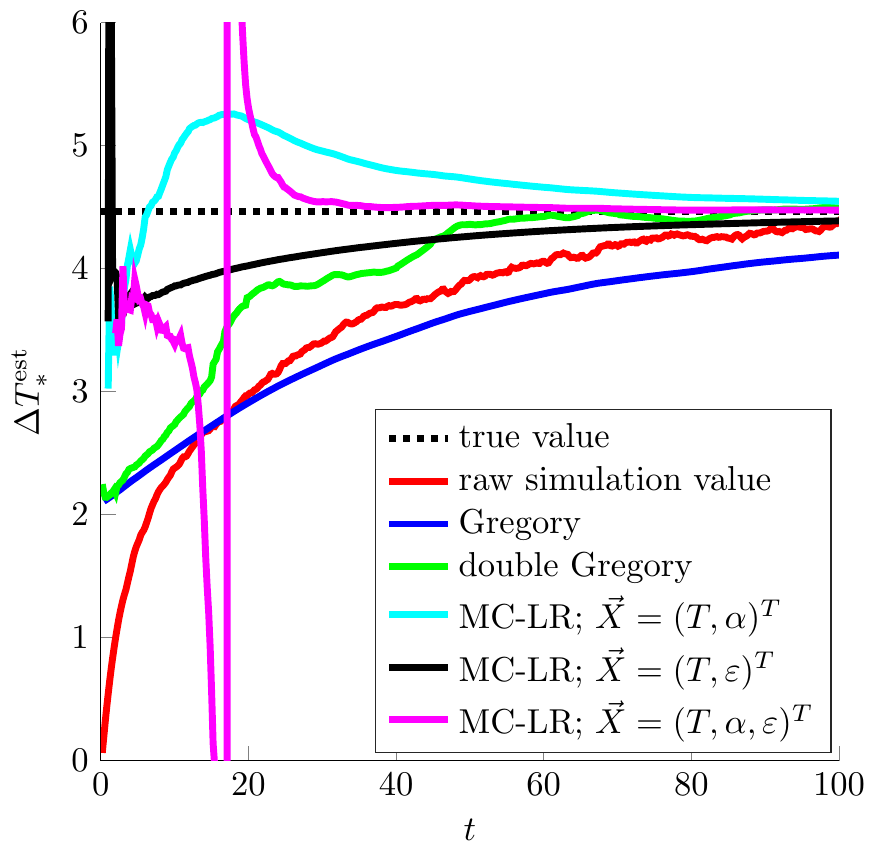}
			\caption{$\nu = 0.25$}
		\end{subfigure}
		~
		\begin{subfigure}[t]{0.48\textwidth}
			\centering
				\includegraphics[width = \textwidth]{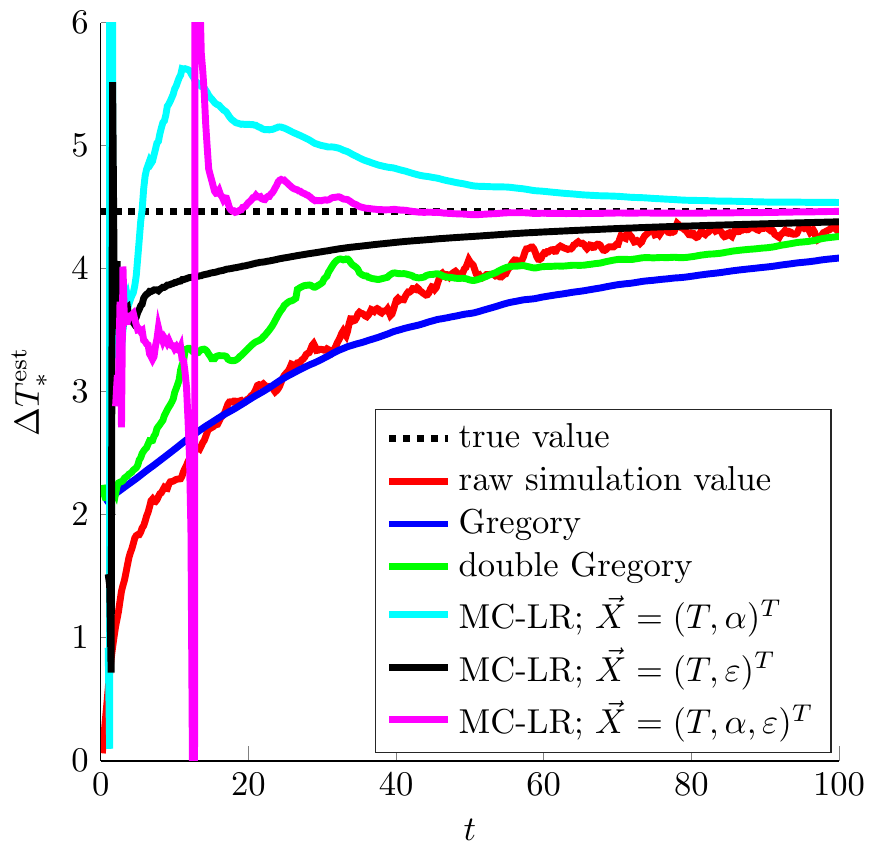}
			\caption{$\nu = 0.5$}
		\end{subfigure}
		~
		\begin{subfigure}[t]{0.48\textwidth}
			\centering
				\includegraphics[width = \textwidth]{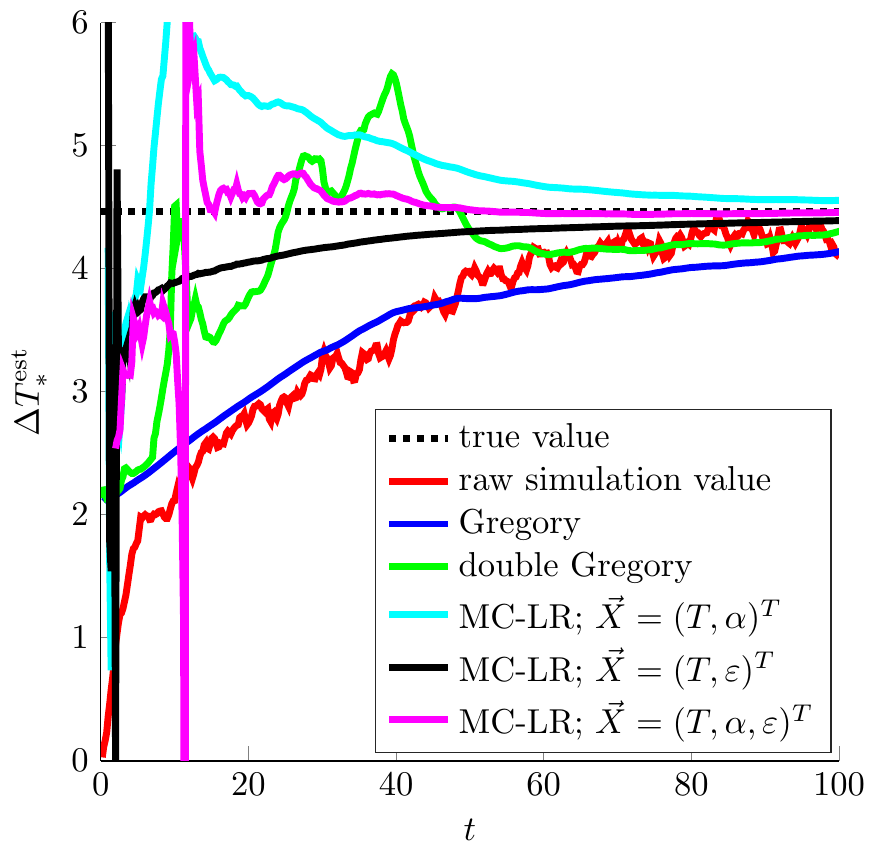}
			\caption{$\nu = 1$}
		\end{subfigure}
	\caption[Estimated warming of conceptual model over time using different estimation techniques]{Estimated values for one single realisation of~\eqref{eq:simple_model_3} using the different estimation techniques described in the text, for various noise strenghts $\nu$.}
	\label{fig:results_simple_model_3}
\end{figure}

Linear regression for all of these techniques has been standard least squares regression. Examples of the results of these estimation techniques at various noise strengths $\nu$ are given in Figure~\ref{fig:results_simple_model_3}. The results are given as function of time $t$: estimates $\Delta T_*^\mathrm{est}(t)$ contain the estimation value when regression is applied only on data up to time $t$.

\subsection{Assessment of Estimation Techniques}
The previously described estimation techniques all converge to the true equilibrium warming $\Delta T_*$. However, they do so at different rates. Therefore, to assess and quantify their power, it is of interest to quantify how fast and how accurate they are. For this, the remaining relative error can be used -- the maximum of the relative errors that occurs for later predictions. More precisely, the relative error at time $t$ is given by
\begin{equation}
	e^\mathrm{rel}(t) := \left| \frac{\Delta T_*^\mathrm{est}(t) - \Delta T_*}{\Delta T_*^\mathrm{est}(t)} \right|,
\end{equation}
and the remaining relative error is
\begin{equation}
	e^\mathrm{rel}_\mathrm{rem}(t) := \max_{s \geq t}\ e^\mathrm{rel}(s).
\end{equation}
This quantification of the error helps to avoid lucky hits: if random fluctuations by chance happen to steer an estimation to the right value for a short amount of time, this should not be incorrectly marked as a success for that estimation technique. So, taking the maximum of the remaining errors remedies this and thus gives a better idea of the (maximum of the) kind of error to expect when working with a prediction technique on that time scale.

\begin{figure}
	\centering
		\begin{subfigure}[t]{0.48\textwidth}
			\centering
				\includegraphics[width = \textwidth]{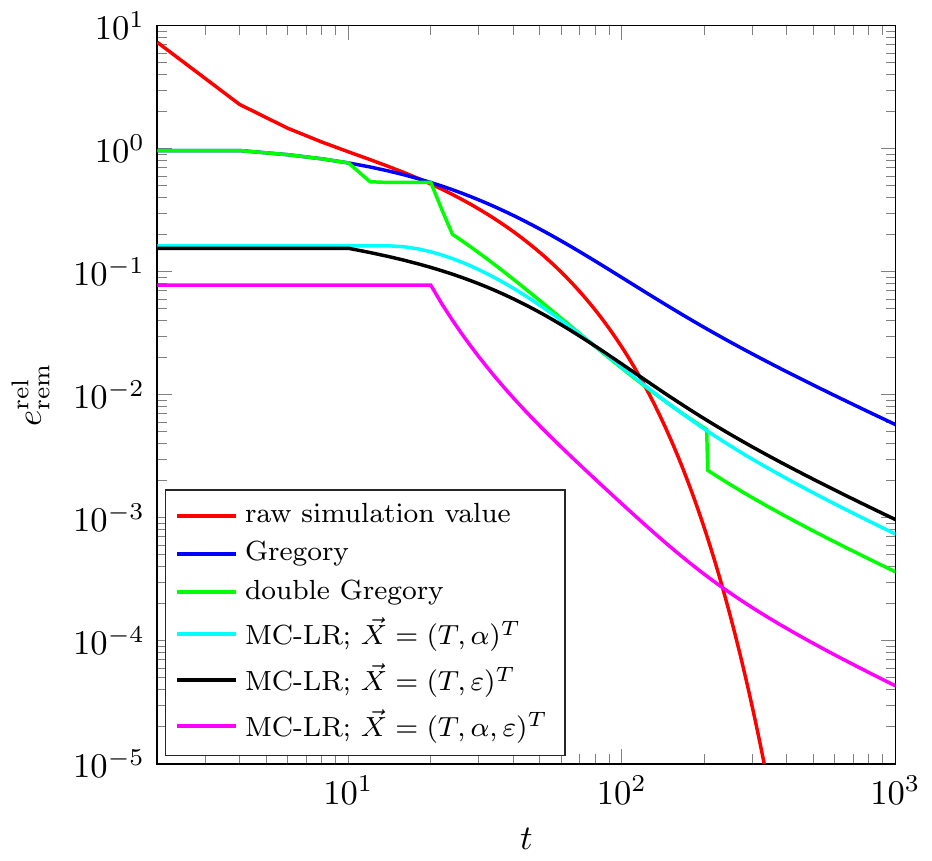}
			\caption{$\nu = 0$}
		\end{subfigure}
		~
		\begin{subfigure}[t]{0.48\textwidth}
			\centering
				\includegraphics[width = \textwidth]{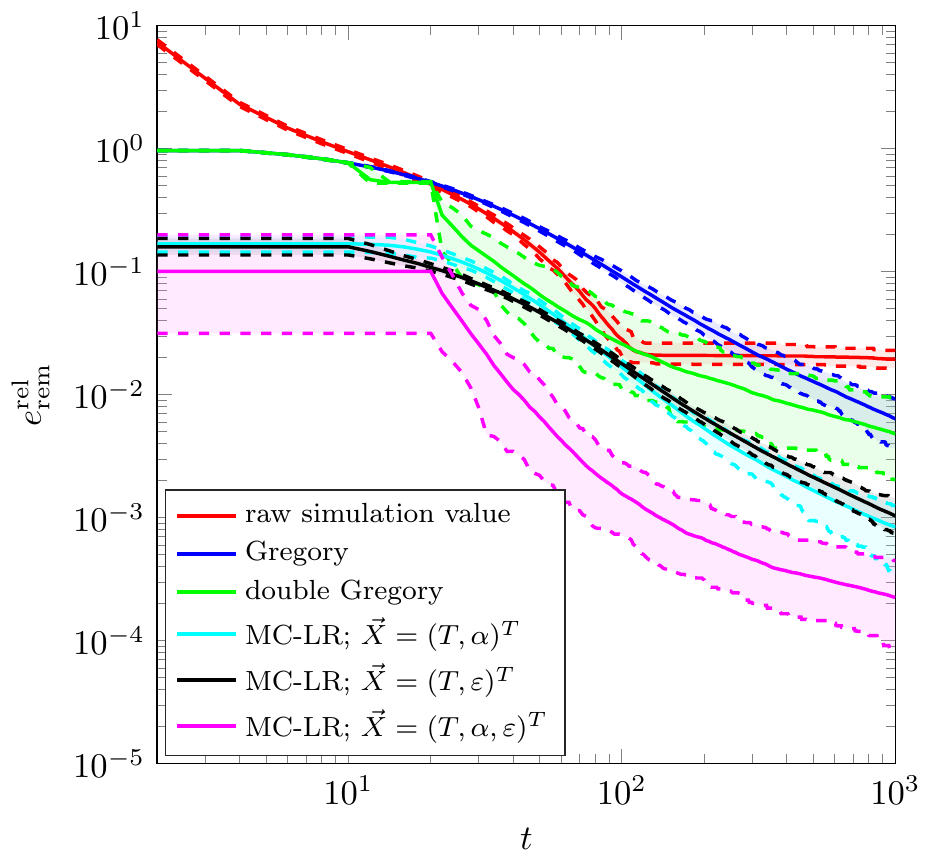}
			\caption{$\nu = 0.25$}
		\end{subfigure}
		~
		\begin{subfigure}[t]{0.48\textwidth}
			\centering
				\includegraphics[width = \textwidth]{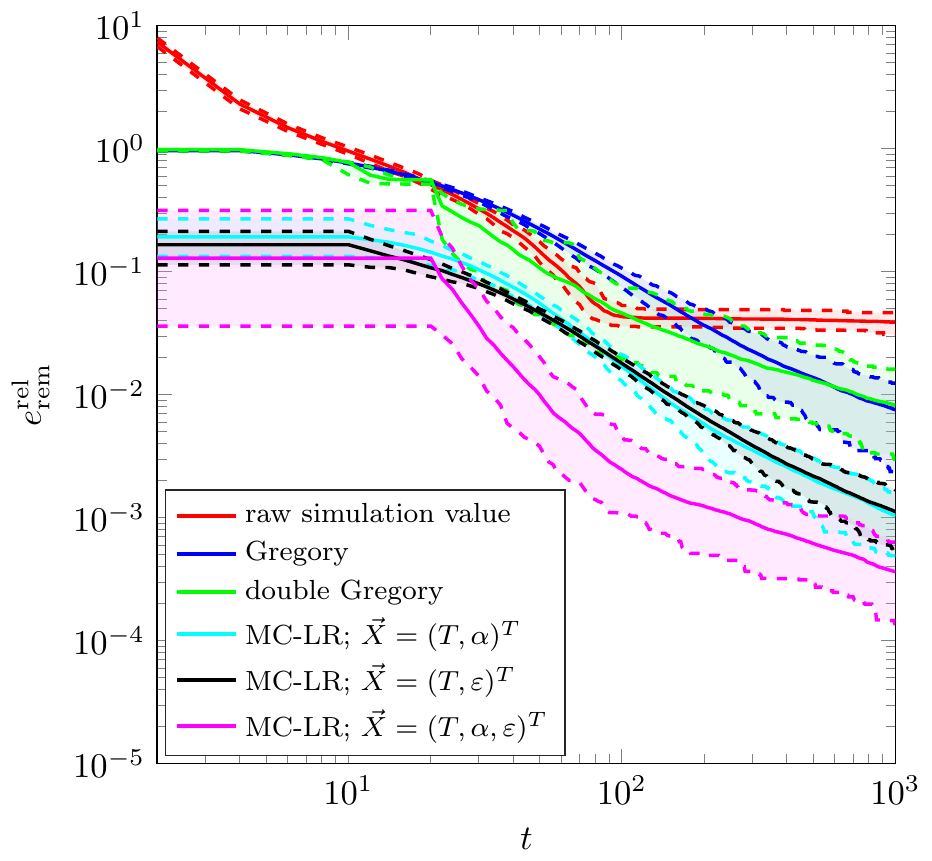}
			\caption{$\nu = 0.5$}
		\end{subfigure}
		~
		\begin{subfigure}[t]{0.48\textwidth}
			\centering
				\includegraphics[width = \textwidth]{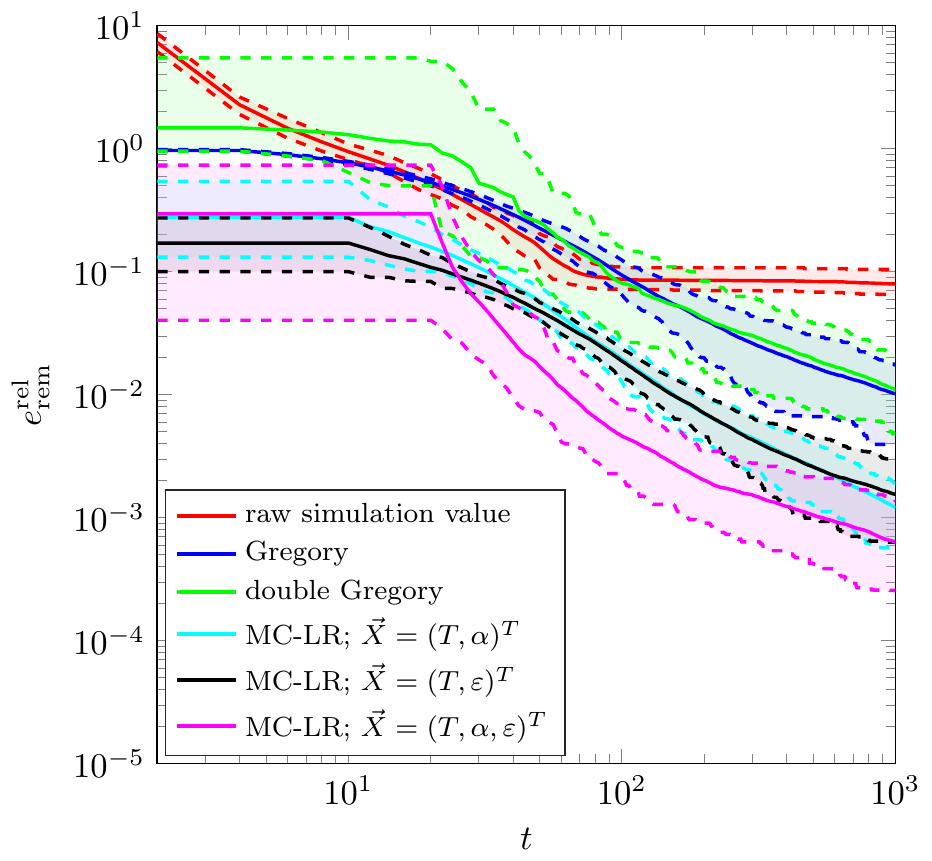}
			\caption{$\nu = 1$}
		\end{subfigure}

	\caption[Statistical results for various estimation techniques on a conceptual model]{Plots of $e_\mathrm{rem}^\mathrm{rel}(t)$ for various noise levels $\nu$. Solid lines indicate the mean values, and dashed lines the $5$ and $95$ percentile values. Colors indicate the type of approximation technique used: red indicates the use of the raw temperature time series, blue is the Gregory method on all data, green is a `double Gregory' method, cyan indicates a multi-component linear regression using temperature and albedo, black using temperature and emissivity, and magenta using temperature, albedo and emissivity. All techniques are run on the same ensemble of $100$ runs of the simulation.}
	\label{fig:simple3_ensemble}
\end{figure}

An ensemble run of $100$ simulations has been performed for noise levels $\nu = 0$, $\nu = 0.25$, $\nu = 0.5$ and $\nu = 1$. The results on the remaining error are given in Figure~\ref{fig:simple3_ensemble}.

\subsubsection{Comparison of Eigenvalues}
\begin{figure}
	\centering
		\includegraphics[width = \textwidth]{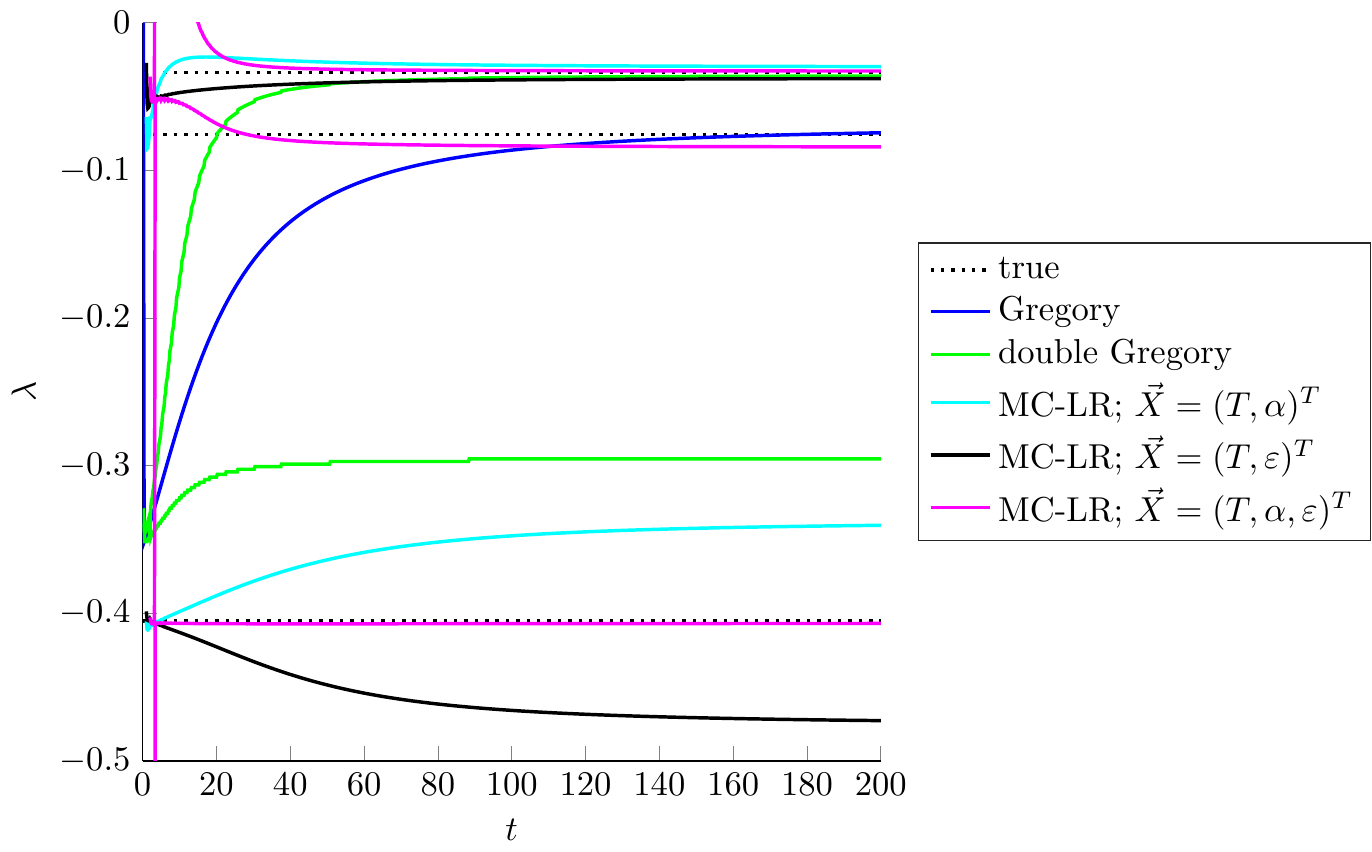}
	\caption[Eigenvalues found by the estimation techniques over time for a conceptual model]{Eigenvalues found by the different methods over time. The black dotted lines indicate the true eigenvalues, the colored lines the eigenvalues found from the different estimation techniques. Note that the multi-component linear regression methods do find multiple eigenvalues, whereas the Gregory methods can only find one. From this, it can be seen that the Gregory method underestimates the smallest eigenvalues and some of the multi-component regressions overestimate this eigenvalue, which is directly related to over- and underestimation of the equilibrium warming. Based on a simulation with no noise ($\nu = 0$).}
	\label{fig:eigenvalueEstimations_3}
\end{figure}

Furthermore, in this simple model it is possible to determine the real Jacobian of the new equilibrium state and its eigenvalues. Although the fitted matrix $A$ does not have the same eigenvalues (since $\frac{dT}{dt} = \frac{\Delta R}{C_T}$), the Jacobian matrix can be recovered from matrix $A$ through a row operation (i.e. by multiplying the `$\Delta R$ row' by $C_T$). Hence, also the values for the eigenvalues can be compared, to see how successful the various methods are in tracking these. Here, especially the rate at which the smallest eigenvalues are captured is relevant for accurate long-term predictions. The results, for the system without noise, are given in Figure~\ref{fig:eigenvalueEstimations_3}.

\subsection{Discussion of Results on Conceptual Models}

The tests on the conceptual models show the increased performance of the multi-component linear regression techniques; the estimates $\Delta T_*^\mathrm{est}(t)$ converge to the real value at a faster rate than the convential `Gregory' method, and they even perform well for high noise levels. Overall, it seems that a multi-component linear regression using $T$, $\alpha$ and $\varepsilon$ gives the best estimates -- which is not surprising since the conceptual model constitutes of these three components. The MC-LR techniques using either $T$ and $\alpha$ or $T$ and $\varepsilon$ perform similar and have similar errors. The classic `Gregory method' is less effective -- often performing worse than the raw temperature time series. Finally, the `double Gregory' method is quite good for low noise levels, but shows inconsistent behaviour at higher noise levels; since that method needs to find the inflection point automatically in the regression, sudden changes, dictated by the noise, tend to lead to wrong determination of this inflection point, thus producing estimates that are way too high or way too low.

The eigenvalue plot (Figure~\ref{fig:eigenvalueEstimations_3}) is also very illuminating. The real system has three eigenvalues ($\lambda_1 = -0.405 < \lambda_2 = -0.076 < \lambda_3 = -0.034$). The different estimation techniques all essentially estimate (some of) these; it depends on the method how many are found (and which). The MC-LR technique with $T$, $\alpha$ and $\varepsilon$ does find all three, more or less. Critically, the largest eigenvalue (closest to zero; $\lambda_3$) is estimated too high, which is the reason estimates $\Delta T_*^\mathrm{est}(t)$ are (slightly) too high. Furthermore, for very small $t$, positive eigenvalues are found, which directly relates to the poles in the estimate for small $t$. The MC-LR techniques that use two observables ($m=2$) fit two eigenvalues: in this case, both of these methods find good approximations of $\lambda_3$, but the other one does not correspond with any of the real eigenvalues. Again, over- and underestimation of $\Delta T_*^\mathrm{est}(t)$ is directly related to over- and underestimation of the largest eigenvalue. The `Gregory method' only fits one eigenvalue. As explained in section~\ref{sec:GregoryMethod}, this is some interpolation of the real system's eigenvalues; in this case, certainly the fitted eigenvalue is nowhere near $\lambda_3$, and thus a large underestimation of the real equilibrium warming is made when using this method. Finally, the `double Gregory' technique does fit two eigenvalues; one turns out to be close to the highest eigenvalue, $\lambda_3$ (and the other an interpolation of the two remaining real eigenvalues), which explains why estimates using this technique are quite good (for low noise levels).

Lastly, in the presented results, standard linear regression (i.e. least squares minimalisation) was used. Hence, the fitted matrix $A$ has no non-zero entries. The real Jacobian, however, does have zero entries. These could be identified when using a slightly different linear regression technique that also minimizes for instance the $L^1$ or $L^2$ norm of the fitted coefficients, such as the Lasso regression \citep[e.g][]{LassoRegression} or Ridge regression \citep[e.g][]{RidgeRegression}. This could result in (slightly) better estimates, but requires some hyper-fitting to obtain adequate fitting settings (which are probably different settings for different models and experiments). As the obtained fitted eigenvalues and estimates $\Delta T_*^\mathrm{est}(t)$ do correspond reasonably well, we have not explored this possibility in any detail, but only state its existence here. Nevertheless, in case many zero entries are expected to occur in the Jacobian, the use of this kind of regularisation might provide better estimates.

\FloatBarrier
\newpage
\section{Additional Details on Global Climate Models in LongRunMIP}

To test the equilibrium estimation techniques on complex models, data from LongRunMIP~\citep{Rugenstein2019} was used. These model runs have a significantly longer simulation time compared to the typically used $150$ year runs (the minimum requirement for models participating in most model intercomparison projects such as CMIP6,~\cite{eyring2016overview}). Hence, the model runs used here are typically relatively equilibrated at the end of the simulation. Therefore, bounds on the real model equilibrium warming can be found, allowing for a quantitative assessment of the equilibrium estimation techniques.

In this study, we have used data from the abrupt $4\times$CO$_2$ experiments on the models in LongRunMIP. That is, eleven models, of various levels of complexity, have been considered. Data for all of these models include at least $1000$ years of simulation. An overview of the models is given in table~\ref{tab:LongRunMIP_models}.

\begin{table}
\caption[Overview of models used in this study]{Overview of models used in this study, alongside the length of their abrupt $4\times$CO$_2$ experiment run and their pre-Industrial run and the obtained best fit value and $5\%$-$95\%$ range for the model's equilibrium warming.}
\centering
\begin{tabular}{rccc}
			& \multicolumn{2}{c}{--------- length of run ---------}									&	\\
\multirow{-2}{*}{model name} & \small abrupt $4\times$CO$_2$ & \small pre-Industrial control & \multirow{-2}{*}{$DT_*^\mathrm{est,best}$ ($\mathcal{R}_{\Delta T_*}$)} \\\hline
CCSM3 		&	2,120	&	1,530	&   $5.54K$ ($5.46K$ --- $5.63K$)\\\rowcolor{gray!10}
CESM 1.0.4	&	5,900	&	1,000	&   $6.76K$ ($6.73K$ --- $6.80K$)\\
CNRM-CM6-1	&	1,850	&	2,000	&   $11.17K$ ($10.94K$ --- $11.45K$)\\\rowcolor{gray!10}
ECHAM5/MPIOM&1,000	&	100		&   $11.65K$ ($11.44K$ --- $11.92K$)\\
FAMOUS		&	3,000	&	3,000	&   $16.30K$ ($15.52K$ --- $17.78K$)\\\rowcolor{gray!10}
GISS-E2-R	&	5,000	&	5,225	&   $4.84K$ ($4.82K$ --- $4.85K$)\\
HadCM3L	&	1,000	&	1,000	&   $7.09K$ ($6.92K$ --- $7.31K$)\\\rowcolor{gray!10}
HadGEM2-ES	&	1,328	&	239		&   $10.09K$ ($9.79K$ --- $10.53K$)\\
IPSL-CM5A-LR&	1,000	&	1,000	&   $9.46K$ ($9.28K$ --- $9.68K$)\\\rowcolor{gray!10}
MPI-ESM-1.1	&	4,459	&	2,000	&   $6.85K$ ($6.83K$ --- $6.86K$)\\
MPI-ESM-1.2	&	1,000	&	1,237	&   $6.69K$ ($6.65K$ --- $6.73K$)\\\hline
\end{tabular}
\label{tab:LongRunMIP_models}
\end{table}

\subsection{Data Acquisition}

For every model, model output on surface temperature and radiative fluxes has been downloaded from the LongRunMIP database. Specifically, we have downloaded the globally averaged yearly datasets for near surface temperature (`tas'), incoming short-wave radiation at top-of-atmosphere (`rsdt'), outgoing short-wave radiation at top-of-atmosphere (`rsut') and outgoing long-wave radiation at top-of-atmosphere (`rlut'). These datasets have been downloaded for the abrupt $4\times$CO$_2$ experiments and the pre-Industrial control runs.

\subsection{Data Processing}

The downloaded (globally averaged yearly) data has been used to compute the following observables:
\begin{itemize}
		\item Temperature $T$ = `tas';
		\item Radiative top of atmosphere imbalance $R$ = `rsdt' - `rsut' - `rlut';
		\item Effective top-of-atmosphere short-wave albedo $\alpha$ = `rsut' / `rsdt';
		\item Effective top-of-atmosphere long-wave emmisivity $\varepsilon \sigma$ = `rlut' / (`tas')$^4$, where $\sigma$ is the Stefan-Boltzmann constant.	
	\end{itemize}
From the pre-Industrial control runs, initial values $T_0$, $R_0$, $\alpha_0$ and $\varepsilon_0 \sigma$ have been computed as the mean of the full run. The abrupt $4\times$CO$_2$ forcing experiment runs have then been used to compute the changes $\Delta T(t) = T(t) - T_0$, $\Delta R(t) = R(t) - R_0$, $\Delta \alpha = \alpha(t) - \alpha_0$ and $\Delta \varepsilon(t) \sigma = \varepsilon(t) \sigma - \varepsilon_0 \sigma$. Derivatives needed for the estimation techniques have been computed numerically as forward differences.

\subsubsection{Exceptions}
Above, the general procedure to process the data has been explained. There are, however, a few exceptions to this general scheme for some of the models -- as per instructions of the LongRunMIP database readme~\citep{Rugenstein2019}. The following list contains the details of these slight alterations per model.

\begin{description}
	\item[GISS-E2-R :] The control run has some discrepancies as months for the different variables do not line up. Therefore, the first 300 years of data for the datasets `tas' and `rlut' have been ignored as there is no corresponding data for `rsut' and `rsdt'.
	\item[CCSM3 :] The model is out of equilibrium in the first years, both for the control and for the experiment run. Therefore -- as per instruction -- we have not used the control mean for the first years, but rather have taken the anomaly year by year for the first $20$ years of the simulation. For the remainder of the simulation, the mean is taken as normal.
\end{description}

\subsection{Estimation Techniques}
For each model output, estimates of the equilibrium warming $\Delta T_*$ have been made with the following techniques:
\begin{itemize}
	\item[(1)] Use of the raw time series, i.e. $\Delta T_*^\mathrm{est}(t) := \Delta T(t)$;
	\item[(2)] Fit $\Delta R = a \Delta T + f$ on the whole data set, and take $\Delta T_*^\mathrm{est}(t)(t) := - a^{-1} f$ (`Gregory method');
	\item[(3)] Fit $\Delta R = a \Delta T + f$ on all but the first 20 years of data. Estimation is then given by $\Delta T_*^\mathrm{est}(t) := - a^{-1} f$;
	\item[(4)] Fit a double-linear function to $\Delta R$ and $\Delta T$.
	\item[(5)] Fit the linear system $\left[ \begin{array}{c} \Delta R \\ \frac{d \Delta \alpha}{dt} \end{array}\right] = A \left[ \begin{array}{c} \Delta T \\ \Delta \alpha \end{array}\right] + \vec{F}$. $\Delta T_*^\mathrm{est}(t)$ is the first component of $-A^{-1} \vec{F}$  (a MC-LR method).
	\item[(6)] Fit the linear system $\left[ \begin{array}{c} \Delta R \\ \frac{d \Delta \varepsilon \sigma}{dt} \end{array}\right] = A \left[ \begin{array}{c} \Delta T \\ \Delta \varepsilon \sigma \end{array}\right] + \vec{F}$. $\Delta T_*^\mathrm{est}(t)$ is the first component of $-A^{-1} \vec{F}$  (a MC-LR method).
	\item[(7)] Fit the linear system $\left[ \begin{array}{c} \Delta R \\ \frac{d \Delta \alpha}{dt} \\ \frac{d \Delta \varepsilon \sigma}{dt} \end{array}\right] = A \left[ \begin{array}{c} \Delta T \\ \Delta \alpha \\ \Delta \varepsilon \sigma \end{array}\right] + \vec{F}$.  $\Delta T_*^\mathrm{est}(t)$ is the first component of $-A^{-1} \vec{F}$  (a MC-LR method).
\end{itemize}

Linear regression for all of these techniques has been standard least squares regression. The results of these estimations are given in subfigures (a) and (c) of Figures~\ref{fig:results_for_CCSM3}-\ref{fig:results_for_MPIESM12}. The results are given as function of time $t$, where an estimation at time $t$ only uses the data up to simulation time $t$. In subplots (a) the results for the first $500$ years of the run are given and in (c) for the whole run.

\subsection{Assessment of Estimation Techniques}

\subsubsection{Determining a Range of `Best Estimates'}

To be able to assess the performance of different estimation techniques, it would be best to have the real equilibrium warming $\Delta T_*$. However, even in these century-long model runs, the models have not equilibrated fully and therefore the real warming is not known. Hence, we resort to a best estimate of the real model equilibrium warming, which is defined as $\Delta T_*^\mathrm{est,best}$. Following \cite{Rugenstein2020}, we define this best estimate as the increment predicted by a linear `Gregory' fit on the last $15\%$ of warming. Concretely: we take the last 50 years of a run and average the radiative imbalance in those years. We then take 85\% of this average and determine the first year (of the whole run) in which the radiative imbalance is below this value. The best estimate is then based on a regression of all years of the simulation run following this year. In subplots (d) of Figures~\ref{fig:results_for_CCSM3}-\ref{fig:results_for_MPIESM12} a plot of $\Delta T$ and $\Delta R$ is given along with the `best estimate' fit.

As the value for $\Delta T_*^\mathrm{est,best}$ significantly incluences the outcome of any comparison and since its value is uncertain even from the long simulation runs, instead of using a single value a range of probably `best estimates' are used. This range is denoted as $\mathcal{R}_{\Delta T_*}$. To find reasonable bounds for this range, we have resampled the data used to compute the best estimate value to construct a distribution of possible estimates for $\Delta T_*$ (we have resampled $N = 10,000$ times and each time have randomly taken $M = 500$ data points). Then, the $5$ and $95$ percentiles of this distribution are taken as the bounds for the range $\mathcal{R}_{\Delta T_*}$. In subplots (e) of figures~\ref{fig:results_for_CCSM3}-\ref{fig:results_for_MPIESM12} the resulting distribution is given along with the best estimate and the $5$ and $95$ percentile values, which are also reported on in table~\ref{tab:LongRunMIP_models}. In subplots (a) and (c) this range is indicated in gray.

\paragraph{Exceptions}\label{sec:IPSLCM5ALR-best_fit_exception}

\begin{description}
	\item[IPSL-CM5A-LR:] The Gregory plot for this model does have an almost horizontal trend near the end of the simulation, making the general resampling process unreliable; it led to a range spanning from values going from $-10^4K$ up to $+10^4K$. As that is not useful in any way, instead of $85\%$ of the average imbalance over the last $50$ years of data, we have used $85\%$ of the average imbalance over \emph{all} years. This does ensure a region is found -- which still is quite large and still lies above the estimated values for all of the estimation techniques.
\end{description}

\subsubsection{Assessment Metrics}

To assess the different estimation techniques, the remaining relative error is used. Since the exact equilibrium warming $\Delta T_*$ is not known for these models, the error is computed as distance to the found range of possible best estimates $\mathcal{R}_{\Delta T_*}$. Thus, the relative error is defined as
\begin{equation}
	e^\mathrm{rel}(t) := \left| \frac{ d\left( \Delta T_*^\mathrm{est}(t), \mathcal{R}_{\Delta T_*} \right) }{ \Delta T_*^\mathrm{est}(t)} \right|,
\end{equation}
where $d\left(x,\mathcal{R}\right)$ denotes the distance between the point $x$ and the range $\mathcal{R}$. That is,
\begin{equation}
	d\left(x,\mathcal{R}\right) := \min_{y \in \mathcal{R}} \| x - y \|.
\end{equation}
The remaining relative error is again given by the maximum of the relative errors that follow at later times:
\begin{equation}
	e^\mathrm{rel}_\mathrm{rem}(t) := \max_{s \geq t} e^\mathrm{rel}(s).
\end{equation}

\begin{figure}
	\centering
	\begin{subfigure}[t]{\textwidth}
		\centering
		\includegraphics{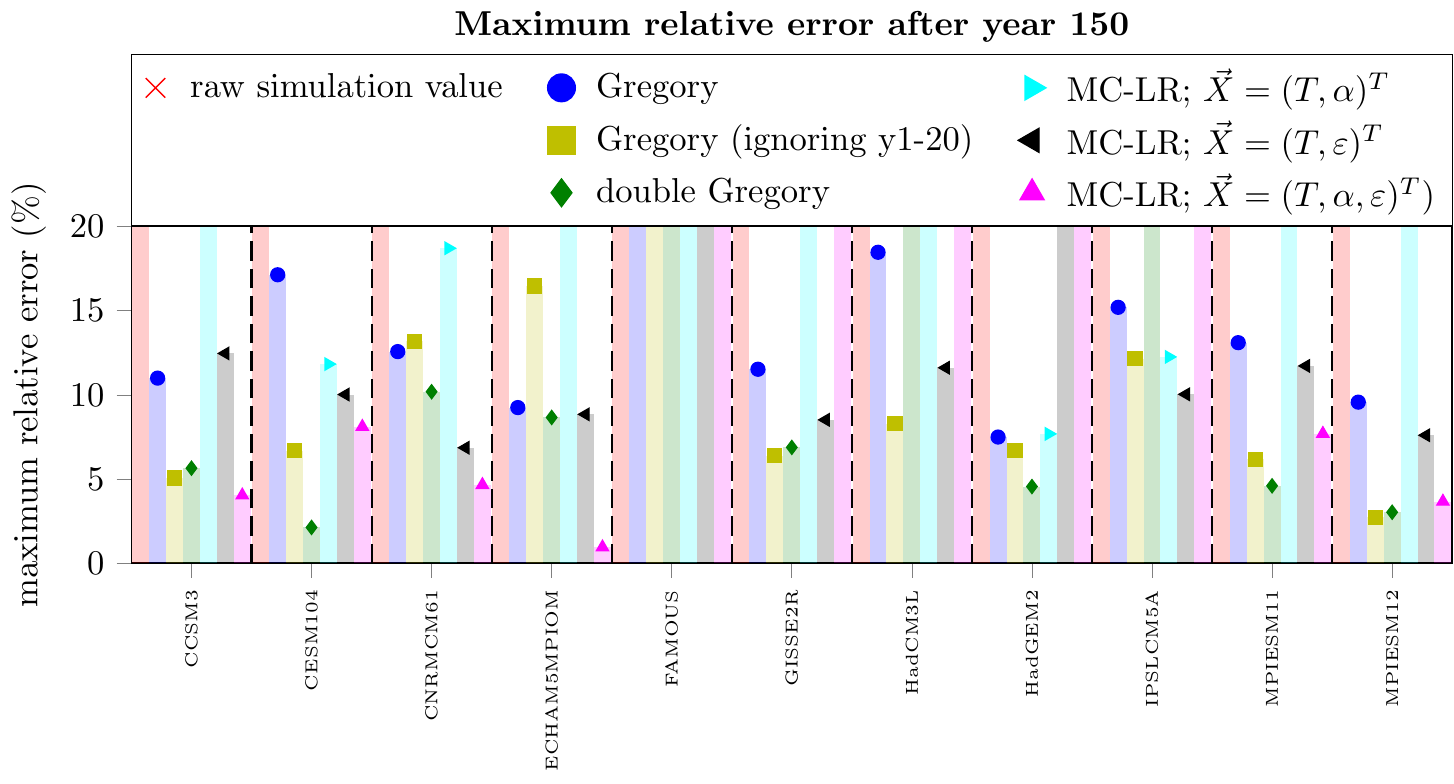}
	\end{subfigure}
	~
	\begin{subfigure}[t]{\textwidth}
		\centering
		\includegraphics{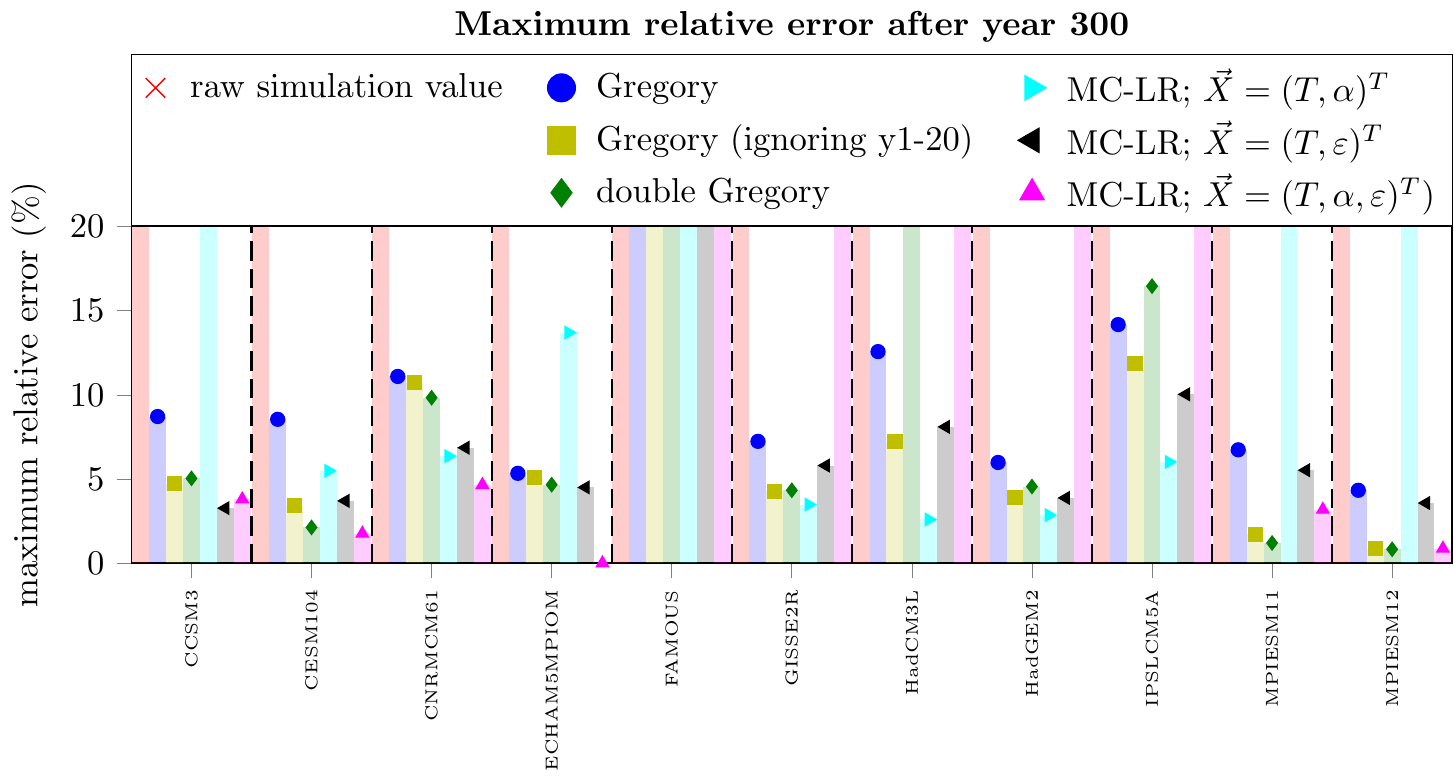}
	\end{subfigure}
	~
	\begin{subfigure}[t]{\textwidth}
		\centering
		\includegraphics{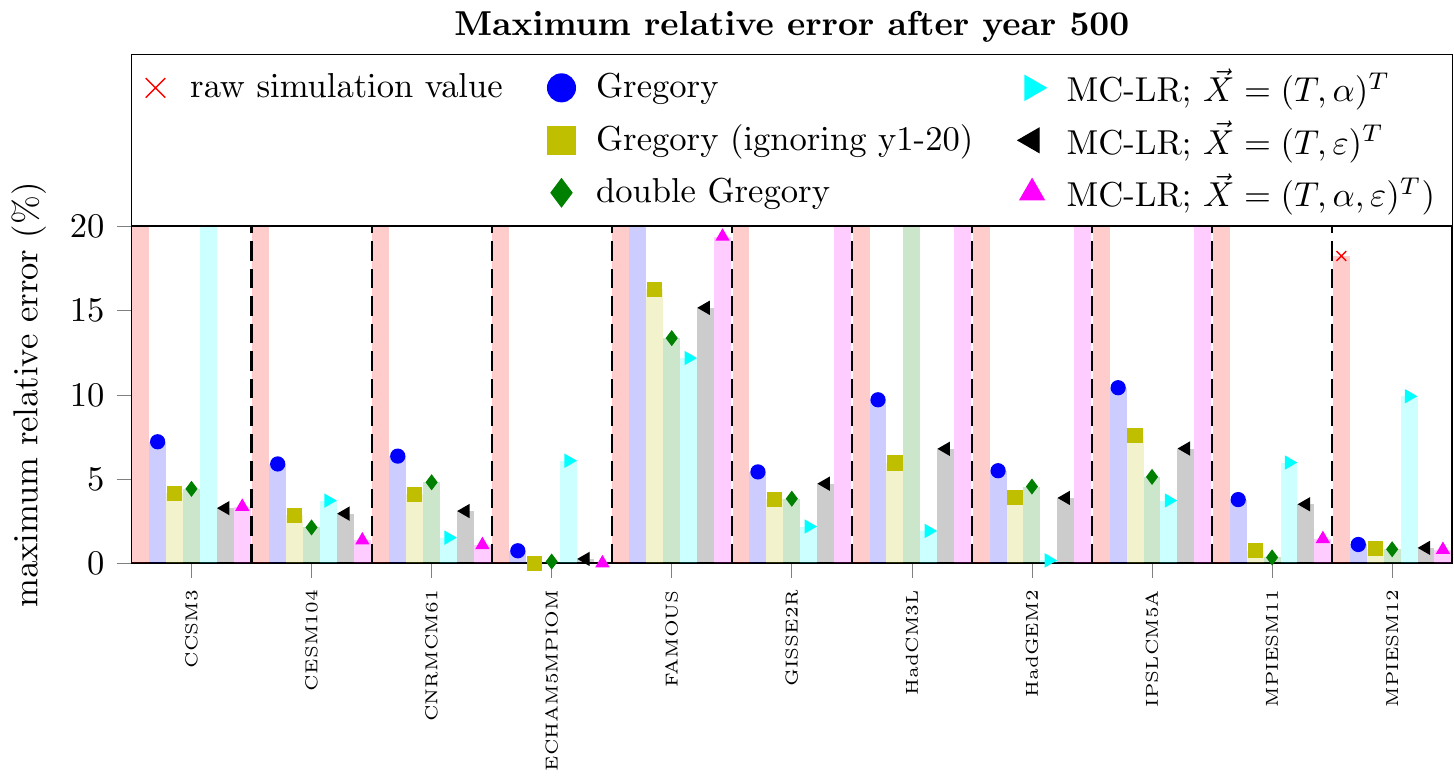}
	\end{subfigure}
	\caption[Remaining relative errors $e_\mathrm{rem}^\mathrm{rel}(t)$ for all estimation techniques on LongRunMIP models]{Remaining relative errors $e_\mathrm{rem}^\mathrm{rel}(t)$ at times $t = 150$ years, $t = 300$ years and $t = 500$ years for the various estimation techniques on the different abrupt $4\times$CO$_2$ experiments in LongRunMIP.}
	\label{fig:remaining_errors_overview}
\end{figure}

The found remaining relative errors for the various estimations are given in subplots (b) of Figures \ref{fig:results_for_CCSM3}-\ref{fig:results_for_MPIESM12}. Moreover, in Figure~\ref{fig:remaining_errors_overview} an overview of the remaining relative error for all models at times $t = 150$ years, $t = 300$ years and $t = 500$ years is given.

\subsection{Discussion of Results on LongRunMIP Models}

\begin{figure}
	\centering
	\begin{subfigure}[t]{0.46\textwidth}
		\centering
		\includegraphics{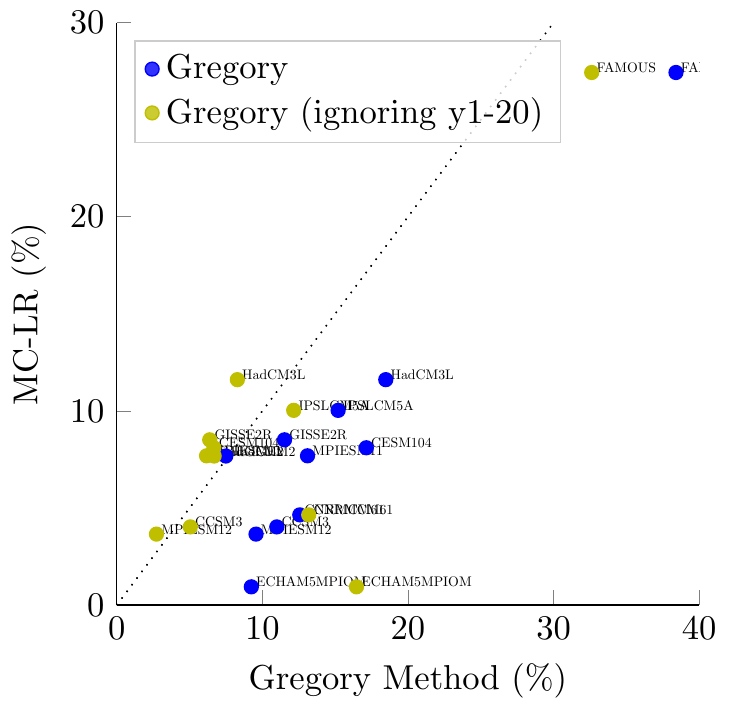}
		\caption{$t = 150y$}
	\end{subfigure}
	~	
	\begin{subfigure}[t]{0.46\textwidth}
		\centering
		\includegraphics{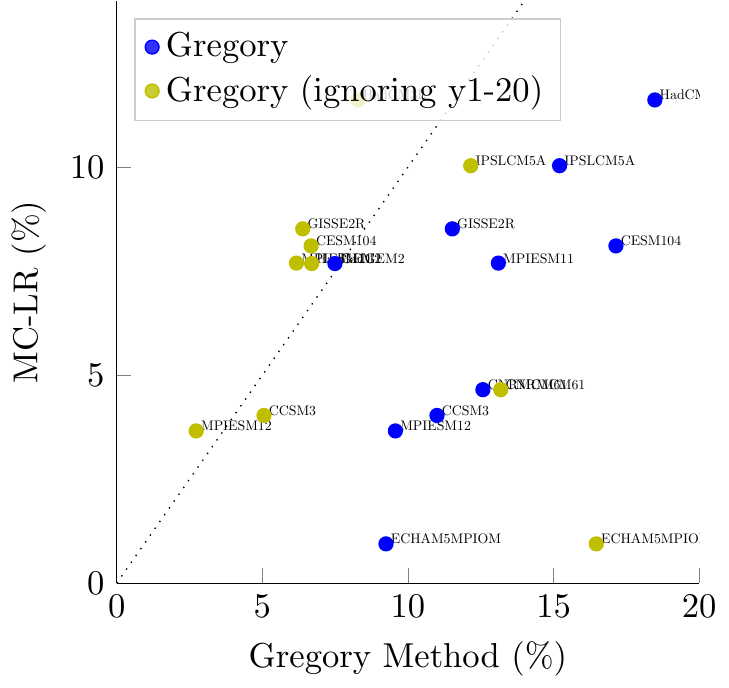}
		\caption{$t=150y$ (zoomed-in)}
	\end{subfigure}
	~	
	\begin{subfigure}[t]{0.46\textwidth}
		\centering
		\includegraphics{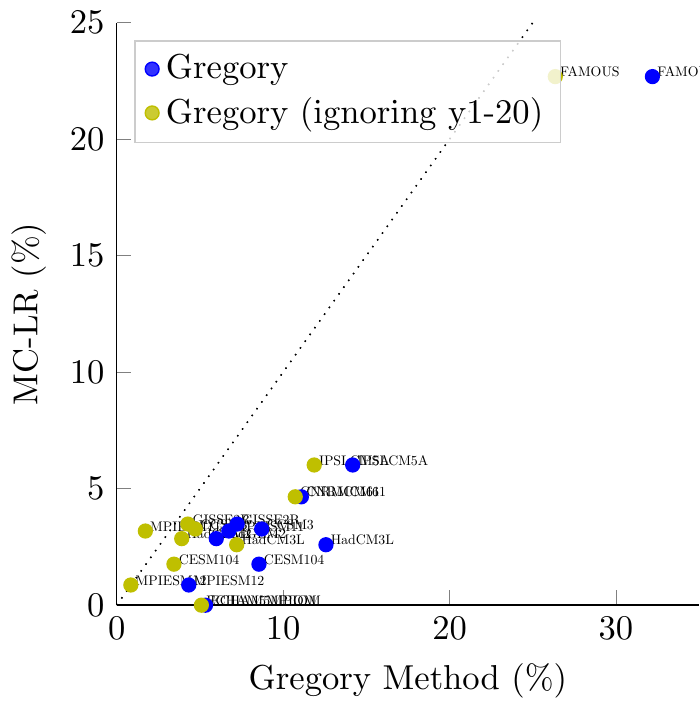}
		\caption{$t = 300y$}
	\end{subfigure}
	~	
	\begin{subfigure}[t]{0.46\textwidth}
		\centering
		\includegraphics{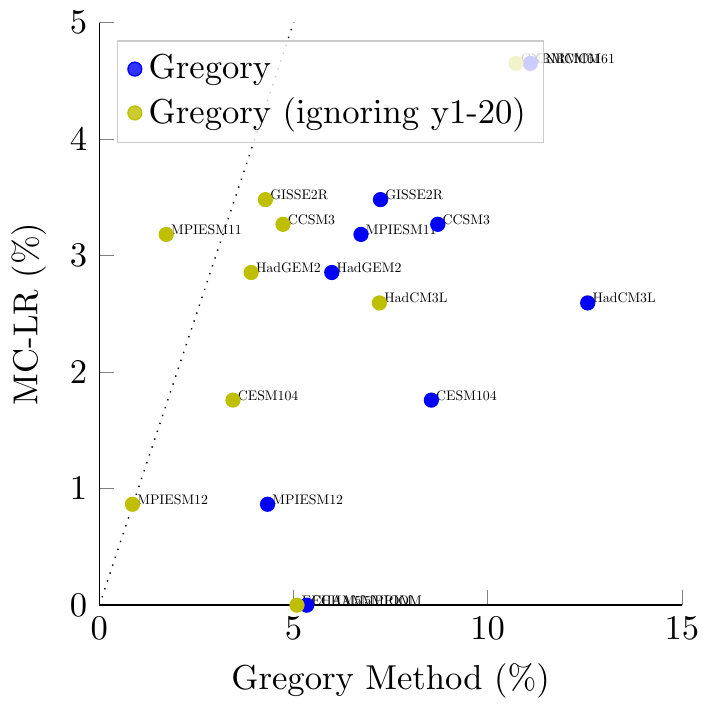}
		\caption{$t=300y$ (zoomed-in)}
	\end{subfigure}
	~
	\begin{subfigure}[t]{0.46\textwidth}
		\centering
		\includegraphics{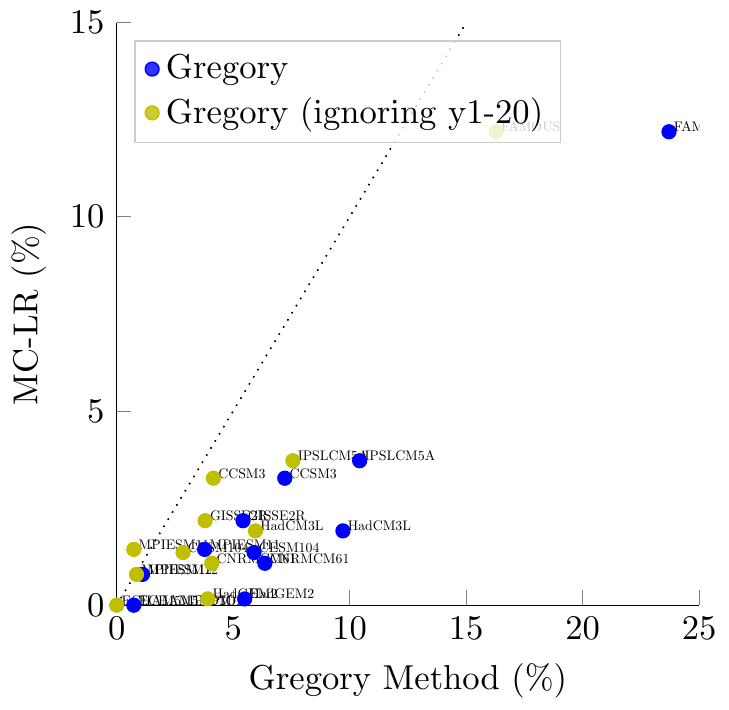}
		\caption{$t = 500y$}
	\end{subfigure}
	~	
	\begin{subfigure}[t]{0.46\textwidth}
		\centering
		\includegraphics{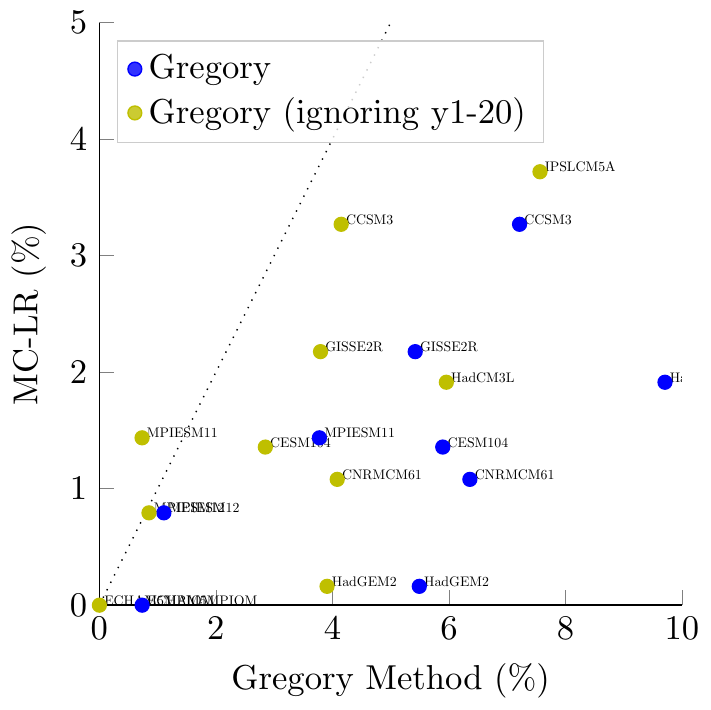}
		\caption{$t=500y$ (zoomed-in)}
	\end{subfigure}
	
	\caption[Comparison of best multi-component linear regression estimate and Gregory estimations]{Comparison of the best multi-component linear regression fit and the classical Gregory estimation techniques. (a-b) time $t = 150 y$, (c-d) time $t = 300 y$ and (e-f) time $t = 500 y$.}
	\label{fig:technique_comparison}
\end{figure}

\begin{figure}
	\centering
	{\bf \huge CCSM3}
	\vspace{1em}
	
	\begin{subfigure}[t]{0.45 \textwidth}
		\centering
		\includegraphics[width=\textwidth]{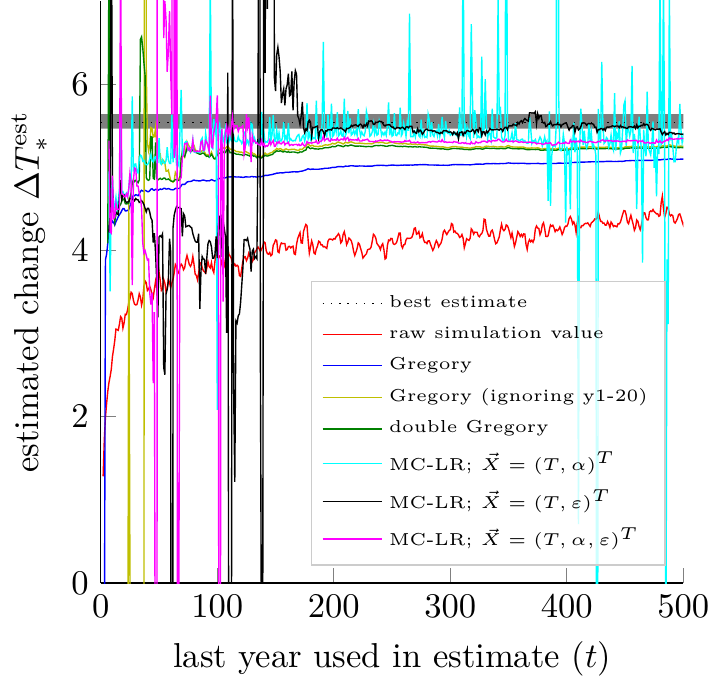}
		\caption{}
	\end{subfigure}
	~
	\begin{subfigure}[t]{0.45 \textwidth}
		\centering
		\includegraphics[width=\textwidth]{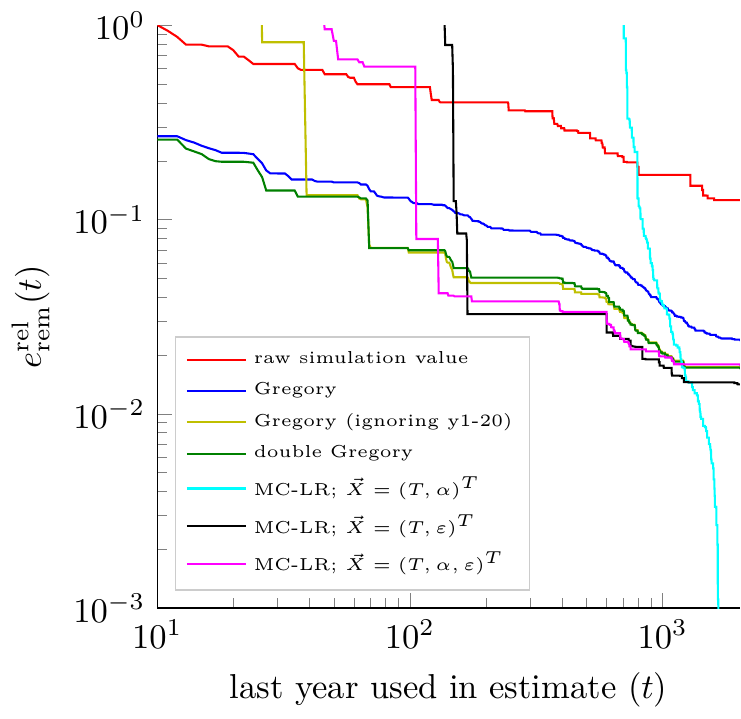}
		\caption{}
	\end{subfigure}
	\\
	\begin{subfigure}[t]{0.32 \textwidth}
		\centering
		\includegraphics[width=\textwidth]{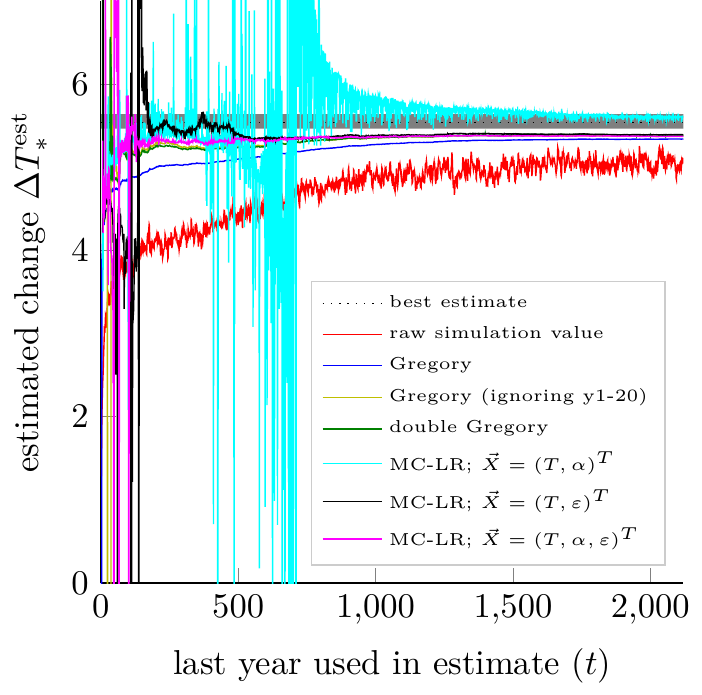}		
		\caption{}
	\end{subfigure}
	\begin{subfigure}[t]{0.32 \textwidth}
		\centering
		\includegraphics[width=\textwidth]{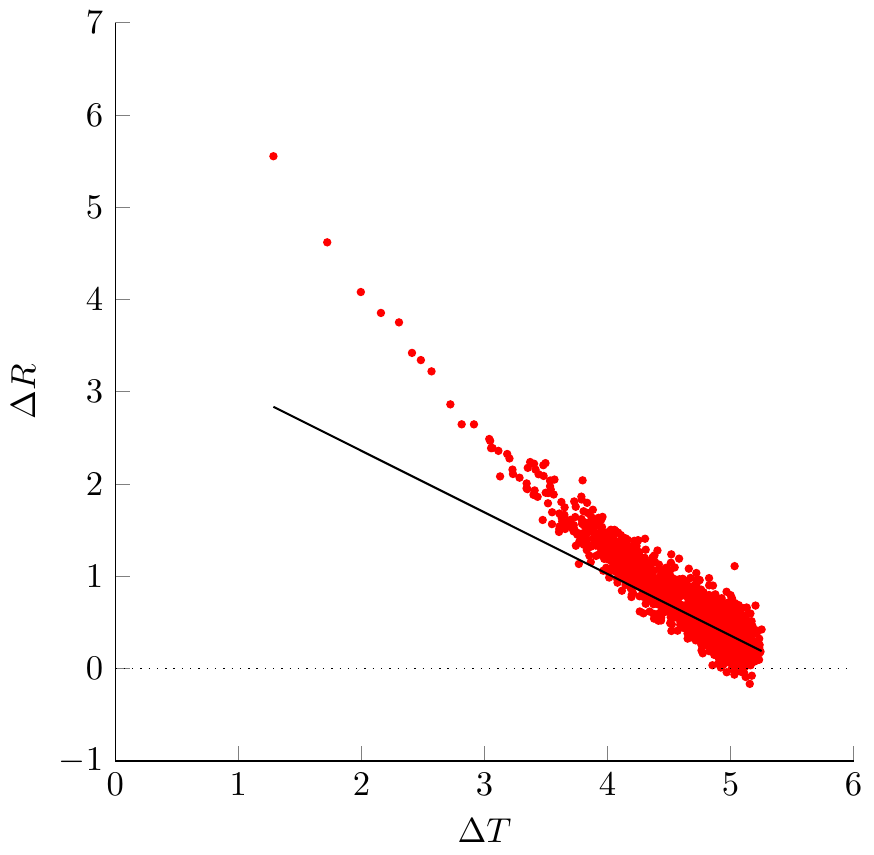}	
		\caption{}
	\end{subfigure}
	\begin{subfigure}[t]{0.32 \textwidth}
		\centering
		\includegraphics[width=\textwidth]{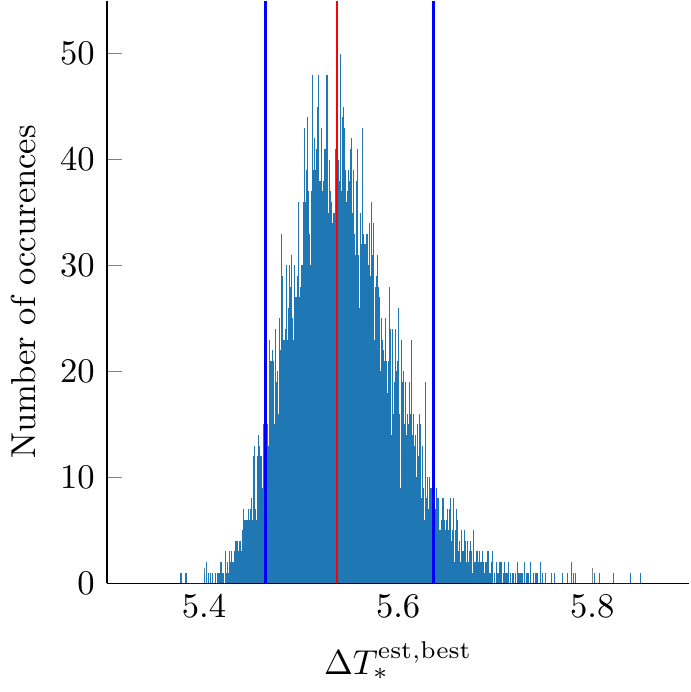}	
		\caption{}
	\end{subfigure}
	
\caption[Results for the model CCSM3]{Results for the model CCSM3. (a) estimated equilibrium warming $\Delta T_*^\mathrm{est}(t)$ for the first $500$ years of data. (b) remaining relative error over time. (c) estimated equilibrium warming for the whole simulation. (d) `Gregory' plot of $\Delta R$ versus $\Delta T$ including fit for the best estimate $\Delta T_*^\mathrm{est,best}$. (e) Histogram for resampling of $\Delta_*^\mathrm{est,best}$.}
\label{fig:results_for_CCSM3}
\end{figure}

\begin{figure}
	\centering
	{\bf \huge CESM 1.0.4}
	\vspace{1em}
	
	\begin{subfigure}[t]{0.45 \textwidth}
		\centering
		\includegraphics[width=\textwidth]{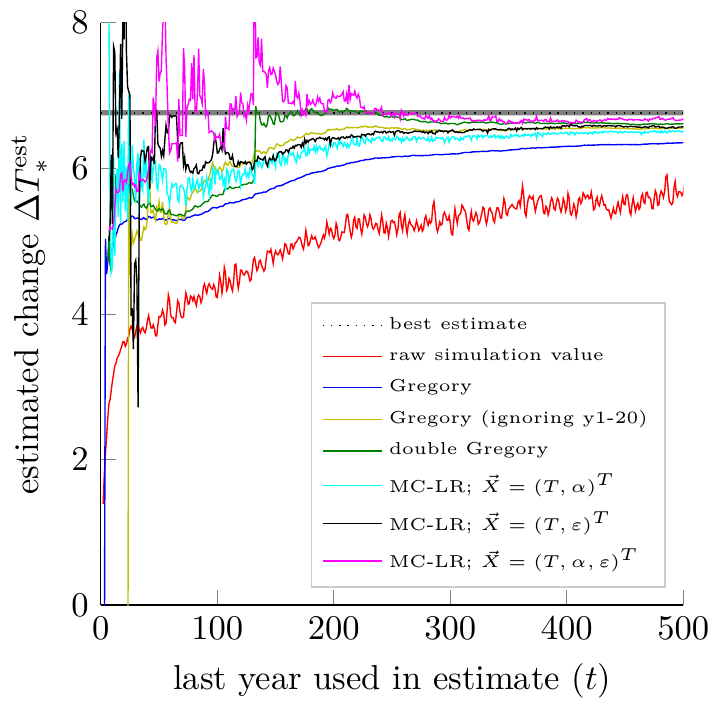}
		\caption{}
	\end{subfigure}
	~
	\begin{subfigure}[t]{0.45 \textwidth}
		\centering
		\includegraphics[width=\textwidth]{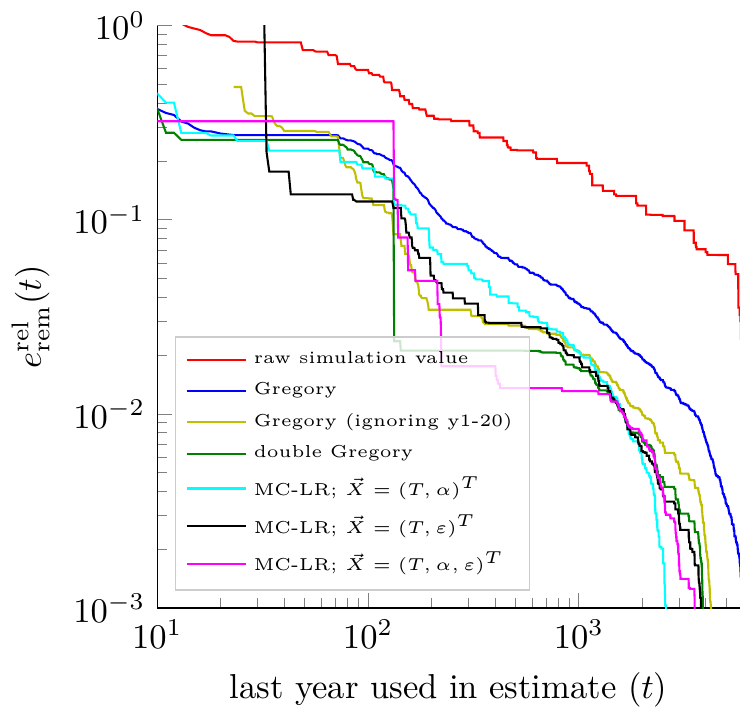}
		\caption{}
	\end{subfigure}
	\\
	\begin{subfigure}[t]{0.32 \textwidth}
		\centering
		\includegraphics[width=\textwidth]{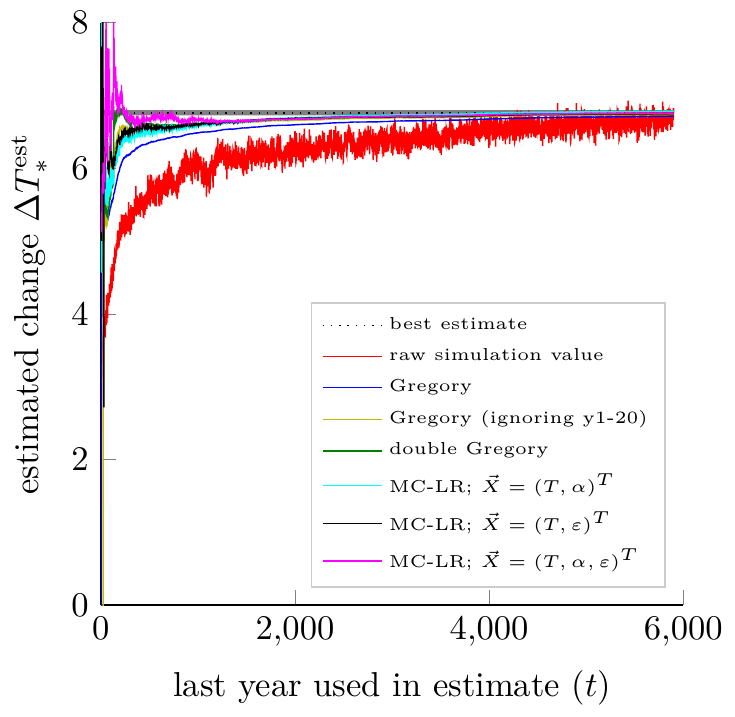}		
		\caption{}
	\end{subfigure}
	\begin{subfigure}[t]{0.32 \textwidth}
		\centering
		\includegraphics[width=\textwidth]{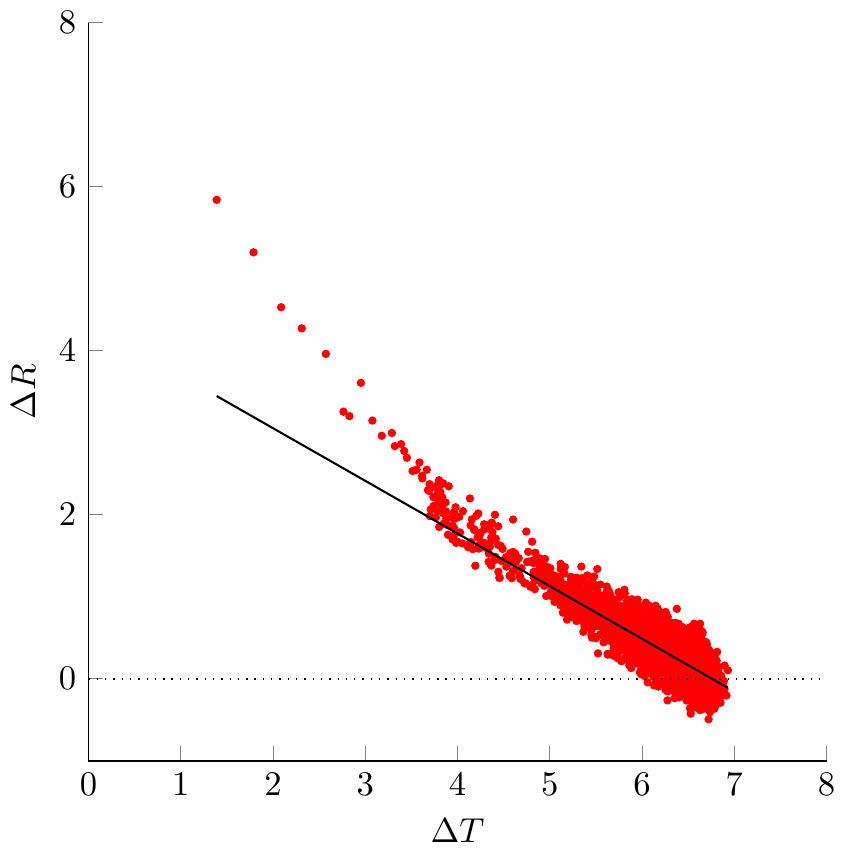}	
		\caption{}
	\end{subfigure}
	\begin{subfigure}[t]{0.32 \textwidth}
		\centering
		\includegraphics[width=\textwidth]{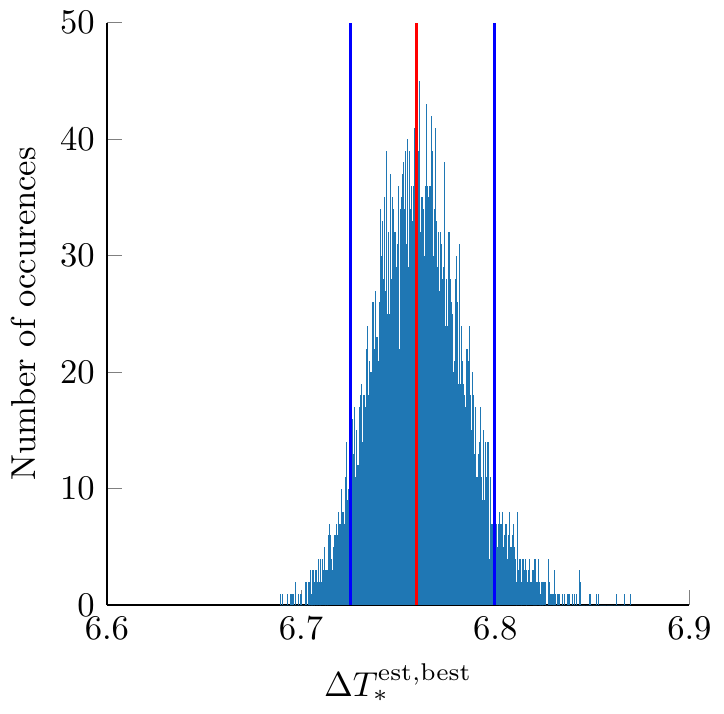}	
		\caption{}
	\end{subfigure}
	
\caption[Results for the model CESM 1.0.4.]{Results for the model CESM 1.0.4. (a) estimated equilibrium warming $\Delta T_*^\mathrm{est}(t)$ for the first $500$ years of data. (b) remaining relative error over time. (c) estimated equilibrium warming for the whole simulation. (d) `Gregory' plot of $\Delta R$ versus $\Delta T$ including fit for the best estimate $\Delta T_*^\mathrm{est,best}$. (e) Histogram for resampling of $\Delta_*^\mathrm{est,best}$.}
\label{fig:results_for_CESM104}
\end{figure}

\begin{figure}
	\centering
	{\bf \huge CNRM-CM6-1}
	\vspace{1em}
	
	\begin{subfigure}[t]{0.45 \textwidth}
		\centering
		\includegraphics[width=\textwidth]{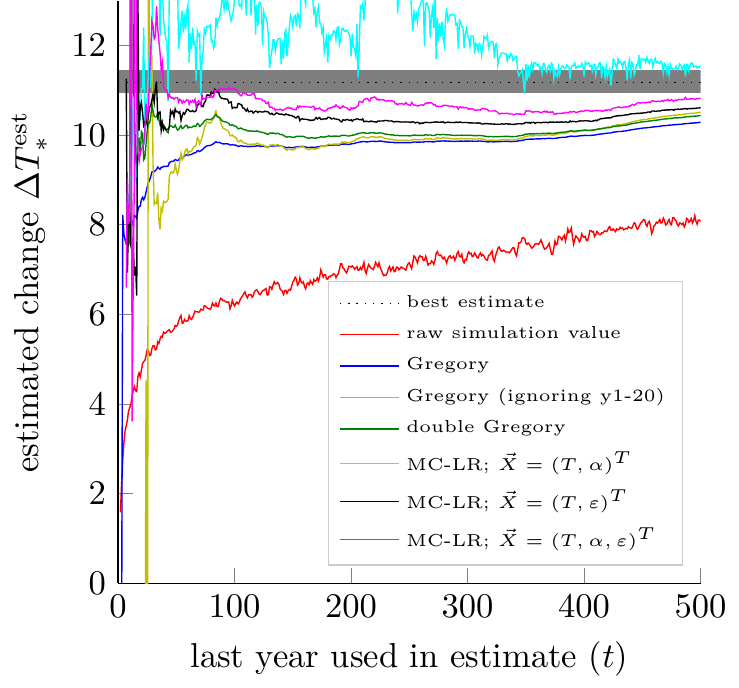}
		\caption{}
	\end{subfigure}
	~
	\begin{subfigure}[t]{0.45 \textwidth}
		\centering
		\includegraphics[width=\textwidth]{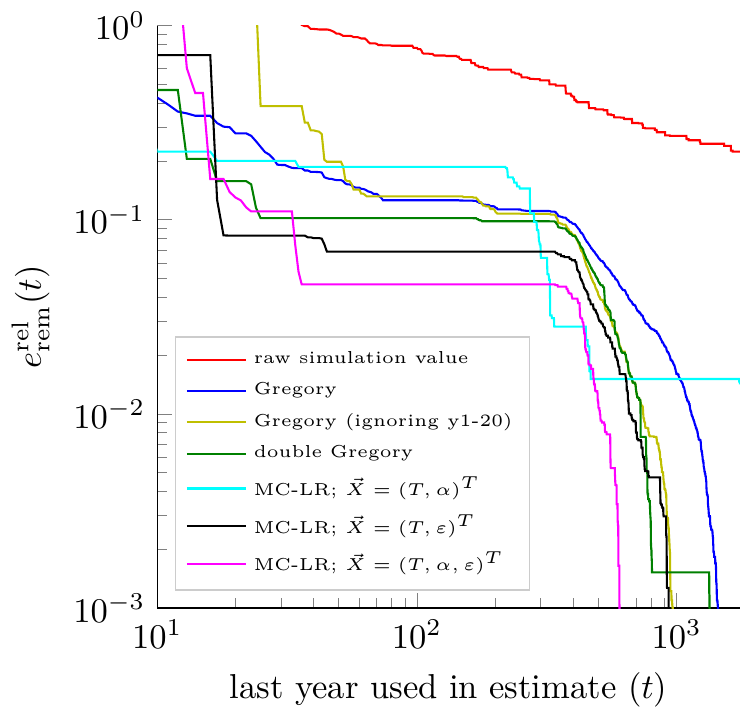}
		\caption{}
	\end{subfigure}
	\\
	\begin{subfigure}[t]{0.32 \textwidth}
		\centering
		\includegraphics[width=\textwidth]{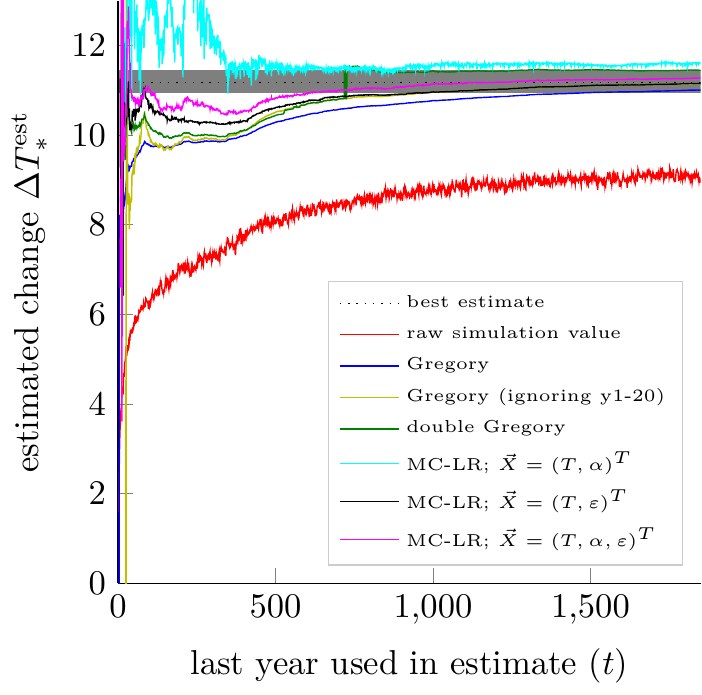}		
		\caption{}
	\end{subfigure}
	\begin{subfigure}[t]{0.32 \textwidth}
		\centering
		\includegraphics[width=\textwidth]{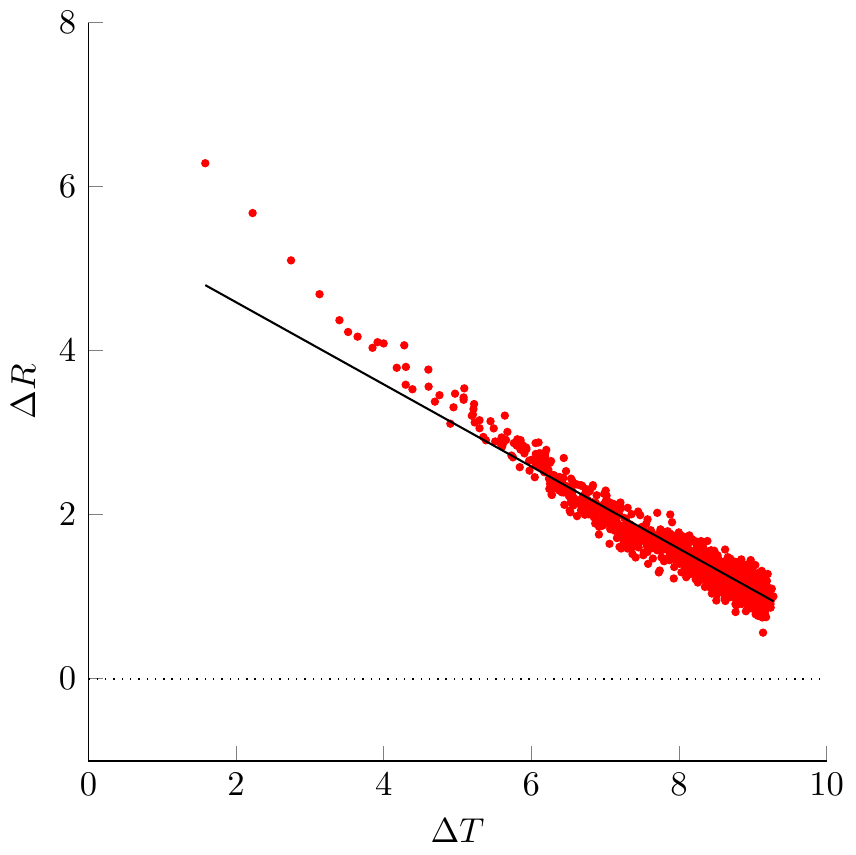}	
		\caption{}
	\end{subfigure}
	\begin{subfigure}[t]{0.32 \textwidth}
		\centering
		\includegraphics[width=\textwidth]{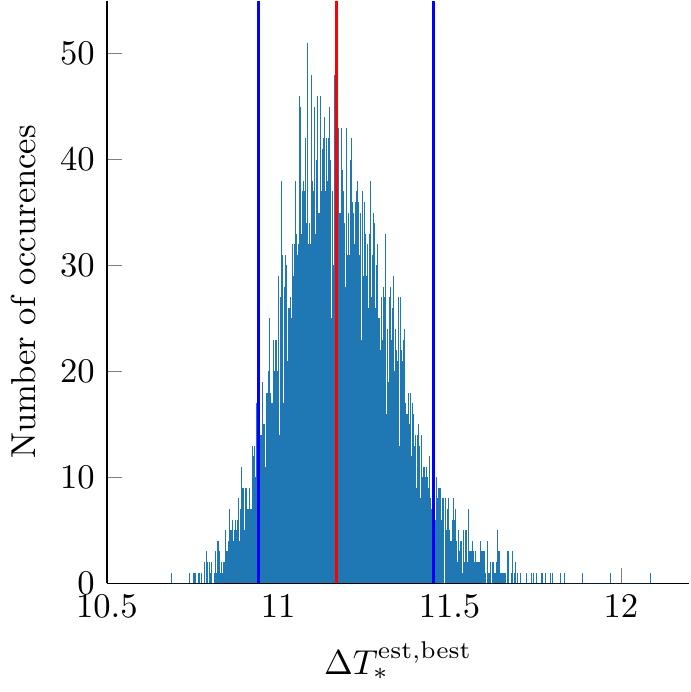}	
		\caption{}
	\end{subfigure}
	
\caption[Results for the model CNRM-CM6-1]{Results for the model CNRM-CM6-1. (a) estimated equilibrium warming $\Delta T_*^\mathrm{est}(t)$ for the first $500$ years of data. (b) remaining relative error over time. (c) estimated equilibrium warming for the whole simulation. (d) `Gregory' plot of $\Delta R$ versus $\Delta T$ including fit for the best estimate $\Delta T_*^\mathrm{est,best}$. (e) Histogram for resampling of $\Delta_*^\mathrm{est,best}$.}
\label{fig:results_for_CNRMCM61}
\end{figure}

\begin{figure}
	\centering
	{\bf \huge ECHAM5/MPIOM}
	\vspace{1em}
	
	\begin{subfigure}[t]{0.45 \textwidth}
		\centering
		\includegraphics[width=\textwidth]{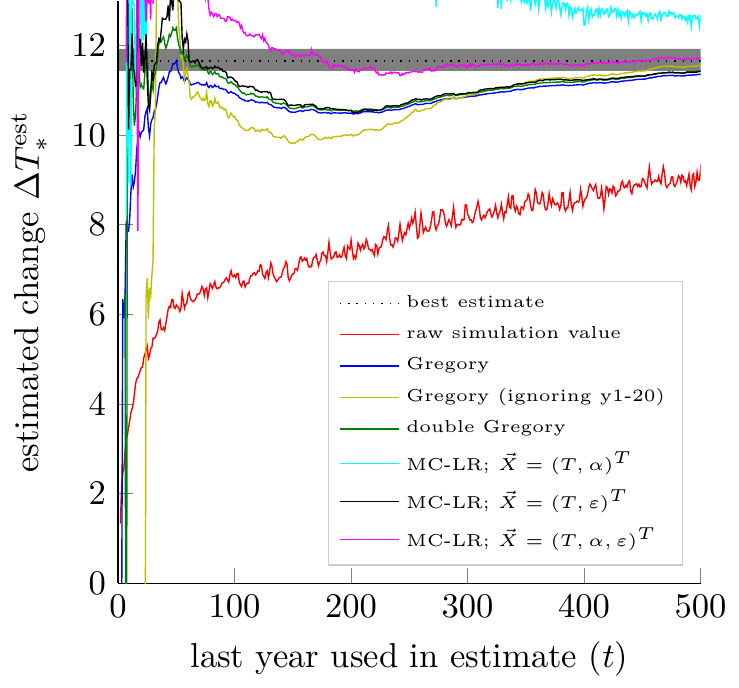}
		\caption{}
	\end{subfigure}
	~
	\begin{subfigure}[t]{0.45 \textwidth}
		\centering
		\includegraphics[width=\textwidth]{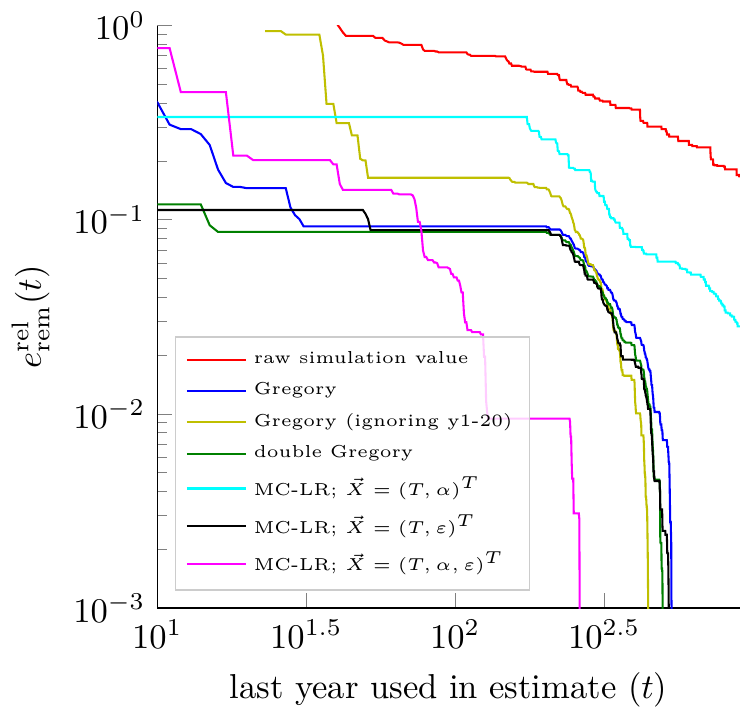}
		\caption{}
	\end{subfigure}
	\\
	\begin{subfigure}[t]{0.32 \textwidth}
		\centering
		\includegraphics[width=\textwidth]{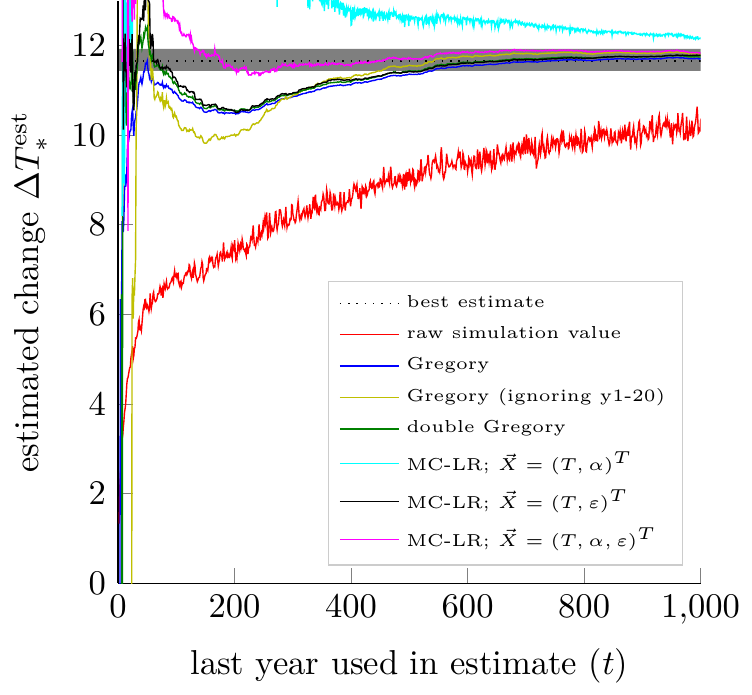}		
		\caption{}
	\end{subfigure}
	\begin{subfigure}[t]{0.32 \textwidth}
		\centering
		\includegraphics[width=\textwidth]{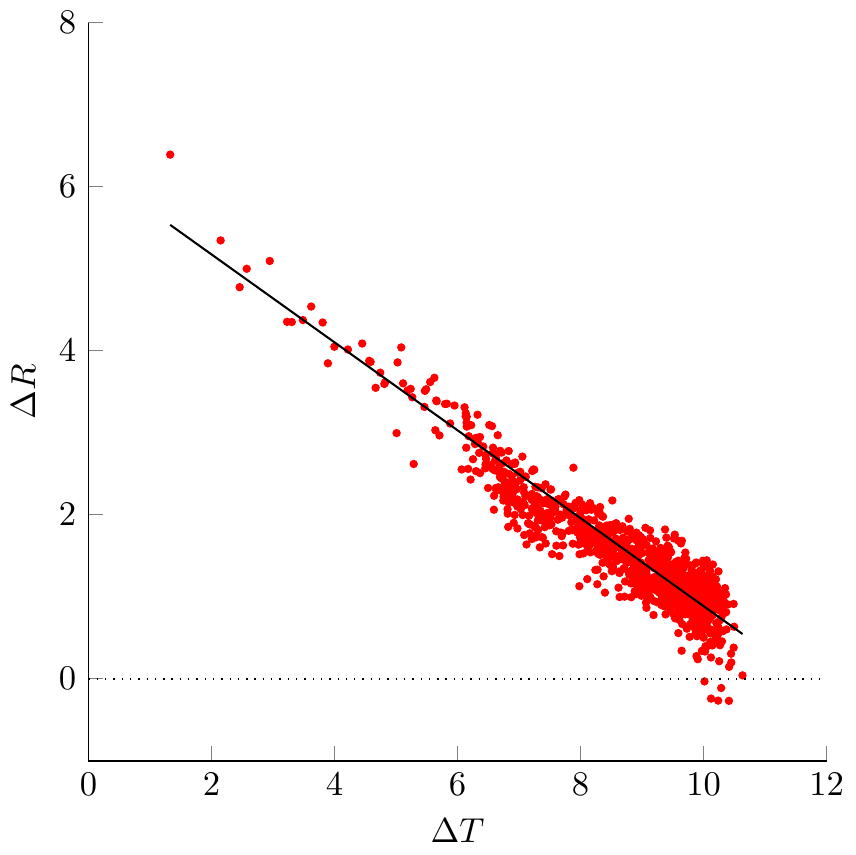}	
		\caption{}
	\end{subfigure}
	\begin{subfigure}[t]{0.32 \textwidth}
		\centering
		\includegraphics[width=\textwidth]{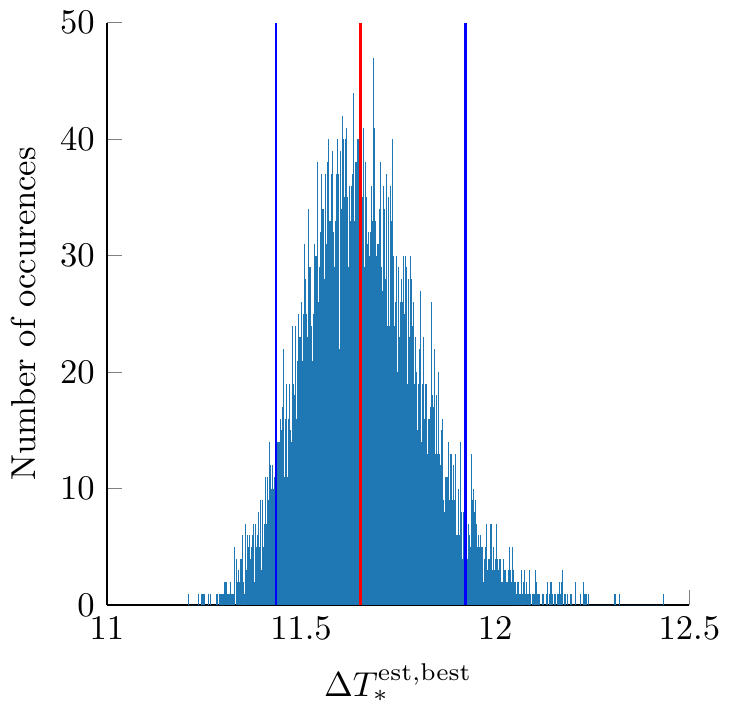}	
		\caption{}
	\end{subfigure}
	
\caption[Results for the model ECHAM/5MPIOM]{Results for the model ECHAM/5MPIOM. (a) estimated equilibrium warming $\Delta T_*^\mathrm{est}(t)$ for the first $500$ years of data. (b) remaining relative error over time. (c) estimated equilibrium warming for the whole simulation. (d) `Gregory' plot of $\Delta R$ versus $\Delta T$ including fit for the best estimate $\Delta T_*^\mathrm{est,best}$. (e) Histogram for resampling of $\Delta_*^\mathrm{est,best}$.}
\label{fig:results_for_ECHAM5MPIOM}
\end{figure}

\begin{figure}
	\centering
	{\bf \huge FAMOUS}
	\vspace{1em}
	
	\begin{subfigure}[t]{0.45 \textwidth}
		\centering
		\includegraphics[width=\textwidth]{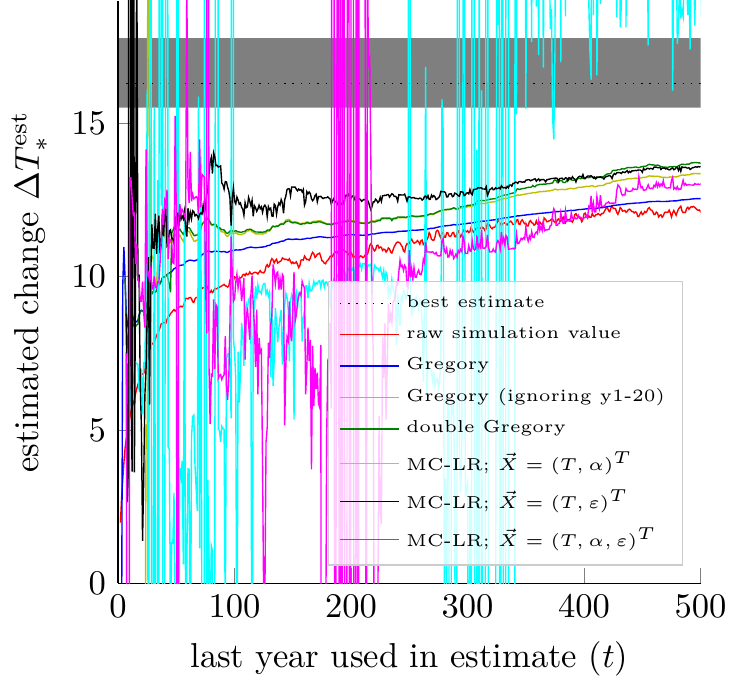}
		\caption{}
	\end{subfigure}
	~
	\begin{subfigure}[t]{0.45 \textwidth}
		\centering
		\includegraphics[width=\textwidth]{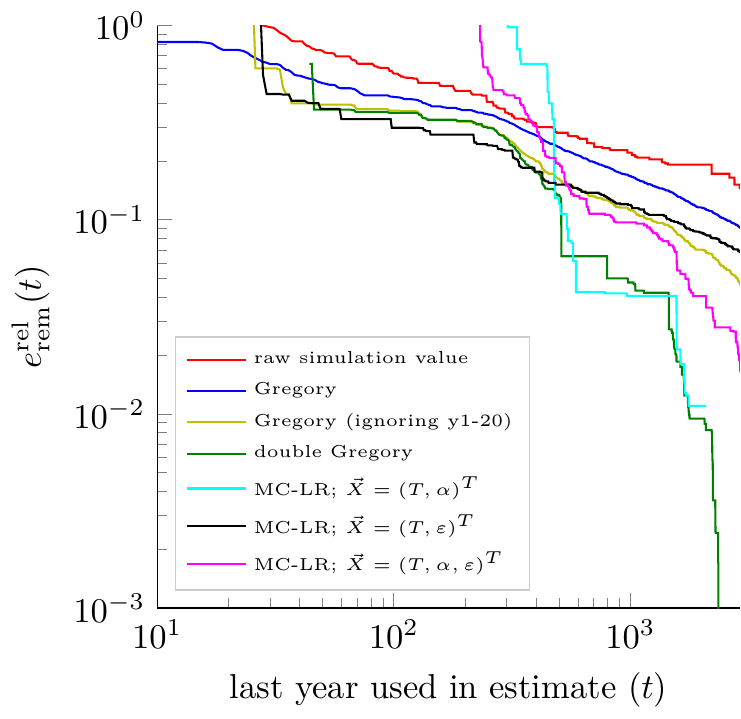}
		\caption{}
	\end{subfigure}
	\\
	\begin{subfigure}[t]{0.32 \textwidth}
		\centering
		\includegraphics[width=\textwidth]{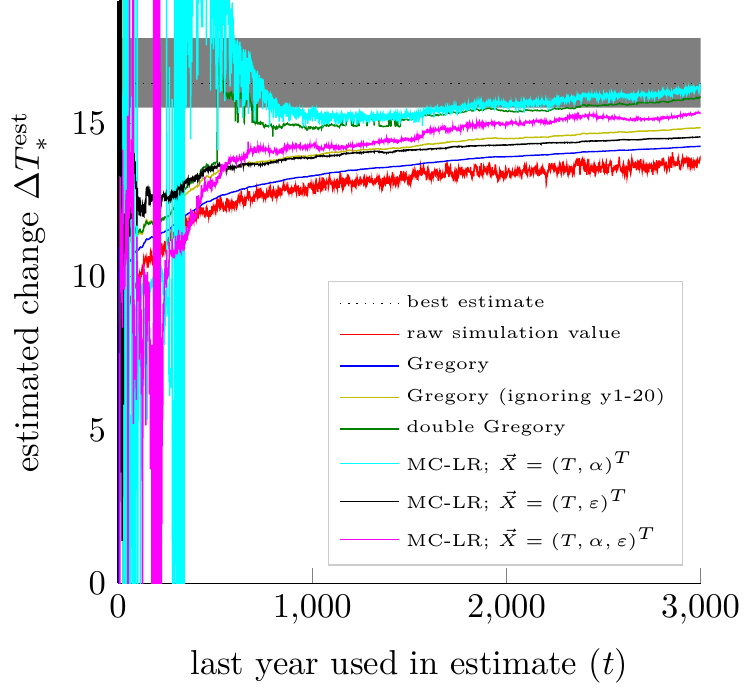}		
		\caption{}
	\end{subfigure}
	\begin{subfigure}[t]{0.32 \textwidth}
		\centering
		\includegraphics[width=\textwidth]{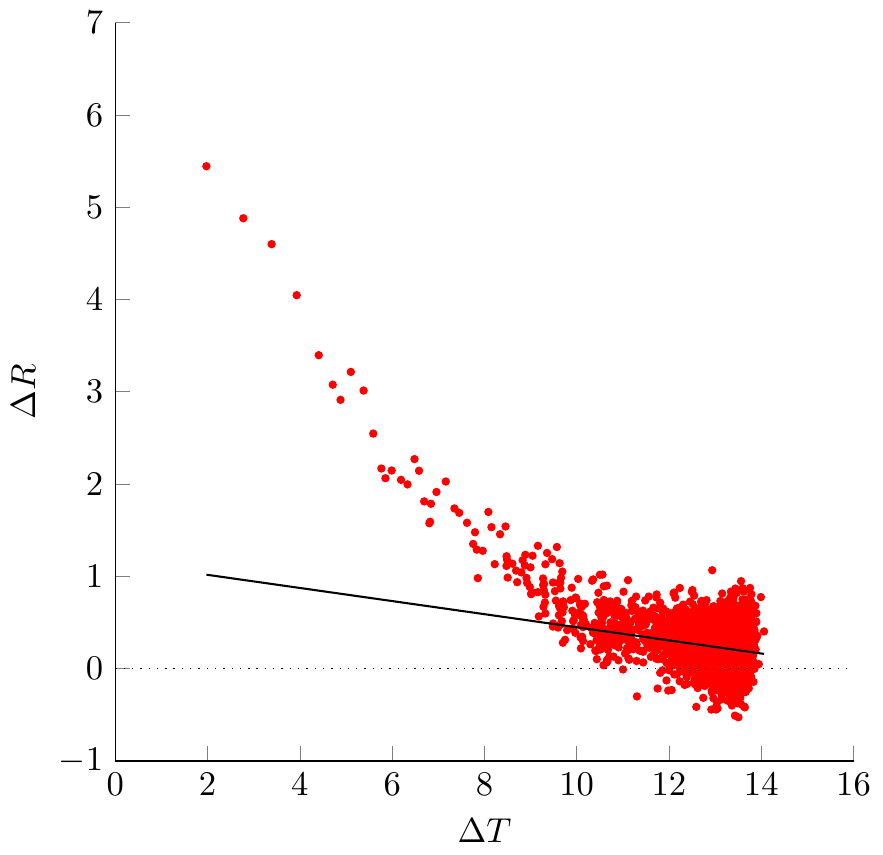}	
		\caption{}
	\end{subfigure}
	\begin{subfigure}[t]{0.32 \textwidth}
		\centering
		\includegraphics[width=\textwidth]{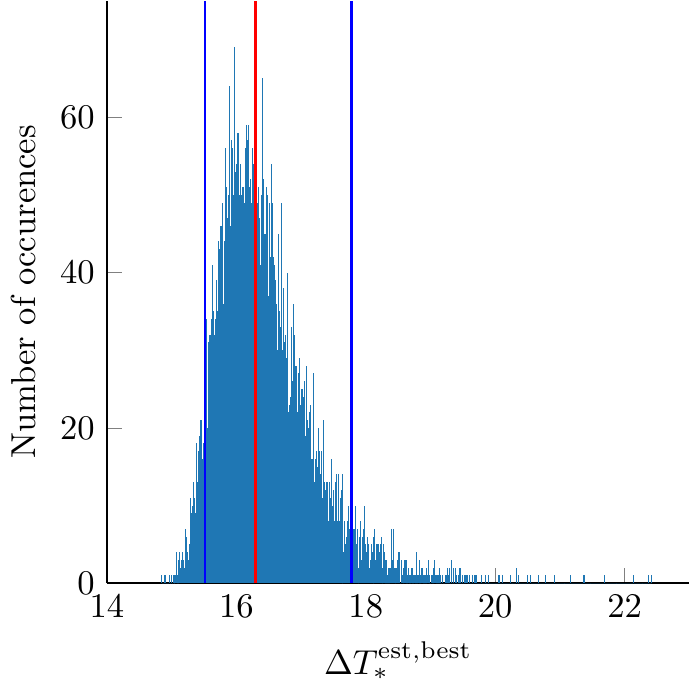}	
		\caption{}
	\end{subfigure}
	
\caption[Results for the model FAMOUS]{Results for the model FAMOUS. (a) estimated equilibrium warming $\Delta T_*^\mathrm{est}(t)$ for the first $500$ years of data. (b) remaining relative error over time. (c) estimated equilibrium warming for the whole simulation. (d) `Gregory' plot of $\Delta R$ versus $\Delta T$ including fit for the best estimate $\Delta T_*^\mathrm{est,best}$. (e) Histogram for resampling of $\Delta_*^\mathrm{est,best}$.}
\label{fig:results_for_FAMOUS}
\end{figure}

\begin{figure}
	\centering
	{\bf \huge GISS-E2-R}
	\vspace{1em}
	
	\begin{subfigure}[t]{0.45 \textwidth}
		\centering
		\includegraphics[width=\textwidth]{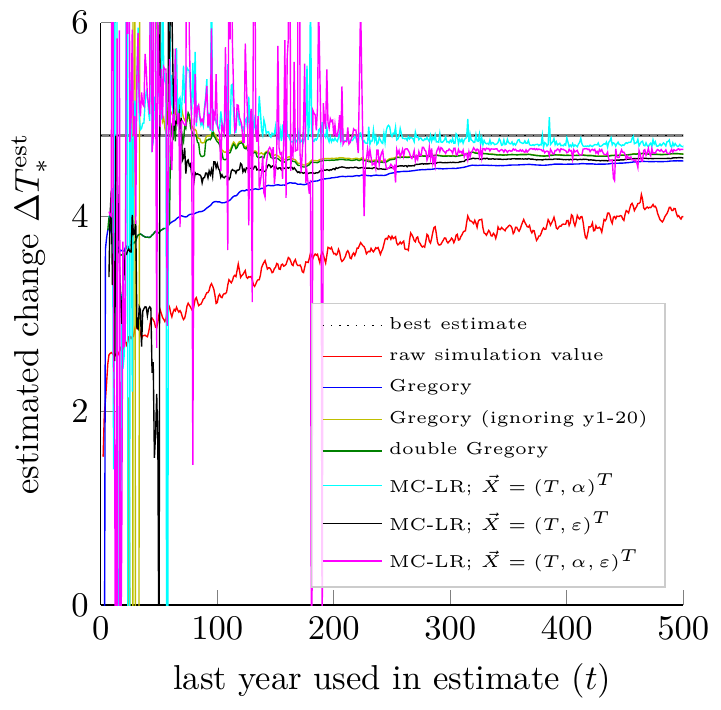}
		\caption{}
	\end{subfigure}
	~
	\begin{subfigure}[t]{0.45 \textwidth}
		\centering
		\includegraphics[width=\textwidth]{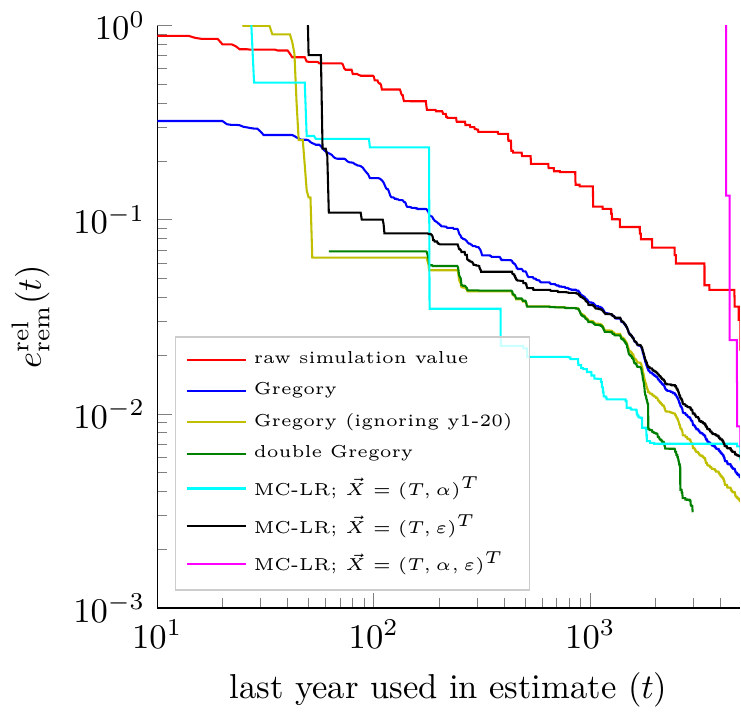}
		\caption{}
	\end{subfigure}
	\\
	\begin{subfigure}[t]{0.32 \textwidth}
		\centering
		\includegraphics[width=\textwidth]{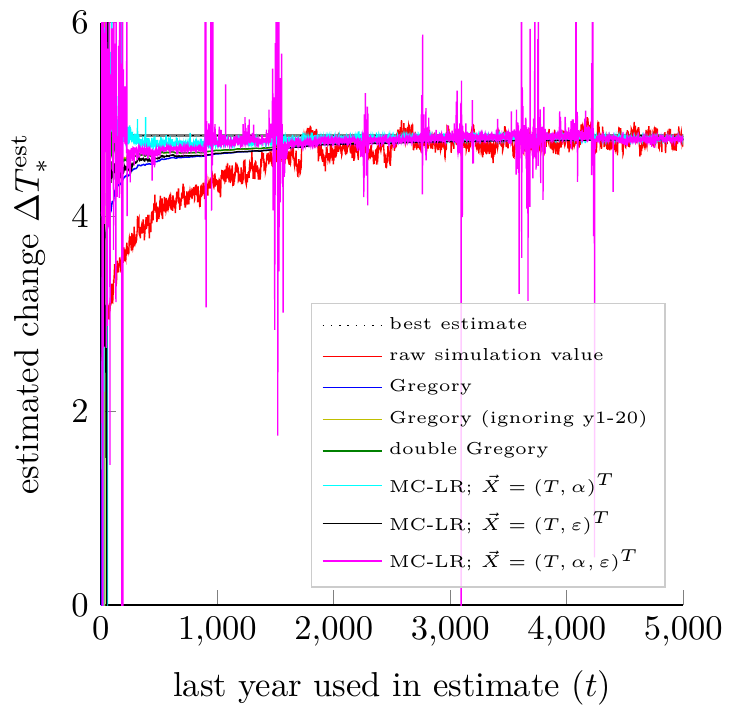}		
		\caption{}
	\end{subfigure}
	\begin{subfigure}[t]{0.32 \textwidth}
		\centering
		\includegraphics[width=\textwidth]{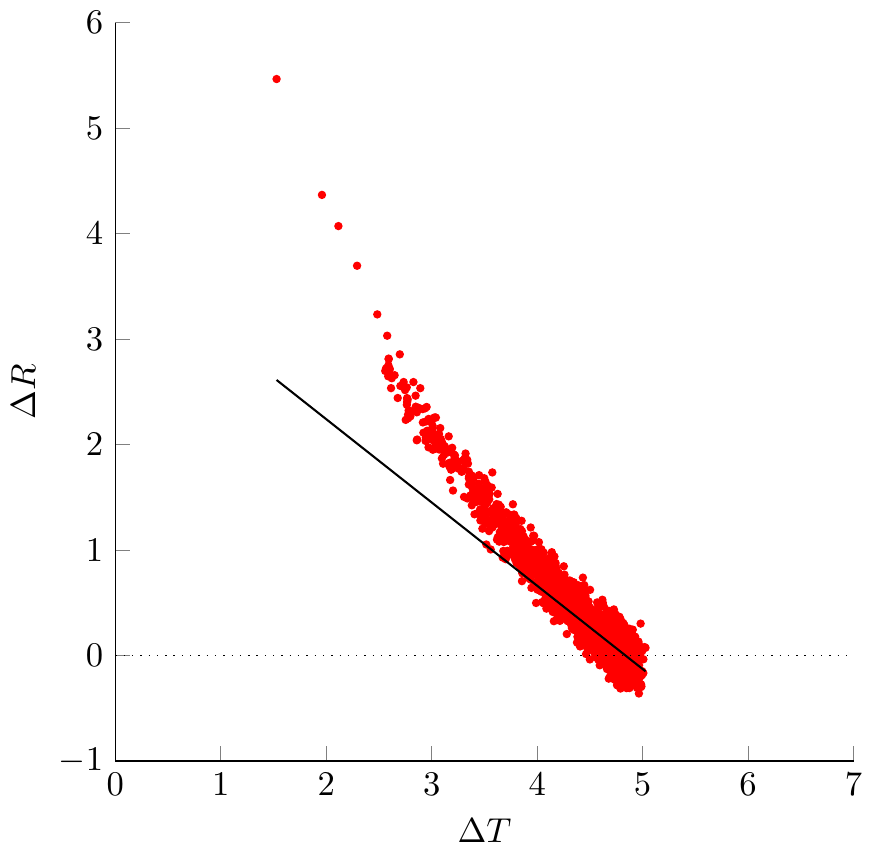}	
		\caption{}
	\end{subfigure}
	\begin{subfigure}[t]{0.32 \textwidth}
		\centering
		\includegraphics[width=\textwidth]{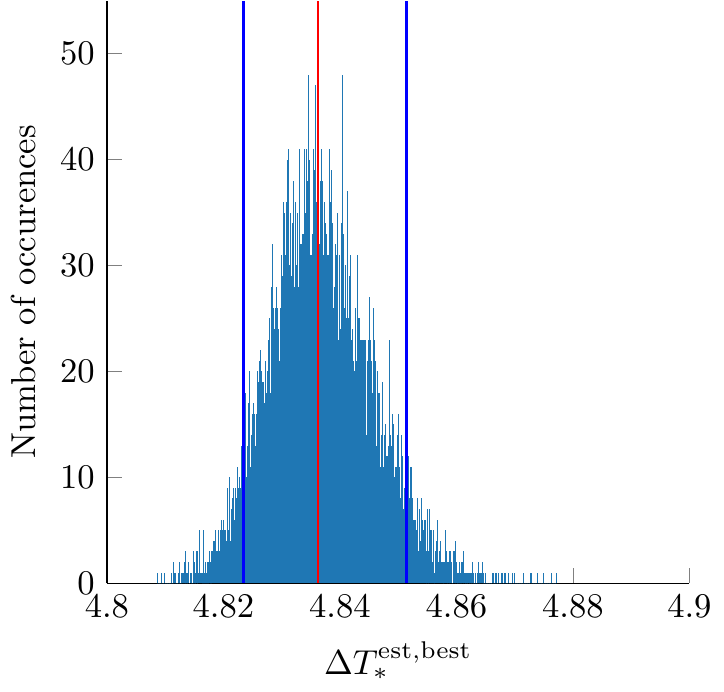}	
		\caption{}
	\end{subfigure}
	
\caption[Results for the model GISS-E2-R]{Results for the model GISS-E2-R. (a) estimated equilibrium warming $\Delta T_*^\mathrm{est}(t)$ for the first $500$ years of data. (b) remaining relative error over time. (c) estimated equilibrium warming for the whole simulation. (d) `Gregory' plot of $\Delta R$ versus $\Delta T$ including fit for the best estimate $\Delta T_*^\mathrm{est,best}$. (e) Histogram for resampling of $\Delta_*^\mathrm{est,best}$.}
\label{fig:results_for_GISSE2R}
\end{figure}

\begin{figure}
	\centering
	{\bf \huge HadCM3L}
	\vspace{1em}
	
	\begin{subfigure}[t]{0.45 \textwidth}
		\centering
		\includegraphics[width=\textwidth]{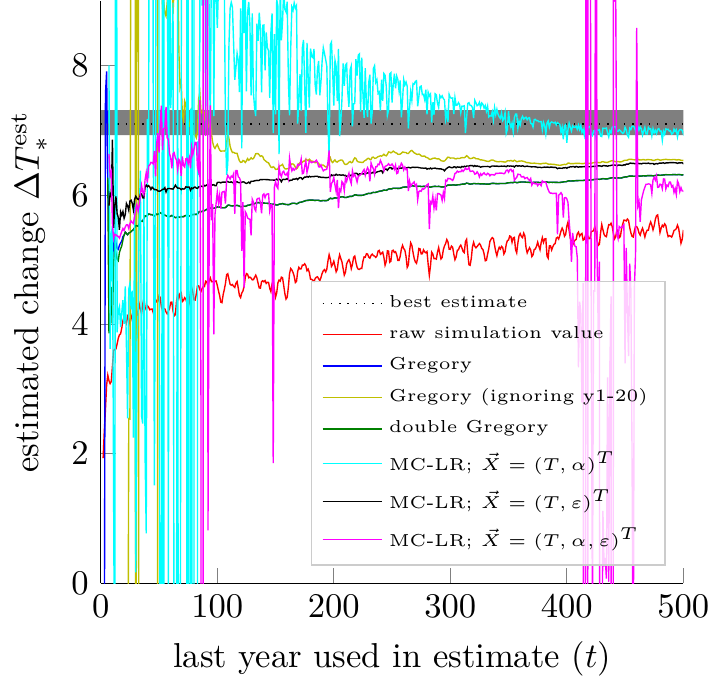}
		\caption{}
	\end{subfigure}
	~
	\begin{subfigure}[t]{0.45 \textwidth}
		\centering
		\includegraphics[width=\textwidth]{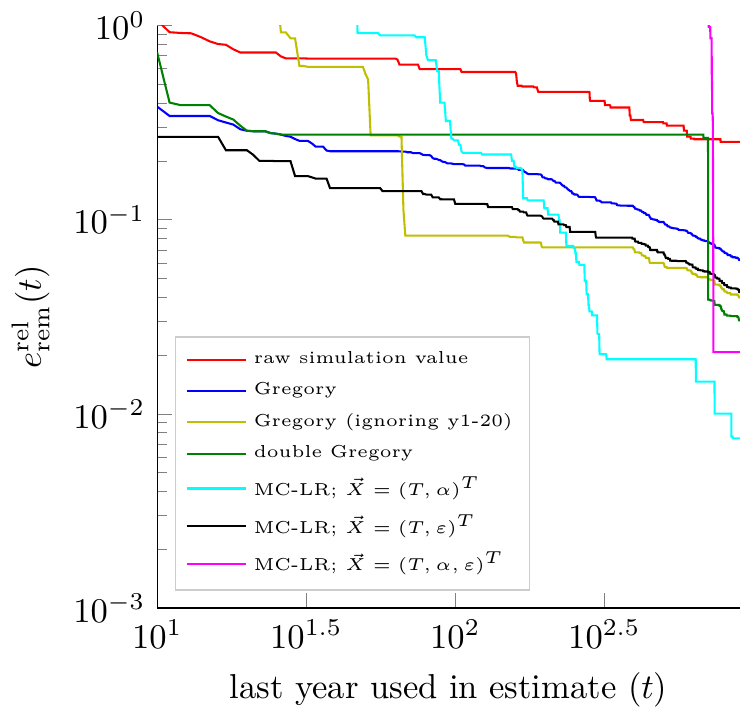}
		\caption{}
	\end{subfigure}
	\\
	\begin{subfigure}[t]{0.32 \textwidth}
		\centering
		\includegraphics[width=\textwidth]{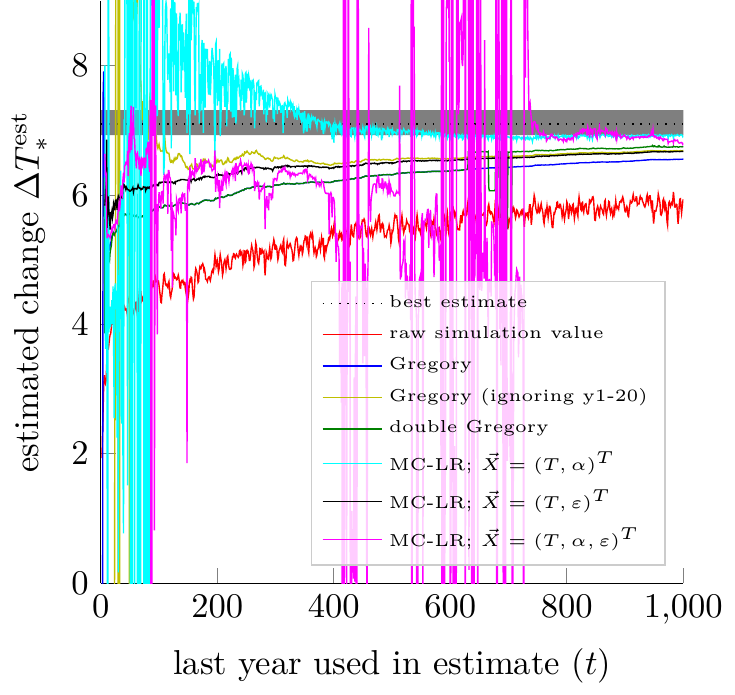}		
		\caption{}
	\end{subfigure}
	\begin{subfigure}[t]{0.32 \textwidth}
		\centering
		\includegraphics[width=\textwidth]{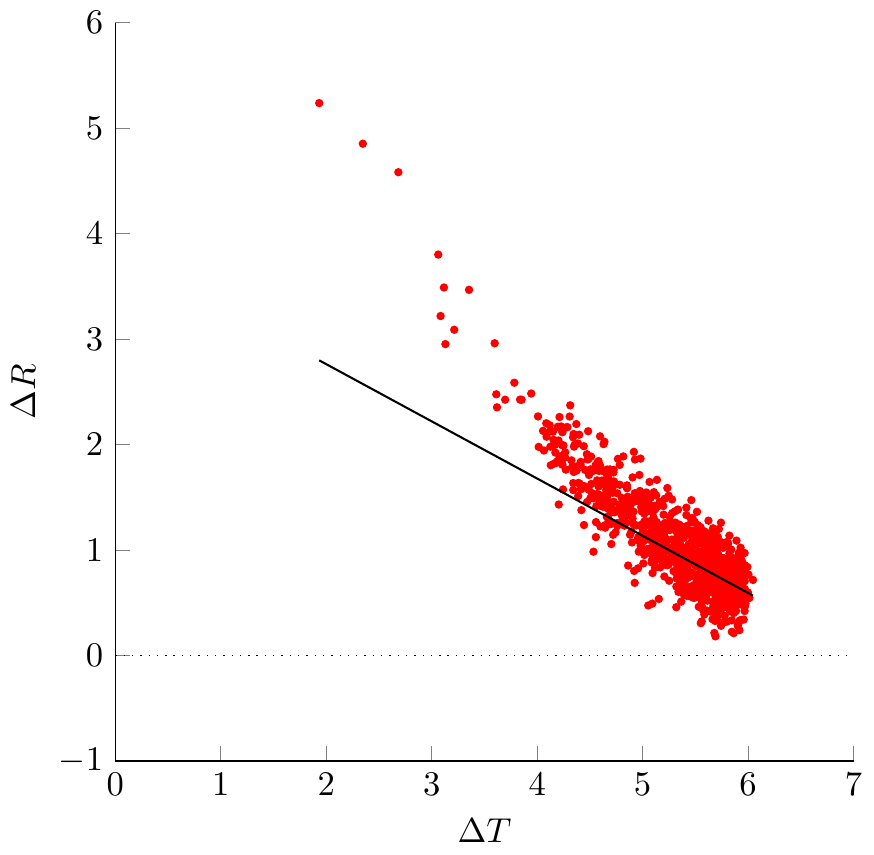}	
		\caption{}
	\end{subfigure}
	\begin{subfigure}[t]{0.32 \textwidth}
		\centering
		\includegraphics[width=\textwidth]{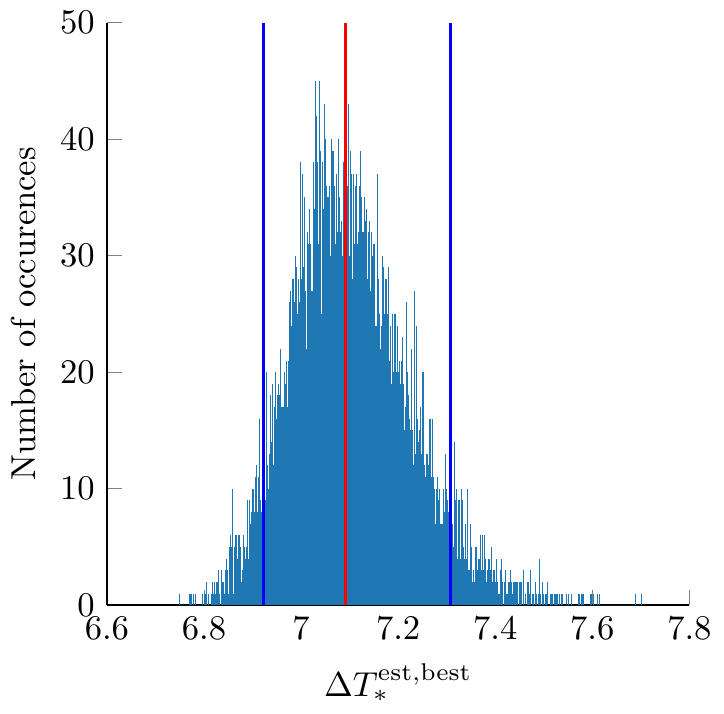}	
		\caption{}
	\end{subfigure}
	
\caption[Results for the model HadCM3L]{Results for the model HadCM3L. (a) estimated equilibrium warming $\Delta T_*^\mathrm{est}(t)$ for the first $500$ years of data. (b) remaining relative error over time. (c) estimated equilibrium warming for the whole simulation. (d) `Gregory' plot of $\Delta R$ versus $\Delta T$ including fit for the best estimate $\Delta T_*^\mathrm{est,best}$. (e) Histogram for resampling of $\Delta_*^\mathrm{est,best}$.}
\label{fig:results_for_HadCM3L}
\end{figure}

\begin{figure}
	\centering
	{\bf \huge HadGEM2-ES}
	\vspace{1em}
	
	\begin{subfigure}[t]{0.45 \textwidth}
		\centering
		\includegraphics[width=\textwidth]{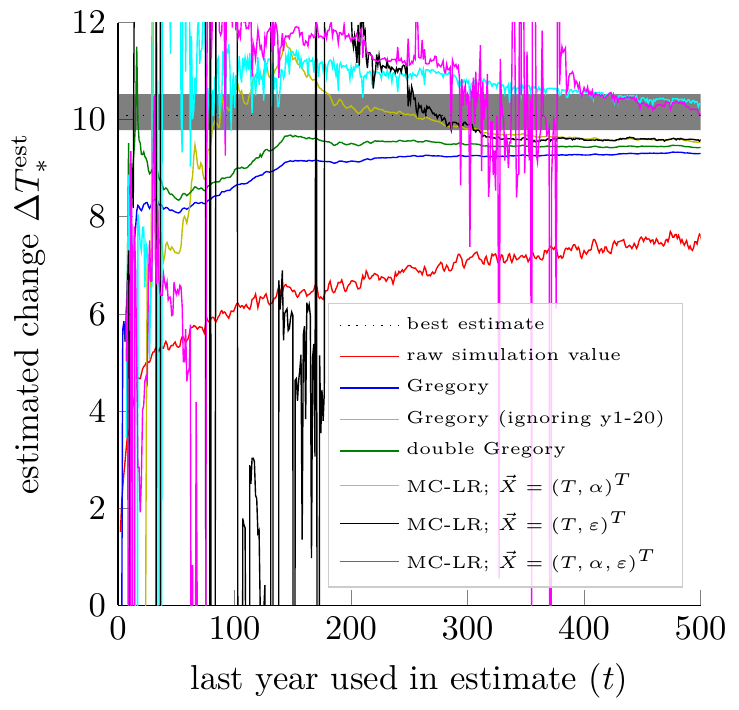}
		\caption{}
	\end{subfigure}
	~
	\begin{subfigure}[t]{0.45 \textwidth}
		\centering
		\includegraphics[width=\textwidth]{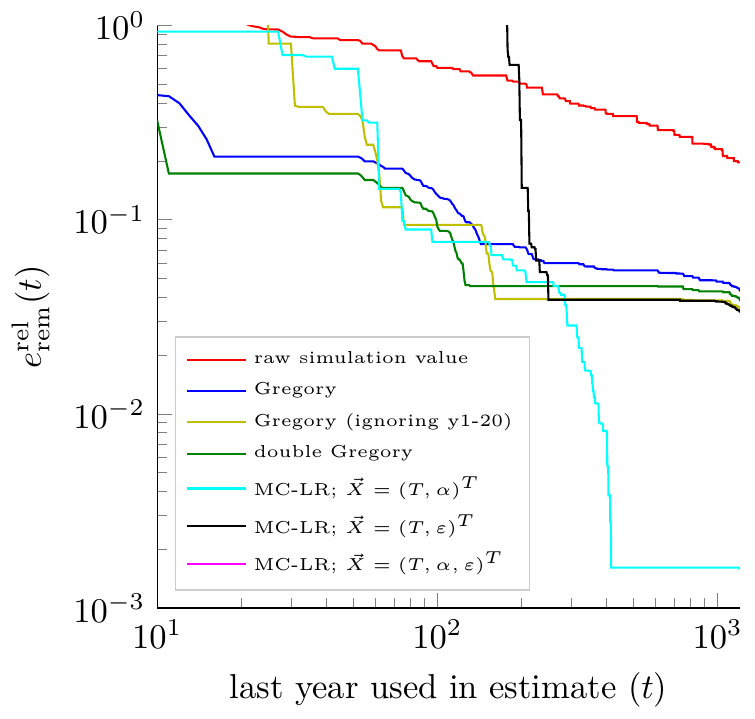}
		\caption{}
	\end{subfigure}
	\\
	\begin{subfigure}[t]{0.32 \textwidth}
		\centering
		\includegraphics[width=\textwidth]{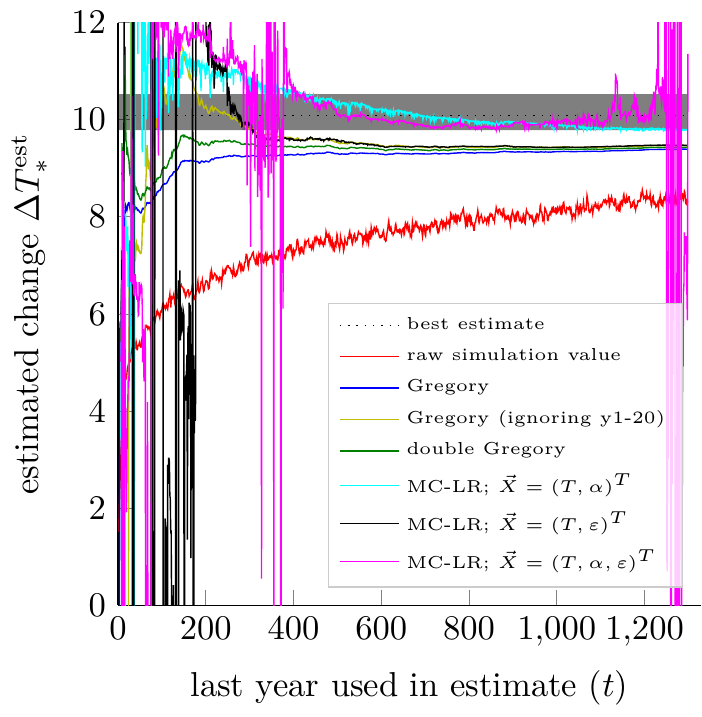}		
		\caption{}
	\end{subfigure}
	\begin{subfigure}[t]{0.32 \textwidth}
		\centering
		\includegraphics[width=\textwidth]{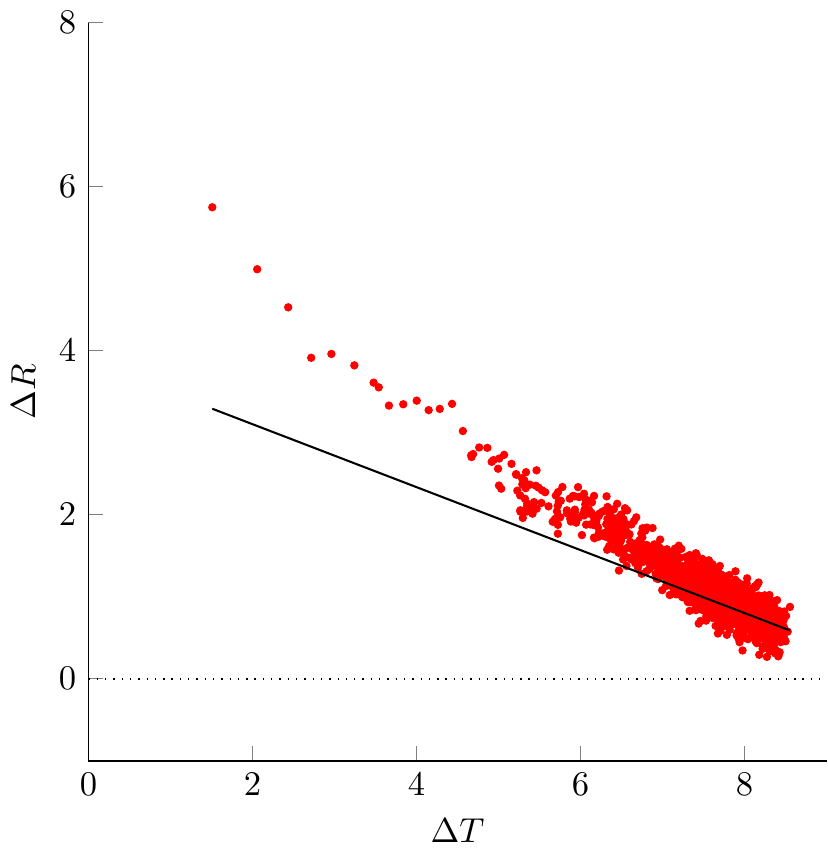}	
		\caption{}
	\end{subfigure}
	\begin{subfigure}[t]{0.32 \textwidth}
		\centering
		\includegraphics[width=\textwidth]{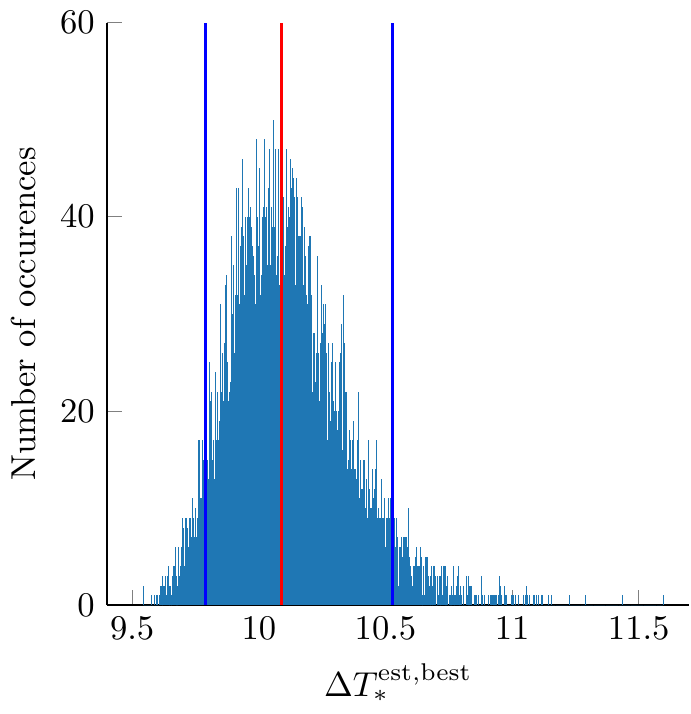}	
		\caption{}
	\end{subfigure}
	
\caption[Results for the model HadGEM2-ES]{Results for the model HadGEM2-ES. (a) estimated equilibrium warming $\Delta T_*^\mathrm{est}(t)$ for the first $500$ years of data. (b) remaining relative error over time. (c) estimated equilibrium warming for the whole simulation. (d) `Gregory' plot of $\Delta R$ versus $\Delta T$ including fit for the best estimate $\Delta T_*^\mathrm{est,best}$. (e) Histogram for resampling of $\Delta_*^\mathrm{est,best}$.}
\label{fig:results_for_HadGEM2}
\end{figure}

\begin{figure}
	\centering
	{\bf \huge IPSL-CM5A-LR}
	\vspace{1em}
	
	\begin{subfigure}[t]{0.45 \textwidth}
		\centering
		\includegraphics[width=\textwidth]{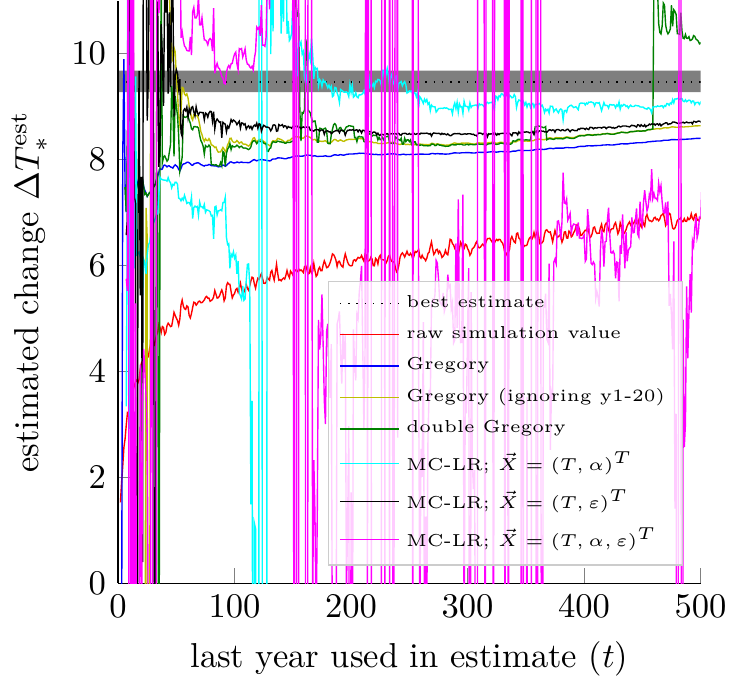}
		\caption{}
	\end{subfigure}
	~
	\begin{subfigure}[t]{0.45 \textwidth}
		\centering
		\includegraphics[width=\textwidth]{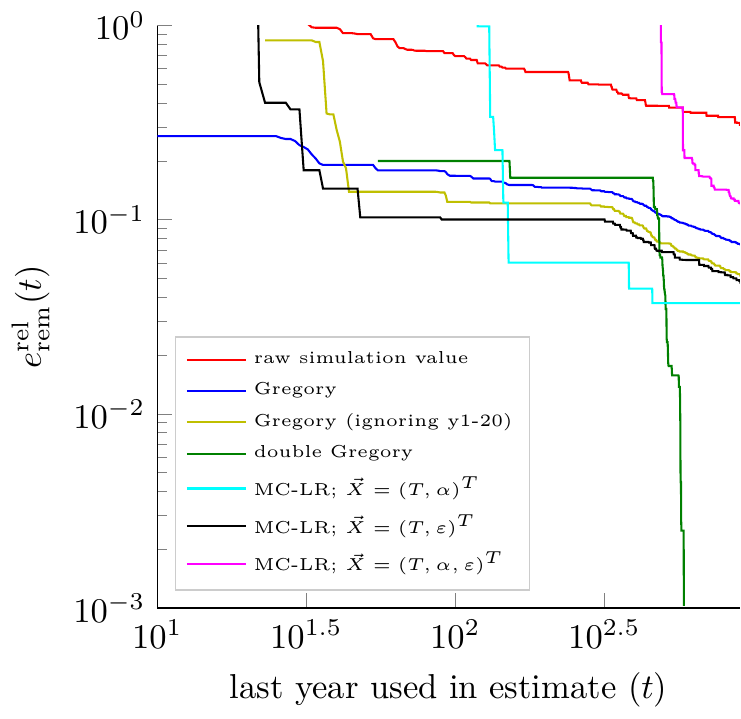}
		\caption{}
	\end{subfigure}
	\\
	\begin{subfigure}[t]{0.32 \textwidth}
		\centering
		\includegraphics[width=\textwidth]{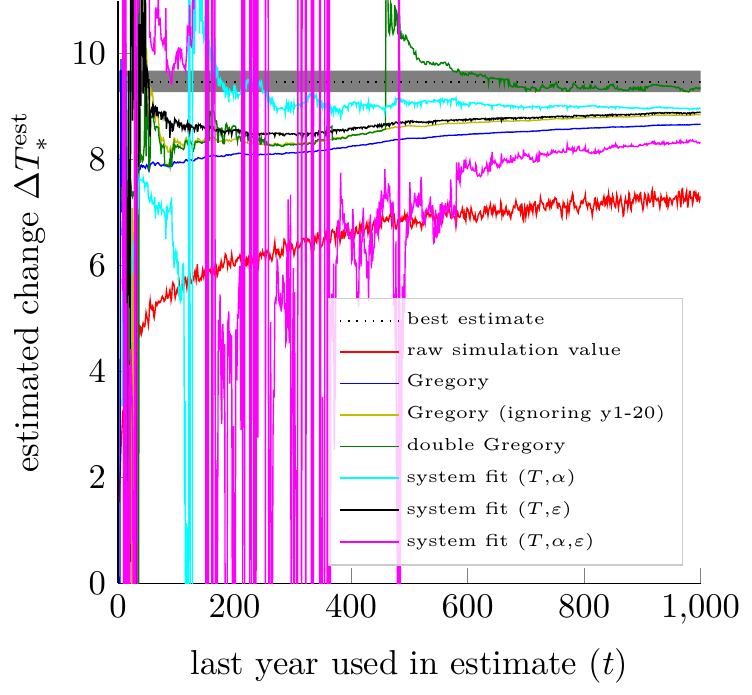}		
		\caption{}
	\end{subfigure}
	\begin{subfigure}[t]{0.32 \textwidth}
		\centering
		\includegraphics[width=\textwidth]{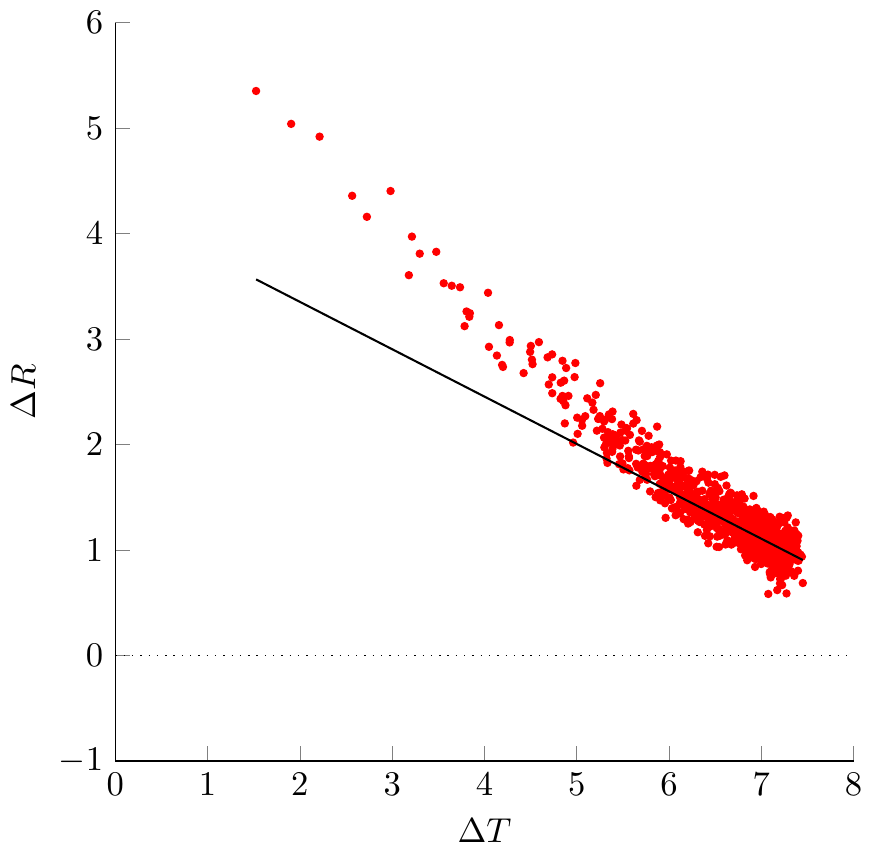}	
		\caption{}
	\end{subfigure}
	\begin{subfigure}[t]{0.32 \textwidth}
		\centering
		\includegraphics[width=\textwidth]{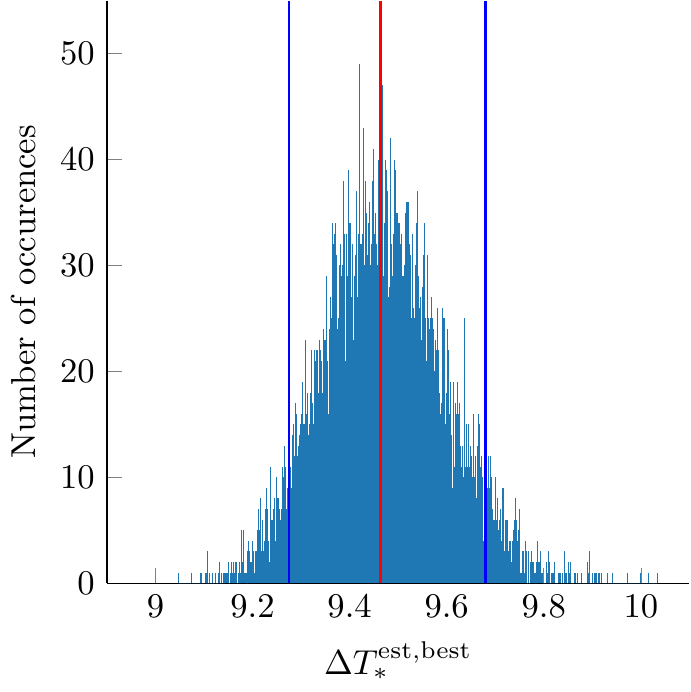}	
		\caption{}
	\end{subfigure}
	
\caption[Results for the model IPSL-CM5A-LR]{Results for the model IPSL-CM5A-LR. (a) estimated equilibrium warming $\Delta T_*^\mathrm{est}(t)$ for the first $500$ years of data. (b) remaining relative error over time. (c) estimated equilibrium warming for the whole simulation. (d) `Gregory' plot of $\Delta R$ versus $\Delta T$ including fit for the best estimate $\Delta T_*^\mathrm{est,best}$. (e) Histogram for resampling of $\Delta_*^\mathrm{est,best}$.}
\label{fig:results_for_IPSLCM5A}
\end{figure}

\begin{figure}
	\centering
	{\bf \huge MPI-ESM 1.1}
	\vspace{1em}
	
	\begin{subfigure}[t]{0.45 \textwidth}
		\centering
		\includegraphics[width=\textwidth]{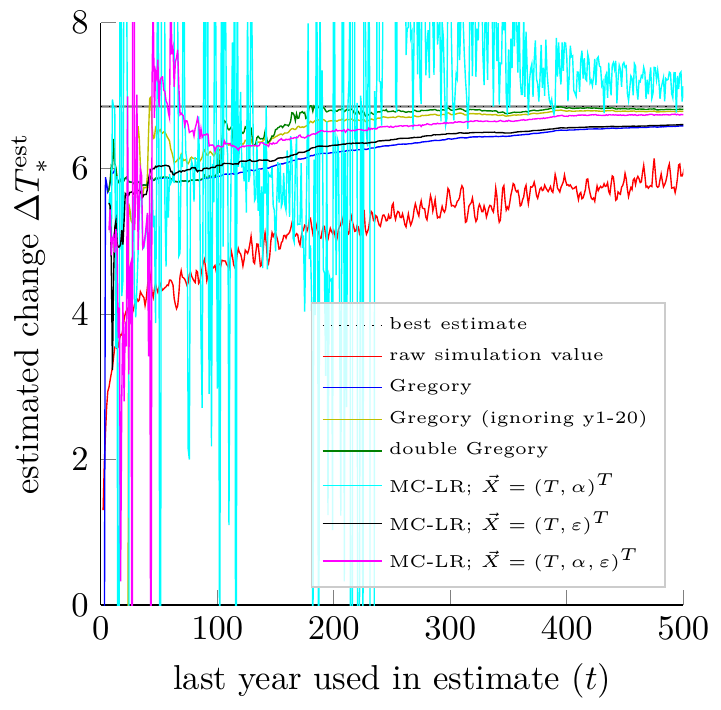}
		\caption{}
	\end{subfigure}
	~
	\begin{subfigure}[t]{0.45 \textwidth}
		\centering
		\includegraphics[width=\textwidth]{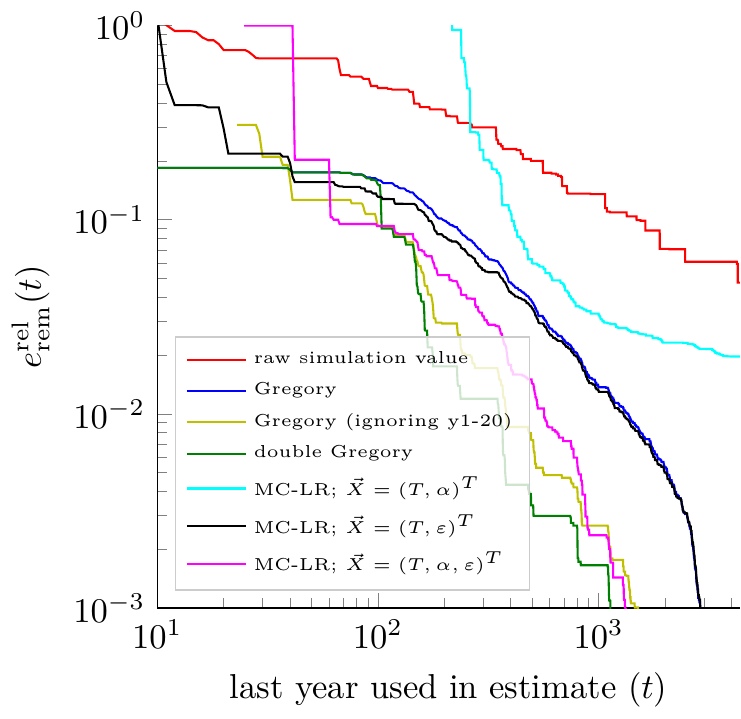}
		\caption{}
	\end{subfigure}
	\\
	\begin{subfigure}[t]{0.32 \textwidth}
		\centering
		\includegraphics[width=\textwidth]{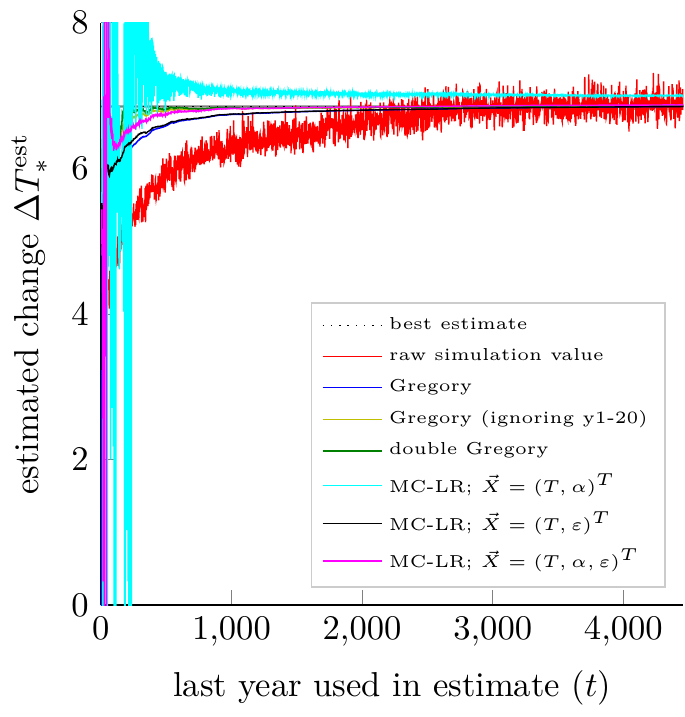}		
		\caption{}
	\end{subfigure}
	\begin{subfigure}[t]{0.32 \textwidth}
		\centering
		\includegraphics[width=\textwidth]{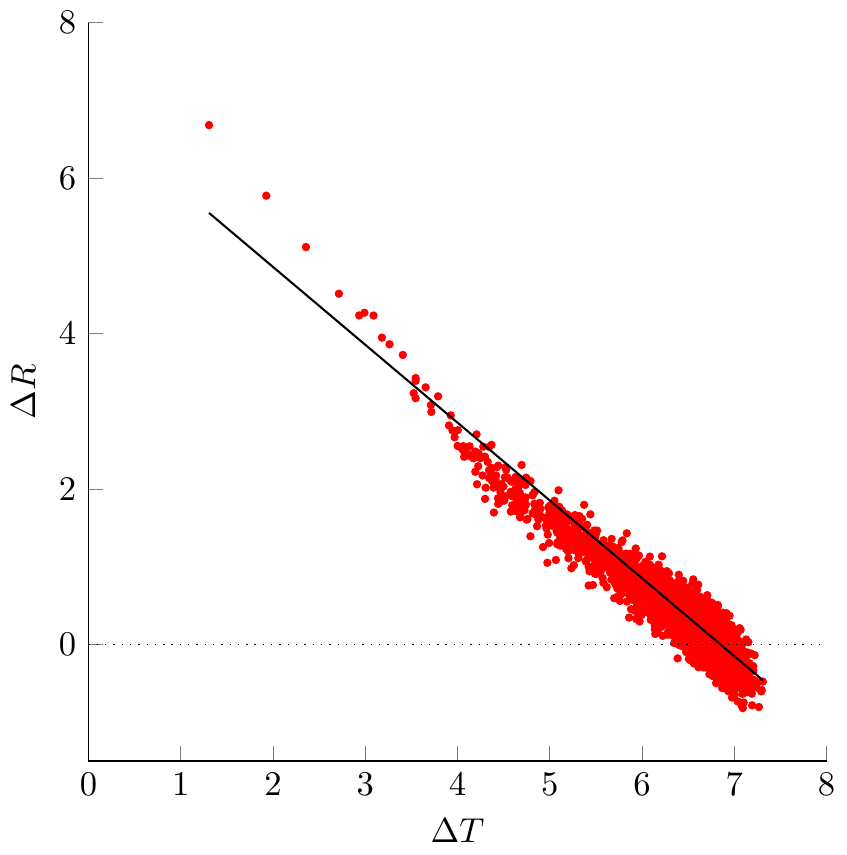}	
		\caption{}
	\end{subfigure}
	\begin{subfigure}[t]{0.32 \textwidth}
		\centering
		\includegraphics[width=\textwidth]{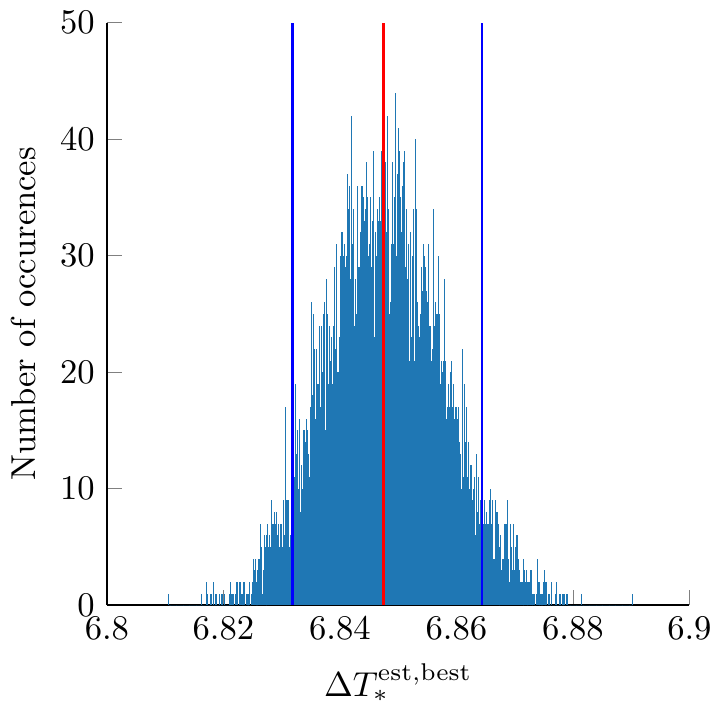}	
		\caption{}
	\end{subfigure}
	
\caption[Results for the model MPI-ESM 1.1]{Results for the model MPI-ESM 1.1. (a) estimated equilibrium warming $\Delta T_*^\mathrm{est}(t)$ for the first $500$ years of data. (b) remaining relative error over time. (c) estimated equilibrium warming for the whole simulation. (d) `Gregory' plot of $\Delta R$ versus $\Delta T$ including fit for the best estimate $\Delta T_*^\mathrm{est,best}$. (e) Histogram for resampling of $\Delta_*^\mathrm{est,best}$.}
\label{fig:results_for_MPIESM11}
\end{figure}

\begin{figure}
	\centering
	{\bf \huge MPI-ESM 1.2}
	\vspace{1em}
	
	\begin{subfigure}[t]{0.45 \textwidth}
		\centering
		\includegraphics[width=\textwidth]{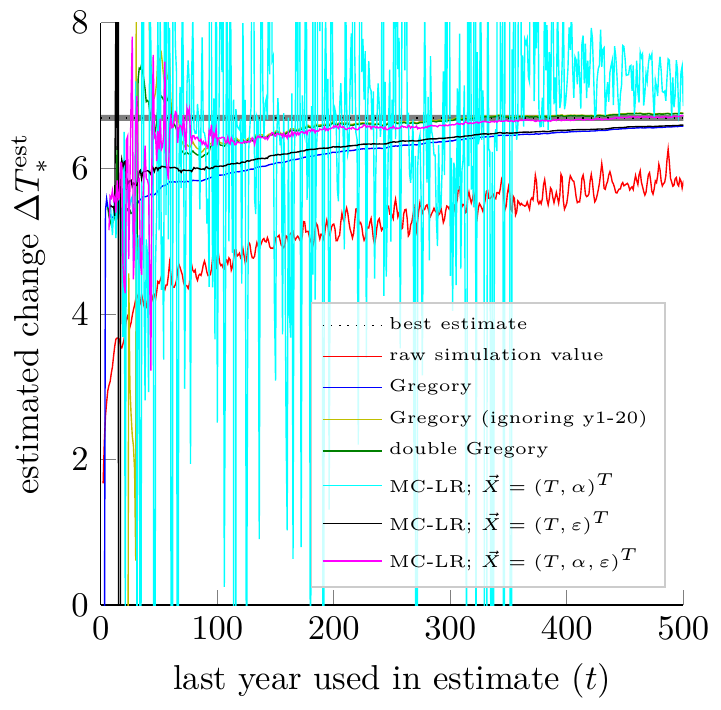}
		\caption{}
	\end{subfigure}
	~
	\begin{subfigure}[t]{0.45 \textwidth}
		\centering
		\includegraphics[width=\textwidth]{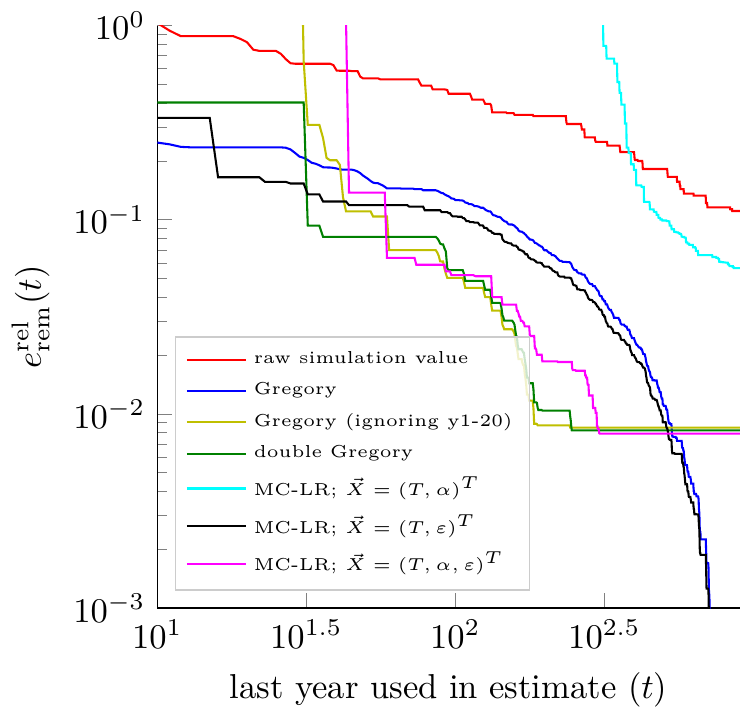}
		\caption{}
	\end{subfigure}
	\\
	\begin{subfigure}[t]{0.32 \textwidth}
		\centering
		\includegraphics[width=\textwidth]{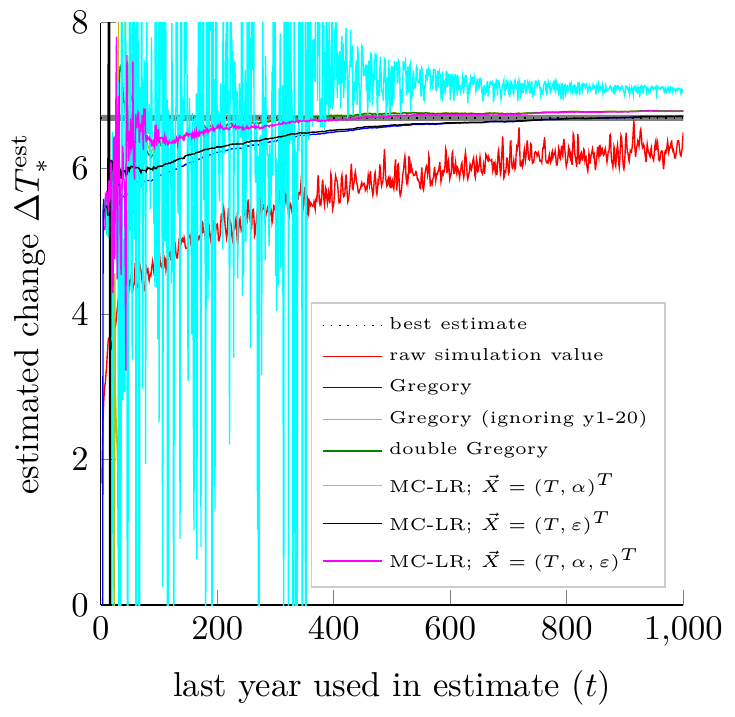}		
		\caption{}
	\end{subfigure}
	\begin{subfigure}[t]{0.32 \textwidth}
		\centering
		\includegraphics[width=\textwidth]{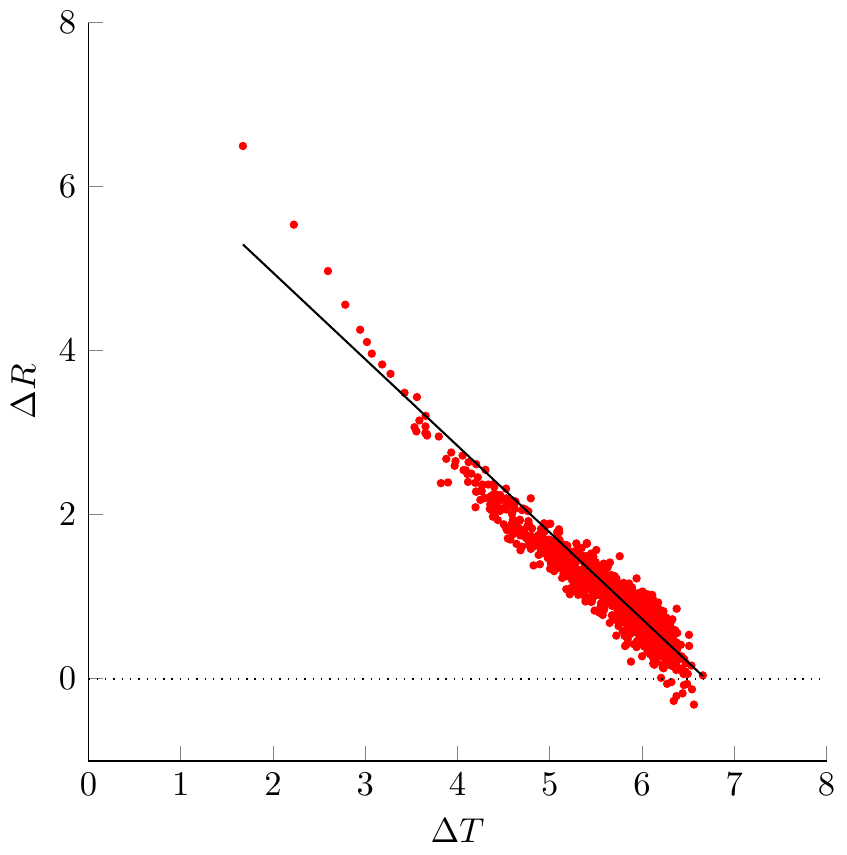}	
		\caption{}
	\end{subfigure}
	\begin{subfigure}[t]{0.32 \textwidth}
		\centering
		\includegraphics[width=\textwidth]{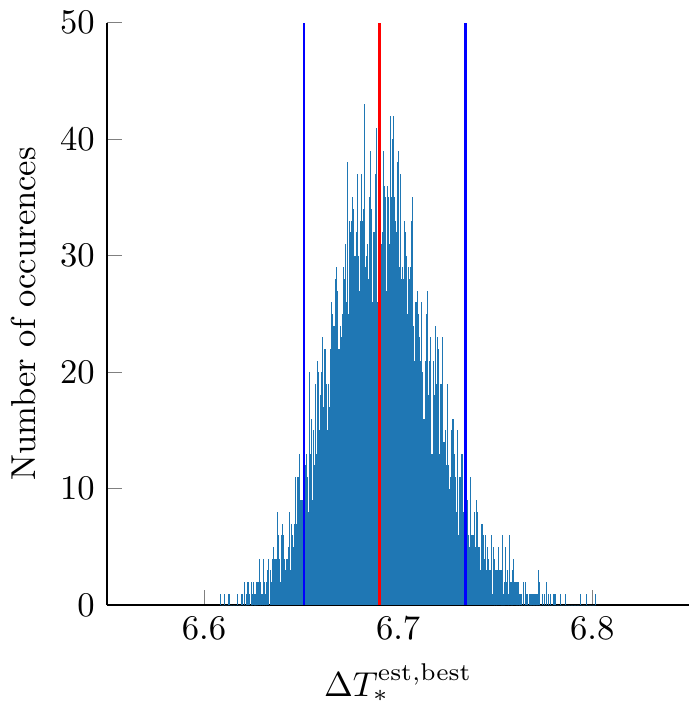}	
		\caption{}
	\end{subfigure}
	
\caption[Results for the model MPI-ESM 1.2]{Results for the model MPI-ESM 1.2. (a) estimated equilibrium warming $\Delta T_*^\mathrm{est}(t)$ for the first $500$ years of data. (b) remaining relative error over time. (c) estimated equilibrium warming for the whole simulation. (d) `Gregory' plot of $\Delta R$ versus $\Delta T$ including fit for the best estimate $\Delta T_*^\mathrm{est,best}$. (e) Histogram for resampling of $\Delta_*^\mathrm{est,best}$.}
\label{fig:results_for_MPIESM12}
\end{figure}

The results for the models in LongRunMIP indicate that there is no overall `one size fits all' best estimation technique: which technique performs best depends a lot on the model under consideration and the time period that is available. That is, it seems to depend a lot on the model and time period whether the used data does adequately contain the relevant dynamics on all of the time scales needed to make accurate estimations of the equilibrium warming. However, in general, typically one of the new multi-component linear regression models tends to work best -- especially when data for at least a few hundred years is available. In particular, the MC-LR estimate outperforms the classical Gregory methods -- as can be seen most clearly from Figure~\ref{fig:technique_comparison} and Figure 4 in the main text. In the rest of this section, the results per model are discussed in slightly more detail.

\begin{description}
	\item[CCSM3:] Although the range $\mathcal{R}_{\Delta T_*}$ is quite small -- and the resampling of the last $15\%$ of warming thus does not show much variance -- the best estimate range seems a bit on the high side; only estimates that take $\vec{X}= (\Delta T, \Delta \alpha)^T$ lie within $\mathcal{R}_{\Delta T_*}$ when using all of the available data. Nevertheless, the Gregory plot shows the plausibility of a late and slow additional warming (that is apparently not picked up completely by the other methods); a longer run of the model is necessary to get a decisive answer on this.
	
	The estimates of the MC-LR method with temperature and albedo do have many poles in years $400$ -- $700$ and overall gives higher estimates. The other techniques do have similar estimates, although the Gregory method needs more than $1,000$ years of data to get close to that value. Other methods perform similarly, and converge on a similar time frame. The MC-LR methods using emissivity (i.e. $\vec{X} = (\Delta T,\Delta \varepsilon \sigma)^T$ or $\vec{X} = (T,\Delta \alpha,\Delta \varepsilon \sigma)^T$) give estimates that are a bit higher -- closer to the range $\mathcal{R}_{\Delta T_*}$ -- and thus have slightly lower errors. Finally, the `double Gregory' and `Gregory (ignoring y1-20)' methods are almost identical from year $50$ onward, indicating that the (first) fast warming happens mostly in the first $20$ years

	\item[CESM 1.0.4:] The found range $\mathcal{R}_{\Delta T_*}$ is very specific, and lines up with estimates (although is a bit higher than most of the estimates). The MC-LR technique that uses both albedo and emissivity is best for $t > 200$ years; the double Gregory method is also quite good -- and also performs well with slightly less years of data. The conventional Gregory method converges very slowly and is still misled by the initial fast warming when using all six centuries of data. Performance of the other methods is similar to each other. The `double Gregory' method is significantly better than the Gregory method that ignores the first $20$ years of data, indicating that initial fast warming happens over a longer time period. Additionally, the double Gregory method tends to underestimate the equilibrium warming suggesting additional late curvature in the Gregory plot, or, equivalently, the presence of a warming that happens over very long time scales.
	
	\item[CNRM-CM6-1:] Spread in the (final) estimates is quite large, and even larger than the range $\mathcal{R}_{\Delta T_*}$; since some of the estimates are higher than this range, this suggest the model is not near equilibrium at the end of the run and the best estimate range might be too low. 
	
	Based on the computations, the MC-LR model with $\vec{X} = (\Delta T,\Delta \alpha,\Delta \varepsilon \sigma)^T$ performs best -- and much better than the Gregory methods. Interestingly, the `double Gregory' estimates are similar to those of the `Gregory (ignoring y1-20)' method up until $t \approx 750$ years. Then the found inflection point changes position and the estimates become similar to the MC-LR method that uses $\vec{X} = (T,\alpha)^T$. This suggest there are two inflection points in the Gregory plot, and warming dynamics happen on three distinct time scales. A fit with a triple linear function reinforces this and finds inflection points around year $5$ and year $200$ (the latter one becomes the dominant one in the double linear fit only when data from more than about $750$ years is used).
	
	\item[ECHAM5/MPIOM:] The obtained range $\mathcal{R}_{\Delta T_*}$ seems reasonable, and all estimates seem to converge to this range. The best method for the ECHAM5/MPIOM model is the MC-LR that uses both albedo and emissivity, which has much lower error values than any other technique from year $100$ onward.
	
	Strangely, for this model the Gregory method performs better when the first $20$ years are kept. This might be caused by the absence of a clear distinct inflection point in the Gregory plot in combination with the presence of a lot of datapoints that lie well below the trend line.
	
	\item[FAMOUS:] The FAMOUS run warms anomalously strong and the Gregory plot shows a strong curvature toward the end of the simulation (a discussion on why this happens can be found in section 3 of the SI of~\cite{Rugenstein2020}). This is also clear from the best estimate range $\mathcal{R}_{\Delta T_*}$ that shows a very large spread ($2.26K$) and the resampling led to outliers that can even go beyond $+20K$ equilibrium warming. None of the estimation techniques are particularly good at estimating this sort of warming. However, when data from $400$ years is used, estimates start to become sensible; from that moment onward the `double Gregory' method and a MC-LR method with $\vec{X} = (\Delta T,\Delta \alpha)^T$ are the best performing.
	
	The `double Gregory' estimates do show a jump in estimation value around year $400$ which suggests that another inflection point is detected at that moment. This is confirmed by a fit with a triple linear function that detects inflection points around year $15$ and year $150$. Especially the last linear part has a very gentle slope, indicating that this model experiences additional warming on a very long time scale and that the equilibrium temperature is high.

	\item[GISS-E2-R:] The best estimate range for GISS-E2-R is small and lines up well with estimated values and temperature time series near the end of the simulation run. The best estimate is obtained with the MC-LR technique that uses temperature and albedo when data for at least $200$ years are used (before that, the `double Gregory' and `Gregory (ignoring y1-20)' are better). The `double Gregory' and `Gregory (ignoring y1-20)' methods align very good initially, but a slight jump can be seen in the estimates by the `double Gregory' method around year $1,800$. A three linear function fit indeed finds inflection points at $t \approx 25$ years and $t \approx 700$ years -- although the slope of the final linear part is similar to the slope of the middle part, thus leading only to a slight additional warming caused by the dynamics on this time scale.
	
	\item[HadCM3L:] The HadCM3L run is relatively short, and the model is not fully equilibrated at the end of the run. Hence estimates are too low, even when all data is used; taking that into account, the best estimate range $\mathcal{R}_{\Delta T_*}$ seems plausible. Up to $t \approx 200$ years, the Gregory method that ignores the first $20$ years of data performs best; if more years of data are available, the best performing estimation technique is a MC-LR method with $\vec{X}= (\Delta T, \Delta \alpha)^T$.
	
	The `double Gregory' method only detects an inflection point when at least $700$ years of data are used (before that, a single linear is the best fit -- that is, the `Gregory' method); at that moment the estimate jumps to slightly above the `Gregory (ignoring y1-20)' method, since the inflection point found is around $t = 35$ years (and hence the `Gregory (ignoring y1-20)' method overestimates the final slope of the Gregory plot).

	\item[HadGEM2-ES:] The range $\mathcal{R}_{\Delta T_*}$ seems a bit high compared to the final estimates, but -- also given the relative short run length -- still plausible. Up to $t \approx 250$ years, the methods `double Gregory' and `Gregory (ignoring y1-20)' have the lowest errors; when more years of data are used, the new MC-LR method with $\vec{X} = (\Delta T,\Delta \alpha)^T$ is best. Noteworthy, estimates made by a fit with $\vec{X} = (\Delta T,\Delta \varepsilon \sigma)^T$ and by the Gregory method that ignores the first $20$ years of data are almost identical from $t = 200$ years onward. Finally, the initial estimates, that use few years, are much higher than the later estimates; this is possibly due to the fact that the initial datapoints in the Gregory plot tend to lie above the long-term linear trend -- hence resulting in too gentle slopes when a limited amount of data is used.	
	
	\item[IPSL-CM5A-LR:] Because of the short run, the model IPSL-CM5A-LR is not equilibrated at the end of the run. This led to problems in finding a good and plausible best fit range (see section \ref{sec:IPSLCM5ALR-best_fit_exception}). The final estimated values also show a large spread, and it indeed seems that the true equilibrium warming of the model is high -- especially considering the late curvature in the Gregory plot, that has been mentioned before in~\cite{Rugenstein2020}.
	
	For times $t < 200$ years, the lowest error is obtained with a MC-LR method that uses emissivity. Then, for larger times $t < 500$ years, the lowest error is obtained with fits that use the albedo; for $t > 500$ years, the `double Gregory' method produces the lowest error values. The estimated value for the `double Gregory' method has a jump around $t = 425$ years; at that moment another inflection point is found. In fact, a fit of a triple linear function indicates a double linear fit is actually best for the whole dataset, and the inflection point is located around $t = 225$ years.
	
	\item[MPI-ESM 1.1:] The model seems well equilibrated at the end of the run. The range $\mathcal{R}_{\Delta T_*}$ is quite specific and lines up well with final estimates. Based on the computed error values, the best method for the MPI-ESM 1.1 model is the `double Gregory' method, closely followed by the `Gregory (ignoring y1-20)' method and the MC-LR method that uses both albedo and emissivity. A fit with a component-wise linear function indicate the complete Gregory plot is best fitted with a double linear function that has an inflection point around $t = 30$ years (thus explaining the slight underestimation of the Gregory method that ignores only the first $20$ years).
	
	\item[MPI-ESM 1.2:] The best estimate range $\mathcal{R}_{\Delta T_*}$ seems a bit too low based on the final estimates (and thus errors for methods that tend to underestimate are too low for high values of $t$). Nevertheless, based on the error computations, for the MPI-ESM 1.2 model, the best estimation methods are the `double Gregory' and `Gregory (ignoring y1-20)' methods -- that both are almost identical, as the only inflection point in the Gregory plot is located around $t = 20$ years. These methods are closely followed by a MC-LR method with $\vec{X} = (\Delta T,\Delta \alpha,\Delta \varepsilon \sigma)^T$, that has similar, but slightly larger error values.
\end{description}

\FloatBarrier

\bibliographystyle{amnatnat}
\bibliography{../sources}